\documentclass[a4paper,11pt]{article}
\pdfoutput=1
\usepackage{jheppub}
\usepackage[T1]{fontenc}
\usepackage{amsmath,amssymb,amsfonts,cases}
\usepackage{amsthm}
\usepackage{enumitem}
\usepackage{graphicx,caption,subcaption}
\usepackage[T1]{fontenc} 
\usepackage{enumitem}  
\usepackage{empheq}
\usepackage{dsfont}
\usepackage{hyperref}
\usepackage[mathscr]{euscript}
\usepackage{dcolumn}
\usepackage{bm}
\usepackage{bbm}
\usepackage{slashed}
\usepackage{wrapfig}
\usepackage{mathtools}
\usepackage[utf8]{inputenc}
\usepackage{sectsty}
\allsectionsfont{\boldmath}

\newcommand{\beq}{\begin{equation}}
\newcommand{\eeq}{\end{equation}}
\newcommand{\bea}{\begin{eqnarray}}
\newcommand{\eea}{\end{eqnarray}}
\newcommand{\bi}{\begin{itemize}}
	\newcommand{\ei}{\end{itemize}}
\newcommand{\ben}{\begin{enumerate}}
	\newcommand{\een}{\end{enumerate}}

\def\Tr{{\rm Tr}}

\renewcommand{\d}{\partial}
\newcommand{\N}{{\mathcal N}}

\renewcommand{\a}{\alpha}

\renewcommand{\d}{\delta}
\newcommand{\e}{\epsilon}

\renewcommand{\k}{\kappa}

\newcommand{\m}{\mu}

\renewcommand{\r}{\rho}

\renewcommand{\t}{\theta}
\newcommand{\z}{\zeta}
\newcommand{\p}{\partial}

\def\a{\alpha}

\def\d{\delta}

\def\h{\eta}

\def\k{\kappa}             

\def\m{\mu}

\def\r{\rho}                                    
                                   
\def\t{\tau}

\def\z{\zeta}

\def\F{\Phi}

\title{\Large Holographic Entanglement Entropy of the Coulomb Branch}
\author[1]{Adam Chalabi,}
\author[2]{S. Prem Kumar,}
\author[1]{Andy O'Bannon,}
\author[3]{Anton Pribytok,}
\author[4]{Ronnie Rodgers,}
\author[1]{Jacopo Sisti}
\affiliation[1]{STAG Research Centre, Physics and Astronomy, University of Southampton, Highfield, Southampton SO17 1BJ, United Kingdom}
\affiliation[2]{Department of Physics, Swansea University, Swansea, SA2 8PP, United Kingdom}
\affiliation[3]{School of Mathematics \& Hamilton Mathematics Institute, Trinity College Dublin, Dublin, Ireland}
\affiliation[4]{Institute  for  Theoretical  Physics,  Utrecht  University, 3584  CC  Utrecht,  The  Netherlands}
\emailAdd{a.chalabi@soton.ac.uk}
\emailAdd{s.p.kumar@swansea.ac.uk}
\emailAdd{a.obannon@soton.ac.uk}
\emailAdd{apribytok@maths.tcd.ie}
\emailAdd{r.j.rodgers@uu.nl}
\emailAdd{J.Sisti@soton.ac.uk}
\abstract{We compute entanglement entropy (EE) of a spherical region in $(3+1)$-dimensional $\mathcal{N}=4$ supersymmetric $SU(N)$ Yang-Mills theory in states described holographically by probe D3-branes in $AdS_5 \times S^5$. We do so by generalising methods for computing EE from a probe brane action without having to determine the probe's back-reaction. On the Coulomb branch with $SU(N)$ broken to $SU(N-1)\times U(1)$, we find the EE monotonically decreases as the sphere's radius increases, consistent with the $a$-theorem. The EE of a symmetric-representation Wilson line screened in $SU(N-1)$ also monotonically decreases, although no known physical principle requires this. A spherical soliton separating $SU(N)$ inside from $SU(N-1)\times U(1)$ outside had been proposed to model an extremal black hole. However, we find the EE of a sphere at the soliton's radius does not scale with the surface area. For both the screened Wilson line and soliton, the EE at large radius is described by a position-dependent W-boson mass as a short-distance cutoff. Our holographic results for EE and one-point functions of the Lagrangian and stress-energy tensor show that at large distance the soliton looks like a Wilson line in a direct product of fundamental representations.}


\begin{document}
	\maketitle
	\flushbottom

\newpage
\section{Introduction}
\label{sec:intro}

Quantum entanglement is of fundamental importance, being a manifestation of purely quantum correlations. For a bipartite system in a pure state, the amount of entanglement between the two complementary subspaces may be quantified by entanglement entropy (EE), defined as the von Neumann entropy of the reduced density matrix of one of the subspaces~\cite{Bombelli:1986rw, Bennett:1995tk}. EE plays a prominent role in disparate areas of physics. For instance, in quantum field theory (QFT) the EE of a spatial subregion has ultraviolet (UV) divergences due to correlations across the subregion's boundary. The leading divergence is proportional to the subregions' surface area~\cite{Srednicki:1993im}. This area law received much attention in attempts to understand the Bekenstein-Hawking entropy of black holes. EE also plays a prominent role in condensed matter physics~\cite{Amico:2007ag, Eisert:2008ur,specialissue}, for example as a probe of quantum phase transitions~\cite{Vidal:2002rm}.

Besides the leading area law term, EE contains subleading divergences and finite contributions that can provide important information~\cite{Callan:1994py,Holzhey:1994we,Solodukhin:1994yz,Solodukhin:1994st,Fursaev:1995ef}. For example, in $(1+1)$-dimensional critical systems the EE of an interval has a logarithmic violation of the area law driven by the central charge~\cite{Holzhey:1994we, Calabrese:2004eu, Calabrese:2009qy}. This feature makes the EE a good quantity from which to infer the universality class of spin chains from numerical computations~\cite{Peschel:2004qbn,Its_2005,Sugino_2018}.

More generally, a logarithmic divergence arises in the EE of any even-dimensional conformal field theory (CFT). Its coefficient, which is universal and regularisation independent, turns out to be a linear combination of the CFT's central charges. This opened a new perspective on monotonicity theorems, such as the well-known $c$-theorem in $1+1$ dimensions~\cite{Zamolodchikov:1986gt}. Indeed, ref.~\cite{Casini:2004bw} proved an entropic version of the $c$-theorem: in any Lorentz invariant QFT describing a renormalisation group (RG) flow from a UV CFT to an infrared (IR) CFT, a certain $c$-function, defined from a spatial interval's EE, decreases monotonically with the interval's length, and agrees with the central charge in the limits of zero length (the UV) and infinite length (the IR). In higher dimensions EE has been used to prove similar monotonicity theorems, such as the $(2+1)$-dimensional $F$-theorem~\cite{Jafferis:2011zi,Casini:2012ei} and the $(3+1)$-dimensional $a$-theorem~\cite{Cardy:1988cwa,Osborn:1989td,Jack:1990eb,Komargodski:2011vj,Solodukhin:2013yha,Casini:2017vbe}.

While string theory has provided methods to compute black hole entropy~\cite{Strominger:1996sh}, the anti-de Sitter/conformal field theory (AdS/CFT) correspondence~\cite{Maldacena,Witten:1998qj,Gubser:1998bc,Aharony:1999ti}, otherwise known as holography, renewed interest in EE in the high-energy physics community. For a gauge theory at large $N$ and strong coupling with a holographic dual, Ryu and Takayanagi (RT) proposed that the EE of a spatial region is proportional to the area of the minimal surface extending in the holographic direction and anchored to the entangling surface at the asymptotic boundary~\cite{Ryu:2006bv,Ryu:2006ef}. Originally a conjecture, this prescription was later proved by Lewkowycz and Maldacena~\cite{Lewkowycz:2013nqa}. The RT formula and its covariant generalisation~\cite{Hubeny:2007xt,Dong:2016hjy} have turned out to be useful for many reasons, including two in particular.

First, holographic EE (HEE) provides a powerful tool to quantify entanglement in strongly-coupled QFTs. In general EE is difficult to compute directly in QFT, even for free theories.\footnote{An exception is $(1+1)$-dimensional CFT, where the Virasoro symmetry provides a systematic approach to computing EE~\cite{Holzhey:1994we, Calabrese:2009qy}.} On the other hand, HEE is relatively straightforward to calculate. Second, the RT formula suggests a deep connection between entanglement and gravity. For example, refs.~\cite{Swingle:2009bg,VanRaamsdonk:2009ar,VanRaamsdonk:2010pw,Maldacena:2013xja} argued that the geometry of an asymptotically AdS space-time should be related to the entanglement structure of the dual QFT's quantum state.

In this paper we use HEE to study RG flows described holographically by probe branes in AdS space-time. In the holographic framework, branes in the gravity theory can describe fields, states, and objects ranging from fields in the fundamental representation of the gauge group (i.e. flavour fields)~\cite{Karch:2002sh}, to baryons~\cite{Witten:1998xy}, to Wilson lines~\cite{Maldacena:1998im,Rey:1998bq,Drukker:2005kx, Hartnoll:2006hr, Yamaguchi:2006tq, Hartnoll:2006is, Gomis:2006im}, and more. Broadly speaking, a brane that reaches the AdS boundary is dual to fields added to the QFT, while a brane that does not reach the AdS boundary describes a state of the QFT.

We employ a common simplification, namely the probe limit, in which the brane's back-reaction on the metric and other bulk fields is neglected. In the dual field theory, the probe limit corresponds to a probe sector with a number of degrees of freedom parametrically smaller in $N$ than the adjoint-representation fields' order $N^2$ degrees of freedom.

How can we calculate HEE from a probe brane? More precisely, how can we holographically compute the leading contribution in $N$ to the EE from the probe sector? Given that the RT approach depends only on the metric, the obvious answer is to compute the brane's linearised back-reaction on the metric, and from that, the resulting change in the minimal surface's area. Although streamlined methods for doing this have been developed in certain cases~\cite{Chang:2013mca,Jones:2015twa}, in general this remains a difficult problem, especially for branes that break symmetries and hence may make the back-reaction complicated.

We will instead use Karch and Uhlemann's method for computing the leading order contribution of a probe brane to HEE directly from a probe brane's action, without computing back-reaction~\cite{Karch:2014ufa}. Their method generalises Lewkowycz and Maldacena's method for proving the RT formula to probe branes (see ref.~\cite{Jensen:2013lxa} for precursor work in conformal cases). Refs.~\cite{Vaganov:2015vpq,Kumar:2017vjv} and~\cite{Rodgers:2018mvq} used the Karch-Uhlemann method to obtain the HEE of non-conformal D-brane and M-brane solutions, respectively.

\begin{figure}
	\begin{center}
		\includegraphics[scale=0.8]{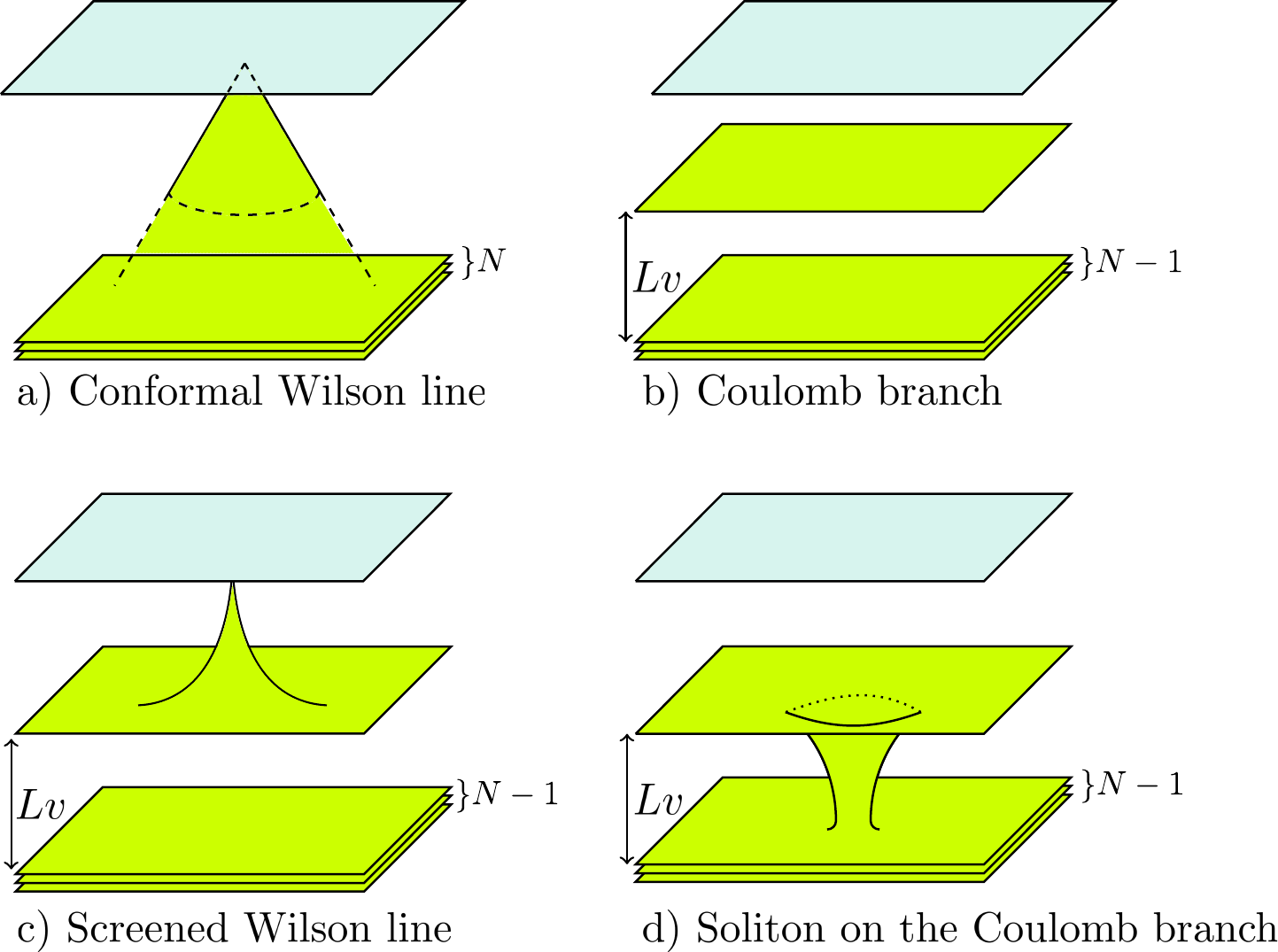}
		\caption{Cartoons of the D3-brane solutions discussed in the paper. The green surfaces depict the D3-branes, while the pale blue planes represent the conformal boundary.} 
		\label{fig:cartoons}
	\end{center}
\end{figure}

In this paper, we use the Karch-Uhlemann method to compute the contribution to HEE from various probe D3-branes in the $AdS_5\times S^5$ background of type IIB supergravity. The holographically dual CFT is $(3+1)$-dimensional \(\mathcal{N}=4\) supersymmetric Yang-Mills (SYM) theory with gauge group \(SU(N)\), with large \(N\) and large 't Hooft coupling, \(\lambda\)~\cite{Maldacena}.

This theory can be realised as the worldvolume theory of a stack of $N$ D3-branes in $\mathbb{R}^{1,9}$. The theory has a Coulomb branch of supersymmetric vacua in which the adjoint scalar superpartners of the gluons acquire non-zero vacuum expectation values (VEVs) and Higgs $SU(N)$ to a subgroup. This corresponds to separating the D3-branes from each other in transverse directions. Separations between D3-branes set the lengths of strings stretched between them, and hence the masses of the W-bosons and their superpartners.

We will focus on the situation with precisely {\em one} D3-brane separated from the stack, thus Higgsing $SU(N)$ to $SU(N-1)\times U(1)$. In the holographic limits $N \to \infty$ and $\lambda \to \infty $ we replace the stack of D3-branes with their near-horizon geometry, $AdS_5 \times S^5$, and treat the single D3-brane as a probe describing an RG flow from $SU(N)$ $\N=4$ SYM to $SU(N-1)\times U(1)$ $\N=4$ SYM. We will also consider D3-branes that describe supersymmetric Wilson lines and spherical solitons in this vacuum.

Cartoons of the relevant D3-brane solutions are shown in figure~\ref{fig:cartoons}, with the $AdS_5$ boundary at the top and the Poincar\'e horizon at the bottom. In the figure we depict the $N$ D3-branes generating $AdS_5 \times S^5$, but of course these D3-branes are not actually present in $AdS_5 \times S^5$, having ``dissolved'' into five-form flux. We show them simply for intuition.

Figure~\ref{fig:cartoons}a depicts the D3-brane holographically dual to a straight Wilson line in the $k$-th rank symmetric representation of $SU(N)$, where the D3-brane carries $k$ units of string charge~\cite{Drukker:2005kx,Gomis:2006sb,Hartnoll:2006is,Gomis:2006im}. We show this solution at the origin of the Coulomb branch, where all $N$ D3-branes generating $AdS_5 \times S^5$ remain coincident. This solution is 1/2-BPS and preserves defect conformal symmetry, so we call it the (conformal) Wilson line D3-brane.

Figure~\ref{fig:cartoons}b depicts the probe D3-brane dual to a point on the Coulomb branch where \(SU(N) \to SU(N-1) \times U(1)\)~\cite{Klebanov:1999tb}. This D3-brane is parallel to the other $N-1$ D3-branes, but is separated from them by a distance $Lv$ in the holographic direction, where $L$ is the $AdS_5$ radius and $v$ sets the VEV of a single adjoint scalar field. If we descend into $AdS_5$ from the boundary, then, when we cross this D3-brane, the five-form flux drops from $N$ to $N-1$. Recalling that the holographic coordinate is dual to the RG scale, with the regions near the boundary and Poincar\'e horizon dual to the UV and IR, respectively, this probe D3-brane clearly describes an RG flow in which \(SU(N) \to SU(N-1) \times U(1)\). We call this the Coulomb branch D3-brane.

Figure~\ref{fig:cartoons}c depicts the D3-brane dual to the Wilson line on the Coulomb branch, now screened by the adjoint scalar VEV. In particular, as we descend into $AdS_5$ this D3-brane initially resembles the Wilson line D3-brane of figure~\ref{fig:cartoons}a, including the $k$ units of string charge, but then interpolates smoothly to the Coulomb branch D3-brane of figure~\ref{fig:cartoons}b. Crucially, this D3-brane is not present below $Lv$. The dual CFT interpretation is that in this Coulomb branch vacuum the Wilson line present in the UV is absent in the IR, that is, the Wilson line has been screened by the adjoint scalar VEV~\cite{Kumar:2016jxy,Kumar:2017vjv,Evans:2019pcs}. This D3-brane is 1/2-BPS but not conformal. We call it the screened Wilson line D3-brane.

Figure~\ref{fig:cartoons}d depicts the D3-brane dual to an excited state on the Coulomb branch, namely a spherically symmetric soliton~\cite{Ghoroku:1999bc,Gauntlett:1999xz,deMelloKoch:1999ui,Schwarz:2014rxa}, interpreted in refs.~\cite{Schwarz:2014rxa,Schwarz:2014zsa} as a phase bubble, or domain wall, separating $SU(N)$ inside from $SU(N-1) \times U(1)$ outside. Indeed, this D3-brane is essentially a Coulomb branch D3-brane with a cylindrical ``spike'' that reaches the Poincar\'e horizon with non-zero radius. This D3-brane carries $k$ units of string charge, dual to $k$ units of $U(1)$ charge uniformly spread over the domain wall. This D3-brane is 1/2-BPS but not conformal: the soliton's mass is $\propto v$. We call it the spherical soliton D3-brane.

In this paper we calculate the contribution of each probe D3-brane mentioned above to the EE of a spherical region of radius $R$, centred on the Wilson line or spherical soliton, using the Karch-Uhlemann method. More precisely, we calculate the change in EE from the case with no probe D3-brane. This difference has no UV divergences.

For all cases except the conformal Wilson line, our results are novel, and revealing. Moreover, they provide a crucial lesson about the Karch-Uhlemann method: for non-conformal probe branes, this method requires a careful accounting for a boundary contribution on the brane worldvolume. Although this boundary term vanishes for the conformal Wilson line, it is non-zero for all our non-conformal probe branes. This boundary term has been neglected in all previous applications of the Karch-Uhlemann method~\cite{Karch:2014ufa,Vaganov:2015vpq,Kumar:2017vjv,Rodgers:2018mvq}. 

The contribution to EE from the conformal, symmetric-representation Wilson line, $S_\mathrm{symm}$, was computed using conformal symmetry and supersymmetric localisation in ref.~\cite{Lewkowycz:2013laa}, and using the Karch-Uhlemann method in ref.~\cite{Kumar:2017vjv}. We reproduce those results.

For the contribution to EE from the Coulomb branch D3-brane, $S_\mathrm{Coul}$, we find a new, analytical (i.e. non-numerical) result. As a non-trivial check, we reproduce our $S_\mathrm{Coul}$ using the RT formula in the fully back-reacted solution for the Coulomb branch. We further show that our $S_\mathrm{Coul}$ obeys the entropic $a$-theorem~\cite{Osborn:1989td,Jack:1990eb,Komargodski:2011vj,Solodukhin:2013yha,Casini:2017vbe} and also the entropic ``area law,'' which states that the coefficient of the area law term must decrease along the RG flow~\cite{Casini:2016udt,Casini:2017vbe}.

We compute the contribution to EE from the screened Wilson line D3-brane, $S_\mathrm{screen}$, numerically. When $R \to 0$ we find $S_\mathrm{screen} \to S_\mathrm{symm}$, as expected, since these two cases coincide in the UV. As $R$ increases, we find that $S_\mathrm{screen}$ decreases monotonically. We know of no physical principle that requires such behaviour. Indeed, this case involves an RG flow of the bulk QFT, triggered by the scalar VEV $\propto v$, which in turn triggers an RG flow on the Wilson line, whose degrees of freedom couple to the scalar such that the non-zero VEV acts as a mass term~\cite{Maldacena:1998im,Gomis:2006sb}. No monotonicity theorem is known for such a situation.

We compute the contribution to EE from the spherical soliton D3-brane, $S_\mathrm{soliton}$, numerically. the result is not monotonic in $R$. When $R \to 0$ we find $S_\mathrm{soliton}\to 0$, as expected. As we increase $R$ we find a maximum near the domain wall, after which $S_\mathrm{soliton}$ decreases.

Curiously, the spherical soliton's mass and charge are both proportional to its radius. In refs.~\cite{Schwarz:2014rxa,Schwarz:2014zsa} Schwarz observed that an asymptotically flat extremal Reissner-Nordstr\"om black hole with sufficiently large charge shares this property, raising the question of whether the soliton reproduces any other features of extremal Reissner-Nordstr\"om. Indeed, in ref.~\cite{Kumar:2020hif}, some of us showed that the spherical soliton supports a spectrum of quasi-normal modes with both qualitative and quantitative similarities to those of an extremal black hole. In this paper we address one of Schwarz's key questions, namely whether at large charge the EE of a sphere coincident with the soliton scales with surface area (after suitable regularisation), similar to a black hole's Bekenstein-Hawking entropy. We find numerically that this EE scales not with surface area, but with a power of the soliton's radius $\approx 1.3$

In the IR limit $R \to \infty$ we expect both $S_\mathrm{screen}$ and $S_\mathrm{soliton}$ to approach $S_\mathrm{Coul}$, since in that limit all the dual D3-branes look like the Coulomb branch D3-brane.  In the large-$R$ asymptotics of $S_\mathrm{screen}$ and $S_\mathrm{soliton}$ we indeed find $S_\mathrm{Coul}$, however we also find other contributions, including a term that grows as $R$, as well as $R$-independent constants. Remarkably, we found a simple and intuitive way to reproduce these terms, as follows.

In figures~\ref{fig:cartoons}b, c and d, a W-boson is dual to a string stretched between the probe D3-brane and the Poincar\'e horizon. In particular, the W-boson mass is the length of such a string times its tension. In $SU(N-1)\times U(1)$ $\N=4$ SYM, the W-boson mass acts as a UV cutoff. Indeed, our result for $S_\mathrm{Coul}$ resembles that of a CFT with this UV cutoff. In figures~\ref{fig:cartoons}c and d the W-boson clearly acquires a position-dependent mass, hence this UV cutoff becomes position dependent. Remarkably, for $S_\mathrm{screen}$ and $S_\mathrm{soliton}$ in the $R \to \infty$ limit, we find that taking $S_\mathrm{Coul}$ and replacing the constant cutoff with the position-dependent cutoff reproduces our numerical results for $S_\mathrm{screen}$ and $S_\mathrm{soliton}$ up to and including order $1/R$. In particular, in $S_\mathrm{soliton}$ this reproduces the term $\propto R$ and an $R$-independent term linear in $k$. In a large-$v$, large-$k$ limit we show that this contribution arises from the part of the probe D3-brane near the Poincar\'e horizon, which behaves as a cylindrical shell of $k$ strings. Specifically, this contribution is precisely that of a Wilson line in the direct product of $k$ fundamental representations.

We also compute the probe sector's contribution to the VEV of the Lagrangian, via the D3-branes' linearised back-reaction on the dilaton. We obtain new, analytical results in various limits. For example, for the screened Wilson line, when $R \to \infty$ our result resembles that of a point charge in Maxwell theory. What begins in the UV as a Wilson line of $SU(N)$ appears in the IR as a point charge of the $U(1)$ and a singlet of $SU(N-1)$. In other words, it is screened, as expected. For the spherical soliton we find a maximum near the soliton's radius. In the large-$v$ and large-$k$ limits we compute the probe sector's contribution to the VEVs of the Lagrangian and stress-energy tensor analytically, and find results consistent with $S_\mathrm{soliton}$. In particular, in each VEV, when $R \to \infty$ we find a term linear in $k$, with the form expected of a Wilson line in the direct product of $k$ fundamental representations.

All of our results for the EE of non-conformal branes rely crucially on the boundary term in the Karch-Uhlemann method, as we discuss in detail in each case. More generally, our results pave the way for pursuing many fundamental questions and applications of HEE directly in the probe limit, without computing backreaction.

This paper is organized as follows. In section~\ref{sec:review_HEE} we review EE, the RT formula, and the Karch-Uhlemann method, and illustrate the method in a simple example, a fundamental representation Wilson line. In section~\ref{sec:D3_brane} we review the probe D3-branes described above, and in section~\ref{sec:HHE_cases} we compute their contributions to HEE. In section~\ref{sec:lagran_dens} we compute the Lagrangian VEVs, and for the soliton, also the stress-energy tensor. Section~\ref{sec:concl} is a summary and discussion of future research. We collect many technical details in four appendices.

\section{Holographic Entanglement Entropy}
\label{sec:review_HEE}

In this section we review the calculation of EE in holography. In section~\ref{sec:rtreview} we review the definition of EE and its computation in holography as proposed by Ryu and Takayanagi (RT)~\cite{Ryu:2006bv,Ryu:2006ef}. Our focus will be on AdS space-times containing one or more probe branes. A convenient method for calculating the leading order contribution of probe branes to EE was given by Karch and Uhlemann in ref.~\cite{Karch:2014ufa}. In section~\ref{sec:probe_ee} we review their method, and highlight the presence of a contribution they neglected. This contribution, which takes the form of a boundary term on the probe brane, is in fact non-zero in all our non-conformal examples. Indeed, it can provide a crucial contribution to the EE as we will see in section~\ref{sec:HHE_cases}.

\subsection{Review: entanglement entropy and holography}
\label{sec:rtreview}

Given a generic quantum state described by a density matrix $\rho$ and a bipartition of the Hilbert space \(H\) into a subspace $A$ and its complement $\bar{A}$, the reduced density matrix on \(A\) is defined as  $\rho_A \equiv \Tr_{\bar{A}} \rho$. The EE of $A$ is defined as the von Neumann entropy of \(\r_A\),
\begin{equation}
S_{A} \equiv - \Tr_A \rho_A \log \rho_A\,.
\end{equation}
If the total state \(\r\) is pure, which will be the case in all that follows, then $S_A = S_{\bar{A}}$ and the EE is a good measure of the amount of quantum entanglement between $A$ and $\bar{A}$. 

In QFT, a natural way to partition the Hilbert space is to do so geometrically, taking \(A\) to be a set of states in a subregion of a Cauchy hypersurface. The EE may then be computed through the replica trick~\cite{Callan:1994py}, an approach which has been particularly successful for two-dimensional CFTs~\cite{Holzhey:1994we,Calabrese:2004eu, Calabrese:2009qy,Calabrese:2009ez,Calabrese:2010he}. The first step of the replica trick is to construct the quantity $\Tr_A \rho_A^n$ with $n\in\mathbb{Z}^+$, which is equal to the partition function \(Z\) of $n$ copies of the original theory glued together along \(A\). This is equivalent to the partition function of the original theory on a manifold $\mathcal{M}_n$ with conical deficit $2\pi /n$ at the entangling surface between \(A\) and \(\bar{A}\), namely $\Tr_A \rho_A^n = Z[\mathcal{M}_n]/Z[\mathcal{M}_1]^n$. Defining the \(n\)-th R\'enyi entropy as
\begin{equation}
S_A^{(n)} \equiv \frac{1}{1-n} \log \Tr_A \rho_A^n\,,
\end{equation}
the EE is given by the limit
\begin{equation}
\label{eq:eelim}
S_A = \lim_{n\rightarrow 1} S_A^{(n)} = - \lim_{n\rightarrow 1}\partial_n \Tr_A \rho_A^n\,.
\end{equation} 
Defining a generating functional $W[\mathcal{M}_n]\equiv-\log Z[\mathcal{M}_n]$, eq.~\eqref{eq:eelim} may be rewritten as
\begin{equation}
\label{eq:EE_repl_trick_action}
S_A = \lim_{n\rightarrow 1} (\partial_n-1)W[\mathcal{M}_n]\,.
\end{equation} 
This form of the EE will be useful in the following.

For a QFT with a holographic dual, the EE of a time-independent state can be computed through the RT formula~\cite{Ryu:2006bv,Ryu:2006ef} 
\begin{equation}
\label{eq:Ryu_Tak}
S_{A} = \frac{\mathcal{A}\left[\gamma_A^{(\mathrm{min})}\right]}{4 G_N}\,,
\end{equation}
where $\gamma_A^{(\mathrm{min})}$ is the minimal surface in the bulk space-time homologous to the region $A$ at the asymptotic boundary, $\mathcal{A}$ is the area of $\gamma_A^{(\mathrm{min})}$, and $G_N$ is Newton's constant of the bulk gravity theory. This result was extended to time-dependent cases in refs.~\cite{Hubeny:2007xt,Dong:2016hjy}. 

The RT formula was proved by Lewkowycz and Maldacena in ref.~\cite{Lewkowycz:2013nqa} by defining a generalised gravitational entropy, in an extension of the usual Gibbons-Hawking thermodynamic interpretation of Euclidean gravity solutions~\cite{Gibbons:1976ue} to cases with no $U(1)$ symmetry.\footnote{See also ref.~\cite{Casini:2011kv} for an earlier proof of the RT formula for the special case of a spherical entangling region in a CFT.} Consider a semiclassical gravitational theory on a Euclidean manifold with a boundary. We assume that the boundary has a direction which is topologically a circle, parametrised by a coordinate \(\t\) with \(\t \sim \t + 2\pi\), but it need not have a $U(1)$ isometry. The boundary conditions are assumed to respect $\tau$'s periodicity. In the context of EE, \(\t\) winds around the boundary of the subregion \(A\). Let \(I(n)\) be the on-shell action of the gravity theory with the period extended to \(\t \sim \t + 2\pi n\), with \(n \in \mathbb{Z}^+\), while maintaining boundary conditions invariant under \(\t \to \t + 2\pi\). The generalised gravitational entropy is then defined as
\begin{equation}
\label{eq_generalised_grav_entropy}
S_\mathrm{grav} = \lim_{n\rightarrow 1} (\partial_n-1) I(n)\,.
\end{equation}
For a theory on a static asymptotically AdS space-time, the on-shell action is equal to the dual QFT's generating functional \(W\). The right-hand side of eq.~\eqref{eq_generalised_grav_entropy} is then identical to the right-hand side of eq.~\eqref{eq:EE_repl_trick_action}, and so \(S_\mathrm{grav} = S_A\).

Evaluating the right-hand side of eq.~\eqref{eq_generalised_grav_entropy} requires analytic continuation of \(I(n)\) to non-integer \(n\). A convenient prescription for this continuation is as follows~\cite{Lewkowycz:2013nqa}. For integer \(n\), assuming the bulk respects the symmetry of the boundary conditions under \(\t \to \t + 2\pi\), we have $I(n)= \left. n I(n)\right|_{2\pi}$, where $\left.I(n)\right|_{2\pi}$ denotes the on-shell action with \(\t\) integrated only over the range \([0,2\pi)\). With the crucial assumption that this result applies also at non-integer \(n\), the right-hand side of eq.~\eqref{eq_generalised_grav_entropy} becomes
\begin{equation}
\label{eq_ML:method}
S_\mathrm{grav}  = \lim_{n\rightarrow 1} \left. \partial_n I(n) \right|_{2\pi}\,.
\end{equation}
In ref.~\cite{Lewkowycz:2013nqa}, the authors showed that the right-hand sides of eqs.~\eqref{eq:Ryu_Tak} and~\eqref{eq_ML:method} are equivalent for asymptotically AdS space-times, thus proving the RT formula. We will henceforth denote the generalised gravitational entropy and EE by the same symbol \(S_A\).

\subsection{Holographic entanglement entropy of probe branes}
\label{sec:probe_ee}

In this paper we compute EE in holographic duals of spacetimes containing branes. We write the total action for the gravitational theory as
\begin{equation}
I = I_\mathrm{bulk} + I_\mathrm{brane}\,,
\end{equation}
where \(I_\mathrm{bulk}\) denotes the bulk action of the \((d+1)\)-dimensional gravitational theory, while \(I_\mathrm{brane}\) denotes the action of a \((p+1)\)-dimensional brane, with $p\leq d$. We assume that \(I_\mathrm{brane}\) is proportional to a tension \(T_p\). We work in the probe limit, defined as the limit in which \(T_p\) is small in units of $G_N$ and the $AdS_{d+1}$ curvature radius, $L$, so that the back-reaction of the brane on the bulk fields is negligible. More precisely, the probe limit is an expansion in the dimensionless parameter \(\e \equiv T_p G_N L^{p+2-d} \ll 1\) to order $\e$.

Before taking the probe limit, the presence of the brane changes the metric of the bulk space-time. This leads to a change in HEE due to a change in the area of the minimal surface in eq.~\eqref{eq:Ryu_Tak}. Our goal is to determine the EE in the probe limit, meaning to order $\epsilon$. In ref.~\cite{Karch:2014ufa}, Karch and Uhlemann proposed a method for computing this without having to compute the full back-reaction of the brane, i.e. directly in the probe limit, by extending Lewkowycz and Maldacena's arguments to probe branes.

We decompose the bulk and brane actions as
\begin{equation}
\label{eq:action_decomposition}
I_\mathrm{bulk} = \int_{\z_h}^\infty d\z \int d^d x \, \mathcal{L}_\mathrm{bulk} + I_\mathrm{ct,bulk}\,,
\qquad
I_\mathrm{brane} = \int_{\z_h}^\infty d\z \int d^p y \, \mathcal{L}_\mathrm{brane} + I_\mathrm{ct,brane}\,.
\end{equation}
Here we have introduced a coordinate \(\z \in [\z_h,\infty)\), where \(\z=  \z_h\) is the locus where the \(\t\)-circle degenerates, and the boundary of $AdS_{d+1}$ is at \(\z \to \infty\). We have denoted the remaining bulk coordinates by \(x\), and the remaining coordinates on the brane by \(y\). The counterterm action \(I_\mathrm{ct,bulk}\) consists of boundary terms at a large-\(\z\) cut-off \(\z_c\), and is needed to render the on-shell action finite and the variational problem well-defined~\cite{deHaro:2000vlm}. The brane counterterms \(I_\mathrm{ct,brane}\) are also needed if the brane reaches the boundary of AdS~\cite{Karch:2005ms}. We collectively denote the bulk fields as \(\F\) and the fields on the brane as \(X\). In general, the bulk Lagrangian \(\mathcal{L}_\mathrm{bulk}\) depends on \(\F\) and its derivatives \(\p \F\), and similarly \(\mathcal{L}_\mathrm{brane}\) depends on \(X\) and \(\p X\). The brane Lagrangian will also depend on \(\F\), but we assume that it is independent of \(\p \F\). This is true generically for D-brane actions in string theory, including the D3-branes we consider below. With this decomposition, we can write the generalised gravitational entropy in eq.~\eqref{eq_ML:method} as
\begin{align}
S_A = \lim_{n \to 1} \biggl[&
\int_{\z_h}^\infty d\z \int d^d x \left(\frac{\delta \mathcal{L}_\mathrm{bulk}}{\delta \F} \p_n \F + \p_\m \Theta^\m_\mathrm{bulk}  \right)
- \int d^d x \, \left.\mathcal{L}_\mathrm{bulk} \right|_{\z =\z_h}\p_n \z_h 
+  \p_n I_\mathrm{ct,bulk}
\nonumber \\ 
&+\int_{\z_h}^\infty d\z \int d^p y \left(\frac{\delta \mathcal{L}_\mathrm{brane}}{\delta \F} \p_n \F + \frac{\delta \mathcal{L}_\mathrm{brane}}{\delta X} \p_n X + \p_\m \Theta^\m_\mathrm{brane} \right)
\label{eq:der_n_tot} \\
& - \int d^p y \, \left. \mathcal{L}_\mathrm{brane} \right|_{\z = \z_h}\p_n \z_h 
+ \p_n I_\mathrm{ct,brane} \biggr]_{2\pi}\,,
\nonumber
\end{align}
where the total derivative terms \(\p_\m \Theta^\m_\mathrm{bulk}\) and \(\p_\m \Theta^\m_\mathrm{brane}\) arise due to the dependence of \(\mathcal{L}_\mathrm{bulk}\) and \(\mathcal{L}_\mathrm{brane}\) on derivatives of \(\F\) and \(X\), respectively. It can be shown that eq.~\eqref{eq:der_n_tot} is equivalent to the RT formula~\eqref{eq:Ryu_Tak} in the presence of the brane~\cite{Lewkowycz:2013nqa,Karch:2014ufa}.

To evaluate eq.~\eqref{eq:der_n_tot} in the probe limit, we imagine solving the equations of motion for \(\F\) and \(X\) as power series in the small parameter \(\e\). Concretely, we write \(\F = \F^{(0)} + \F^{(1)} + \dots\) where \(\F^{(n)} \sim \mathcal{O}(\e^n)\), and similarly for \(X\) and \(S_A\). Crucially, \(\F^{(0)}\) is the solution of the bulk equations of motion in the absence of the brane, and \(X^{(0)}\) is the solution of the brane equations of motion when \(\F = \F^{(0)}\), so
\begin{align}
\label{eq:eom_satisfied}
\left. \frac{\delta \mathcal{L}_\mathrm{bulk}}{\delta \F} \right|_{\F = \F^{(0)}} &= 0\,, & 
\left. \frac{\delta \mathcal{L}_\mathrm{brane}}{\delta X} \right|_{\F = \F^{(0)},\,X=X^{(0)}} &= 0\,.
\end{align}
In the small \(\e\) expansion, the bulk contributions to \(S_A\) (the terms on the first line of eq.~\eqref{eq:der_n_tot}) are dominated by an \(\mathcal{O}(\e^{0})\) term that depends only on \(\F^{(0)}\). This is the gravitational entropy in the absence of the brane. Since by definition \(\F^{(0)}\) extremises the bulk action, there is no \(\mathcal{O}(\e)\) contribution to the first line of eq.~\eqref{eq:der_n_tot}. 

The \(\mathcal{O}(\e)\) piece of \(S_A\), which we denote \(S_A^{(1)}\), is thus obtained purely from the brane contribution, i.e. the second and third lines of eq.~\eqref{eq:der_n_tot}, evaluated on the leading order solutions \(\F^{(0)}\) and \(X^{(0)}\). We can simplify this contribution by noting that \(\d \mathcal{L}_\mathrm{brane}/\d X\) vanishes by eq.~\eqref{eq:eom_satisfied}. Furthermore, the total derivative term \(\p_\m \Theta^\m\) may be integrated over \(\z\), yielding boundary terms at \(\z=\z_h\) and \(\z = \z_c\). The \(\z_c\) boundary term must cancel part of the counterterm contribution. Concretely, if we write \(I_\mathrm{ct,brane} = \int d^p y \left. \mathcal{L}_\mathrm{ct,brane} \right|_{\z=\z_c}\), then
\begin{equation} \label{eq:S_ct_brane_variation}
\p_n I_\mathrm{ct,brane} = \int d^p y \, \left[ \frac{\d \mathcal{L}_\mathrm{ct,brane}}{\d X} \p_n X + \frac{\d \mathcal{L}_\mathrm{ct,brane}}{\d \F} \p_n \F \right]_{\z=\z_c}\,.
\end{equation}
The boundary term at \(\z=\z_c\) from \(\p_\m \Theta^\m\) must cancel the term containing \(\d \mathcal{L}_\mathrm{ct,brane}/\d X\) in eq.~\eqref{eq:S_ct_brane_variation}, since the variational problem demands that we choose boundary conditions for \(X\) such that the whole action, including boundary terms, is stationary on a solution of the equations of motion. In total, the \(\mathcal{O}(\e)\) contribution to the EE is then
\begin{align}
S_A^{(1)} = \lim_{n \to 1} \biggl[
\int_{\z_h}^\infty d\z \int  d^p y & \frac{\delta \mathcal{L}_\mathrm{brane}}{\delta \F} \p_n \F
+ \int d^p y  \frac{\delta \mathcal{L}_\mathrm{ct,brane}}{\delta \F} \p_n \F
\nonumber \\
&- \int d^p y \left. \mathcal{L}_\mathrm{brane}\right|_{\z = \z_h} \p_n \z_h 
+ \int d^p y \left. N_\mu \Theta^\mu_\mathrm{brane} \right|_{\z = \z_h}
\biggr]_{2\pi}\,,
\label{eq:EE_prob_brane_gen} 
\end{align}
where \(N_\m\) is the unit normal vector to the surface at \(\z=\z_h\), and this expression is to be evaluated on the leading order solutions \(\F = \F^{(0)}\) and \(X = X^{(0)}\).

In ref.~\cite{Karch:2014ufa} Karch and Uhlemann argued that the final term in eq.~\eqref{eq:EE_prob_brane_gen} generically vanishes. However, we find that this is not the case. Indeed, we find that this term is non-zero for all the probe D3-branes we study that break conformal symmetry in the dual CFT. We will see that this boundary term will be crucial to obtaining the correct expression for the EE of the Coulomb branch D3-brane in section~\ref{sec:EE_Coul_branch}, and a finite and continuous expression for the EE of the spherical soliton in section~\ref{sec:EE_W_bos}.

\subsubsection{HEE of spherical entangling regions}

A significant challenge when using eq.~\eqref{eq:EE_prob_brane_gen} is that we need to know the solution for the metric $g_{\mu\nu}$ and other bulk fields when $n \ne 1$, in order to evaluate the derivatives \(\p_n \F\) and \(\p_n \z_h\). In general, this solution can be very difficult to find. However, for a spherical entangling region $A$ of radius $R$, and in the CFT vacuum state dual to $AdS_{d+1}$ space-time, the appropriate metric was found in refs.~\cite{Emparan:1999gf,Casini:2011kv}. We will make extensive use of this solution, so we provide a brief review of it here.

To begin, consider the Euclidean $AdS_{d+1}$ metric in Poincar\'e coordinates,
\begin{equation}
\label{eq:metricAdS}
ds^2 =  \frac{L^2}{r^2} \, d r^2 + \frac{r^2}{L^2} \left( d t_E^2  + d\rho^2 + \rho^2 d \Omega_{d-2}^2  \right).
\end{equation}
Here $r$ is the $AdS_{d+1}$ radial coordinate, with the Poincar\'e horizon at $r=0$ and the boundary at $r \to \infty$, $t_E$ is the Euclidean time, $\rho$ denotes the CFT radial coordinate, and \(d \Omega_{d-2}^2\) is the round metric on $ S^{d-2} $. We now make a coordinate transformation to hyperbolic slicing, defining coordinates \((\zeta,u,\t)\) such that
\begin{gather}
r = \frac{L^2}{R} \left(\zeta \cosh u +\sqrt{\zeta^2 -1} \cos \tau \right) \,, \qquad 
t_E =R\frac{\sqrt{\zeta^2 -1} \sin \tau}{\zeta \cosh u + \sqrt{\zeta^2 -1} \cos \tau}\,, \qquad
\nonumber \\
\rho =R\frac{ \zeta \sinh u}{\zeta \cosh u + \sqrt{\zeta^2 -1} \cos \tau}\,.
\label{eq:CHM_mapping}
\end{gather}
The new coordinates are dimensionless and take values in the following intervals: $u \in [0, +\infty)$, $\zeta \in [1, + \infty)$ and $\tau \in [0,2\pi) $. The metric in eq. \eqref{eq:metricAdS} now becomes
\begin{align}
\label{eq:hyperb_metric}
ds^2 &= L^2 \left[ \frac{d\zeta^2}{f(\z)} + f(\z) d \tau^2 + \zeta^2  du^2 + \z^2 \sinh^2u \, d \Omega^2_{d-2}\right] \,, &  f(\z) &= \z^2 - 1\,.
\end{align}
At the $AdS_{d+1}$ boundary, the coordinate transformation in eq.~\eqref{eq:CHM_mapping} implements a conformal transformation that maps the sphere's causal development to $S^1$ (parametrised by $\tau$) times a hyperbolic plane~\cite{Casini:2011kv}. Eq.~\eqref{eq:CHM_mapping} thus maps the reduced density matrix $\rho_A$ for spherical region $A$ to a thermal density matrix on the hyperbolic plane. Indeed, the metric in eq.~\eqref{eq:hyperb_metric} has a Euclidean horizon at $\zeta_h = 1$, where the radius of the circle parametrised by \(\t\) shrinks to zero, with inverse Hawking temperature $2 \pi $. From eq.~\eqref{eq:CHM_mapping} we find that in Poincar\'e coordinates the horizon is located at $r^2 + \rho^2 =R^2$, which is precisely the RT surface of region $A$~\cite{Ryu:2006bv,Ryu:2006ef}. The EE thus maps to the Bekenstein-Hawking entropy of the horizon, as expected. Of course, we merely changed coordinates, so the space is still precisely $AdS_{d+1}$, and the horizon is simply that due to the observer's acceleration~\cite{Emparan:1999gf,Casini:2011kv}. For the gravitational theories of interest in this paper, the period of the Euclidean time can be modified to $\tau \sim \tau + 2\pi n$ by replacing \(f(\z)\) in eq.~\eqref{eq:hyperb_metric} with
\begin{align}
\label{eq:hyperb_metric_n}
f_n(\zeta)  &= \zeta^2 - 1 - \frac{\left( \zeta_h^d- \zeta_h^{d-2}\right)}{\zeta^{d-2}}\,, &
\zeta_h &= \frac{\sqrt{1+n^2 d (d-2)  }+1}{n d  }\,,
\end{align}
where the horizon is now at \(\z = \z_h\). The metric in eq.~\eqref{eq:hyperb_metric} with \(f(\z)\) in eq.~\eqref{eq:hyperb_metric_n} is the topological black hole in hyperbolic slicing studied in ref.~\cite{Emparan:1999gf}. 

Given a solution for the worldvolume fields on a probe brane in $AdS_{d+1}$, we will compute $S_A^{(1)}$ by mapping the solution to hyperbolic coordinates via eq.~\eqref{eq:CHM_mapping}, plugging the result into eq.~\eqref{eq:EE_prob_brane_gen}, and then performing the integrals. As mentioned in section~\ref{sec:intro}, EE generically has UV divergences arising from correlations across $A$'s boundary. In eq.~\eqref{eq:CHM_mapping}, we reach $A$'s boundary by fixing $\zeta$ and sending $u \to \infty$: these limits send $r \to \infty$, taking us to the $AdS_{d+1}$ boundary, and $\rho \to R$, taking us to the surface of $A$. The UV divergences of the EE would thus appear on a probe brane as large-$u$ divergences. These will not be cancelled by the probe brane's counterterms, $I_\mathrm{ct,brane}$ in eq.~\eqref{eq:action_decomposition}, as those cancel divergences that are independent of the choice of $A$. To be explicit, $I_\mathrm{ct,brane}$ cancels divergences that arise when we fix $u$ and send $\zeta \to \infty$: these limits send $r \to \infty$ but with $\rho$ determined by $u$ and $\tau$, such that $\rho$ need not be at $A$'s surface. These are the usual near-boundary divergences of $AdS_{d+1}$, which could be regulated by a Fefferman-Graham cutoff $r = L^2/\varepsilon$ with $\varepsilon \ll 1$.

None of our probe D3-branes below will exhibit large-$u$ divergences, either because they do not reach the $AdS_5$ boundary at all, like the Coulomb branch or spherical soliton D3-branes in figures~\ref{fig:cartoons}b and d, or because they reach the $AdS_5$ boundary only at a point, which does not produce divergences near $\rho = R$, like the conformal or screened Wilson line D3-branes in figures~\ref{fig:cartoons}a and c. As a result, all of our results for $S_A^{(1)}$ below will be finite for any finite $R$, i.e. they will require no UV regulator.

\subsubsection{A simple example: the probe string}
\label{sec:string}

In order to illustrate the Karch-Uhlemann method of computing probe brane EE, we now apply it to a straight Wilson line in the fundamental representation of \(\mathcal{N}=4\) SYM theory with gauge group \(SU(N)\). A result for the spherical EE of a Wilson line in the fundamental representation, valid in the large $N$ limit and for any value of the 't Hooft coupling $\lambda$, was obtained in ref.~\cite{Lewkowycz:2013laa} using only conformal symmetry and supersymmetric localisation. We will reproduce the holographic computation of this EE using the Karch-Uhlemann method in ref.~\cite{Kumar:2017vjv}, which agrees with ref.~\cite{Lewkowycz:2013laa} when $\lambda \gg 1$. We choose this example both for its simplicity and because the final result will be useful in section~\ref{sec:EE_W_bos}.

In the Maldacena limit of large-$N$ followed by $\lambda \gg 1$, the holographic dual of the fundamental representation Wilson line is a string anchored to the boundary of $AdS_5 \times S^5$~\cite{Maldacena:1998im,Rey:1998bq}, which can be treated as a probe. The action for the probe string is
\begin{equation}
I_\mathrm{string} = \frac{1}{2\pi \alpha'} \int d t_\mathrm{E} \, dr \sqrt{P[g]} + I_\mathrm{ct,string}\,,
\end{equation} 
where $1/(2\pi\alpha')$ is the string tension, $P[g]$ is the induced metric on the string, and we have chosen \((t_E,r)\) as coordinates on the string. The solution dual to the Wilson line is located at $\rho=0$ and at an arbitrary point on the $S^5$.

Mapping to hyperbolic slicing using eq.~\eqref{eq:CHM_mapping}, we find that we can parametrise the string by \((\z,\t)\), and the solution becomes \(u=0\). The on-shell action for generic $n$, with \(\t\) restricted to the range \([0,2\pi)\), becomes
\begin{equation}
\left. I_\mathrm{string}(n)\right|_{2\pi} = \frac{L^2}{2\pi \alpha'} \int_0^{2\pi} d\tau \int_{\zeta_h}^{\infty} d\zeta  + I_\mathrm{ct,string}\,,
\end{equation}
and the EE is $\left.\frac{\partial}{\partial n} I_\mathrm{string}(n)\right|_{2\pi}$ evaluated at $n=1$. In QFT terms, the counterterm action $I_\mathrm{ct,string}$ is only sensitive to UV physics, and so is independent of $n$. Denoting the EE contribution of the fundamental representation Wilson line as \(S_A^{(1)} = S_{\square}\), we find
\begin{equation}
\label{eq:EE_fund_quark}
S_{\square} = \left. \partial_n \left. I_\mathrm{string}(n)\right|_{2\pi}  \right|_{n=1} = - \frac{L^2}{2\pi \alpha'} \left. \partial_n \, \left( 2\pi  \, \zeta_h \right) \right|_{n=1} = \frac{\sqrt{\lambda}}{3}\,,
\end{equation}
where we used $\alpha' = L^2 /\sqrt{\lambda}$. This result agrees with that of ref.~\cite{Lewkowycz:2013laa} when $\lambda \gg 1$.

In this simple case, the probe string's contribution to the EE arises entirely from the variation of the lower endpoint of integration $\zeta_h$ with respect to $n$, i.e. the third term in eq.~\eqref{eq:EE_prob_brane_gen}. This will not be the case in the more complicated examples below.

\section{Probe D3-brane Solutions}
\label{sec:D3_brane}

In this section we briefly review the probe D3-brane solutions in $AdS_5 \times S^5$ that we will consider in this paper. We will use the $AdS_5$ coordinates of eq.~\eqref{eq:metricAdS}, but in Lorentzian signature, so our $AdS_5 \times S^5$ metric is
\begin{equation}
\label{eq:metricAdS_spher}
ds^2 = \frac{L^2}{r^2} \, d r^2 + \frac{r^2}{L^2} \left[ -d t^2  +  d\rho^2 + \rho^2 (d\theta^2 + \sin^2 \theta \, d\phi^2) \right]+ L^2 d \Omega_5^2\,,
\end{equation}
with $\theta \in [0,\pi)$ and $\phi\in[0,2\pi)$. In these coordinates, we choose a gauge in which the Ramond-Ramond 4-form field takes the form\footnote{In eq.~\eqref{eq:C4_gauge_Poinc} we neglect a contribution to \(C_4\) with components only in the \(S^5\) directions. This term is necessary to ensure that \(F_5 \equiv d C_4\) is self-dual, but it will play no role in our calculations since it has vanishing pullback on all D3-brane solutions we consider.}
\begin{equation}
\label{eq:C4_gauge_Poinc}
C_4 = \frac{r^4}{L^4} \,\rho^2 \sin \theta \, dt \wedge d\rho \wedge d\theta \wedge d\phi \, .
\end{equation}
The D3-brane's action is
\begin{equation}
\label{eq:D3brane_general}
I_{D3} = - T_{D3} \int d^4 \xi \sqrt{- \text{det}(P[g]_{ab}+ F_{ab})} + T_{D3} \int P[C_4]+I_\mathrm{ct}\,,
\end{equation}
where $T_{D3} = N/(2\pi^2)$ is the D3-brane tension, $\xi^a$ with $a=1,2,3,4$ are the worldvolume coordinates, $P[g]_{ab}$ is the pullback of the metric to the brane, $F_{ab}$ is the worldvolume $U(1)$ field strength, and $P[C_4]$ is the pullback of $C_4$ to the brane. We assume that the D3-brane is static and spherically symmetric, and spans the $t$ direction, wraps $S^2$ inside $AdS_5$, and sits at a point on $S^5$. The D3-brane will then trace a curve in the $(r,\rho)$ directions, which we parametrise as $r= r(\rho)$. We also assume that the only non-trivial field strength component is $F_{t\r}(\r)$. Plugging this ansatz into eq.~\eqref{eq:D3brane_general} and integrating over the $S^2$ then gives
\begin{equation}
\begin{split}
\label{eq:D3brane_anstaz}
I_{D3}
= -4 \pi T_{D3} \int dt \, d\rho \, \frac{\rho^2\,r^4}{L^4} \left[  \sqrt{  1+ \frac{L^4}{r^4} \left[\left(\frac{\partial r}{\partial\rho}\right)^2 - F_{t\r}^2\right]} - 1 \right] + I_\mathrm{ct}\,.
\end{split}
\end{equation}
The counterterm action $I_\mathrm{ct}$ is non-zero only when the D3-brane reaches the $AdS_5$ boundary, and can be split as  $I_\mathrm{ct}= I_{\text{UV}}+I_{U(1)}$. Here, $I_{\text{UV}}$ are the boundary terms needed to make the action finite and $I_{U(1)}$ is a finite Lagrange multiplier,
\begin{align}
\label{eq:u1}
I_{U(1)} &= - \kappa \, (4 \pi L \, T_{D3})  \int dt \, d\rho \, F_{t\rho}\,, & \kappa &\equiv k \, \frac{\sqrt{\lambda}}{4 N}\,,
\end{align}
which enforces the  condition that the \text{D3}-brane  is endowed with $k>0$ units of string charge. We will henceforth ignore $I_{\text{UV}}$, which is independent of $n$ and thus will not contribute to the probe brane EE in eq.~\eqref{eq:EE_prob_brane_gen}. As shown in refs.~\cite{Schwarz:2014rxa,Schwarz:2014zsa} and references therein, the equations of motion coming from the action in eq.~\eqref{eq:D3brane_anstaz} have BPS solutions with $F_{t\r} = \partial r/\partial \r$ and
\begin{equation}
\label{eq:D3_genric_solution}
r(\r) = v\,L  \pm \frac{\kappa\,L^2 }{\rho }\,,
\end{equation}
where $v>0$ is an integration constant. The solution is thus determined by the two integration constants, $\k$ and $v$, and by the sign in eq.~\eqref{eq:D3_genric_solution}. Different choices for $v$, $\k$, and the sign lead to the solutions described in figure~\ref{fig:cartoons}, with very different interpretations in the dual CFT, as we will now summarise.

\subsection{Conformal Wilson line} 
\label{sec:D3_conf_wils}

While a Wilson line in the fundamental representation corresponds to a single string ending at the $AdS_5$ boundary, higher-rank representations correspond to multiple coincident strings. When the number of strings is of order $N$, a convenient holographic description becomes D-branes carrying string charge~\cite{Callan:1997kz, Gibbons:1997xz}. In this description, the type of brane depends on the representation of the Wilson line.

For example, the $k$-th antisymmetric representation is described by a D5-brane along $AdS_2 \times S^4$ endowed with $k$ units of string charge~\cite{Yamaguchi:2006tq,Gomis:2006sb}. The $AdS_2$ factor implies the dual Wilson line preserves $(0+1)$-dimensional conformal symmetry. Indeed, the Wilson line breaks $\N=4$ SYM's $SO(4,2)$ conformal symmetry group to $SO(1,2) \times SO(3)$, where $SO(1,2)$ are conformal transformations preserving the line and $SO(3)$ is the $S^2$ isometry.

The solution in eq.~\eqref{eq:D3_genric_solution} with $v=0$ and $\kappa\neq0$ with a plus sign describes a conformal Wilson line in the $k$-th symmetric representation of $SU(N)$~\cite{Drukker:2005kx}. In that case,
\begin{equation}
\label{eq:Conf_Wils_emb}
r(\r) = \frac{\kappa \, L^2}{\r}\,,
\end{equation}
so that $r \to \infty$ as $\rho \to 0$ and the D3-brane reaches the $AdS_5$ boundary at a point. The solution describes a D3-brane shaped like a cone with apex at the $AdS_5$ boundary as depicted in figure~\ref{fig:cartoons}a. Both the opening angle and $k$ are determined by $\kappa$, as the D3-brane carries $k=\kappa \,4 N/\sqrt{\lambda}$ units of string charge~\cite{Kumar:2020hif}. In this case the D3-brane's worldvolume is $AdS_2 \times S^2$, so like the D5-branes mentioned above, this D3-brane describes a conformal Wilson line preserving $SO(1,2) \times SO(3) \subset SO(4,2)$.

The EE contribution to a spherical region in ${\cal N}=4$ SYM from a Wilson line in a generic representation, at large $N$ and any $\lambda$, was related to the expectation value of the circular Wilson loop in ref.~\cite{Lewkowycz:2013laa} using conformal symmetry and supersymmetry. Simple closed form expressions exist for the specific cases of  Wilson lines in symmetric and antisymmetric tensor representations, as the corresponding circular Wilson loops can be computed explicitly using  matrix model techniques \cite{Hartnoll:2006is}. In section~\ref{sec:EE_Conf_Wilson} we compute the conformal D3-brane's contribution to the EE of a spherical region, finding agreement with the result for the symmetric representation Wilson line in ref.~\cite{Lewkowycz:2013laa} when $\lambda \gg 1$.

\subsection{Coulomb branch} 
\label{sec:coul_branch}

The simplest non-conformal solution coming from eq.~\eqref{eq:D3_genric_solution} is $v>0$ and $\kappa=0$, so that
\begin{equation}
\label{eq:Coulb_branch_solution}
r(\r) = L v \,.
\end{equation}
This describes a D3-brane sitting at fixed $r$, as depicted in figure~\ref{fig:cartoons}b. We can imagine that such a solution comes from a stack of $N$ D3-branes at $r=0$ after one D3-brane has been pulled to $r=Lv$. The probe approximation is then clearly justified since $N \gg 1$. 

The field theory interpretation of this solution is $\mathcal{N}=4$ $SU(N)$ SYM at large $N$ and strong coupling at a point on the Coulomb branch where $SU(N)\rightarrow SU(N-1) \times U(1)$. More specifically, one of the six adjoint-valued scalar fields of $\N=4$ SYM, $\Psi$, acquires a non-zero VEV, $\langle \Psi \rangle \propto v$~\cite{Klebanov:1999tb}. Such a state describes an RG flow from a UV CFT, $SU(N)$ $\N=4$ SYM, to an IR CFT, $SU(N-1)$ $\N=4$ SYM and a decoupled $U(1)$ $\N=4$ SYM. The two CFTs in the IR interact only via massive degrees of freedom, namely the W-boson supermultiplet, which is bi-fundamental under $SU(N-1) \times U(1)$. A W-boson is dual to a string stretched between the Poincar\'e horizon and the probe D3-brane.

A solution of type IIB supergravity describing fully back-reacted Coulomb branch D3-branes is known~\cite{Klebanov:1999tb}, and has been used to calculate EE using the RT formula for example in refs.~\cite{Aprile:2014iaa,Karch:2014pma}. In appendix~\ref{app:EE_coul_backReac} we use the RT formula to compute the EE at the point on the Coulomb branch with the breaking $SU(N) \to SU(N-1) \times U(1)$, and then take a probe limit. We find perfect agreement with our probe limit result below, obtained using the Karch-Uhlemann method, eq.~\eqref{eq:S_coul_analytic}. Crucially, the agreement occurs only if we include the last term in eq.~\eqref{eq:EE_prob_brane_gen}, which Karch and Uhlemann overlooked in ref.~\cite{Karch:2014ufa}.

\subsection{Screened Wilson line}
\label{sec:screend_wils_line}

Taking $v>0$, $\k>0$, and the plus sign in eq.~\eqref{eq:D3_genric_solution} results in a solution that interpolates between the symmetric-representation Wilson line and Coulomb branch D3-branes,
\beq
\label{eq:screenedsol}
r(\r) = v\,L + \frac{\k \, L^2}{\r}\,.
\eeq
In particular, as $\r \to 0$ the solution approaches the Wilson line solution of eq.~\eqref{eq:Conf_Wils_emb}, $r = \k \, L^2/\r$, and as $\r \to \infty$ the solution approaches the Coulomb branch solution of eq.~\eqref{eq:Coulb_branch_solution} $r = v\,L$. The solution in eq.~\eqref{eq:screenedsol} is depicted in figure~\ref{fig:cartoons}c.

Ref.~\cite{Evans:2019pcs} argued that the solution in eq.~\eqref{eq:screenedsol} describes a symmetric-representation Wilson line \textit{screened} by the adjoint scalar $\Psi$ that acquires a VEV. In the language of condensed matter physics, the Wilson line is an ``impurity.'' Thinking of the adjoint of $SU(N)$ as the combination of fundamental and anti-fundamental, the scalar VEV $\langle \Psi \rangle$ acts as a collection of colour dipoles. In the presence of the impurity these dipoles are polarised, and form a spherically-symmetric screening cloud around the impurity. Such screening is clear qualitatively in figure~\ref{fig:cartoons}c: the impurity is present in the UV (near the $AdS_5$ boundary) but absent in the IR (near the Poincar\'e horizon). Ref.~\cite{Evans:2019pcs} provided quantitative evidence for screening by showing the worldvolume fields support quasi-normal modes dual to quasi-bound states localised at the impurity, a clear signature of screening.

In section~\ref{sec:EE_Screen_Wilson} we compute the EE of a spherical region centered on the screened Wilson line. This case is not conformal, so the EE can have non-trivial dependence on the sphere's radius $R$. More precisely, since $R$ and the scalar VEV $\propto v$ are the only scales available, and EE is dimensionless, the EE can have non-trivial dependence on the dimensionless combination $Rv/L$. We find that as $Rv/L$ increases the impurity's contribution to the EE decreases monotonically, although to our knowledge no physical principle requires such behaviour. In particular, we know of no monotonicity theorem for a case like this, where the bulk RG flow, triggered by $\langle \Psi \rangle \propto v$, in turn triggers an RG flow of Wilson line degrees of freedom, which couple to $\Psi$ in such a way that $\langle \Psi \rangle \propto v$ gives them a mass~\cite{Maldacena:1998im,Gomis:2006sb}.

In section~\ref{sec:Fvev} we find a more detailed description of the screening. We compute the VEV of the probe sector's Lagrangian, which at large $R$ has the form of a point charge with a Coulomb potential, as in Maxwell theory. We thus learn that what begins in the UV as a Wilson line of $SU(N)$ appears in the IR as a point charge in the $U(1)$ sector, but is absent in the $SU(N-1)$ sector, i.e. it is screened in the $SU(N-1)$ sector.

\subsection{Spherical soliton}
\label{sec:W_bos}

Finally, taking $v>0$, $\k>0$, and the minus sign in eq.~\eqref{eq:D3_genric_solution} results in a solution discussed in detail by Schwarz in refs.~\cite{Schwarz:2014rxa,Schwarz:2014zsa},
\begin{equation}
\label{eq:solitonsol}
r(\r) = v\,L - \frac{\k \, L^2}{\r}\,.
\end{equation}
As $\rho \rightarrow \infty$ this solution reduces to the Coulomb branch solution eq.~\eqref{eq:Coulb_branch_solution} like the previous case, but now as $\rho \rightarrow 0$ the D3-brane bends towards the Poincar\'e horizon. Indeed, when $\rho = \k L/v$ the D3-brane intersects the Poincar\'e horizon $r=0$, as depicted in figure~\ref{fig:cartoons}d.

This behaviour implies very different physics compared to the screened Wilson line discussed above. In particular, in refs.~\cite{Schwarz:2014rxa,Schwarz:2014zsa} Schwarz interpreted this solution as a spherically-symmetric soliton ``phase bubble,'' separating $\N=4$ SYM with gauge group $SU(N)$ inside from $\N=4$ SYM with gauge group $SU(N-1)\times U(1)$ outside. The soliton is a spherical shell charged under the $U(1)$ of $SU(N-1) \times U(1)$, and being BPS thus has non-zero mass: it is an excited state of the Coulomb branch. The soliton radius is $R_0\equiv \k L/v$, which is proportional to its total charge \(k\), and its mass is
\begin{equation}\label{M_sol}
4 \pi T_{\text{D3}} \, L^3 \, v \, \k = 4 \pi T_{\text{D3}} \, L^2 \, v^2\, R_0\,,
\end{equation}
where on the right-hand side we emphasised that the soliton's mass is proportional to its radius, just like an extremal black hole. As discussed in section~\ref{sec:intro}, based on this similarity, Schwarz asked in ref.~\cite{Schwarz:2014zsa} whether the EE of a sphere of radius $R_0$ centred on the soliton scales with the soliton's surface area, i.e. as $R_0^2$, when $\k$ is large. In section~\ref{sec:EE_W_bos} we compute this EE, finding that it scales approximately as $R_0^{1.3}$, in contrast to area-law scaling $R_0^2$.

In sections~\ref{sec:EE_W_bos} and~\ref{sec:lagran_dens} we show that in the limits $\kappa \to \infty$ and $v \to \infty$ with $R_0$ fixed, a spherical region's EE, the Lagrangian's VEV, and the stress-energy tensor's VEV all diverge at $R_0$, and as $\rho \to \infty$ approach the result for a Wilson line in a direct product of $k$ fundamental representations. The natural interpretation is that in these limits the spherical soliton is an infinitely thin shell at $R_0$ that at large distances looks like a Wilson line in a direct product of $k$ fundamental representations.

\section{Holographic Entanglement Entropy of D3-branes}
\label{sec:HHE_cases}

In this section we compute the holographic EE contribution of the probe D3-branes reviewed in section~\ref{sec:D3_brane}. In the first part of this section, we consider a generic probe D3-brane, finding all the terms we need to obtain their contribution to the EE. Subsequently, we obtain explicit results for the conformal Wilson line in section~\ref{sec:EE_Conf_Wilson}, the Coulomb branch in section~\ref{sec:EE_Coul_branch}, the screened Wilson line in section~\ref{sec:EE_Screen_Wilson}, and finally the spherical soliton in section~\ref{sec:EE_W_bos}. For the remainder of the paper, we use units in which the $AdS_5$ radius is unity, $L\equiv 1$.

\subsection{General case} 
\label{sec:EE_D3_general}

As discussed in section~\ref{sec:review_HEE}, we consider (Euclidean) $AdS_5$ in hyperbolic slicing whose metric was given in eq.~\eqref{eq:hyperb_metric}. The D3-brane's worldvolume scalars are the $AdS_5$ coordinates whose profile determines the embedding. Parametrising the embedding as $u(\zeta,\tau)$ gives a D3-brane action
\begin{equation}
\begin{aligned}
\label{eq:D3_action_hyper}
I_{D3}  =\;&  T_{D3} \int d\theta \, d \phi \,  d\tau \, d  \zeta \, \sin \theta  \, \zeta^2 \, \sinh^2 u\sqrt{\zeta ^2  (\partial_\zeta u)^2
	f_n(\zeta )+\frac{\zeta ^2  ( \partial_\tau u)^2}{f_n(\zeta )}-F_{\tau \zeta}^2+ 1}  \\
& - T_{D3} \int P[C_4] + I_\mathrm{ct}\,.
\end{aligned}
\end{equation}
A delicate part of writing eq.~\eqref{eq:D3_action_hyper} is finding a convenient gauge for $C_4$. To this end, we point out an observation first made in ref.~\cite{Drukker:2005kx}. The gauge in eq.~\eqref{eq:C4_gauge_Poinc} gives the correct result for the expectation value of a straight Wilson line. However, after performing the mapping  to hyperbolic slicing in eq.~\eqref{eq:CHM_mapping}, one does not obtain the correct value for a circular Wilson loop. The reason is that the action in eq.~\eqref{eq:D3_action_hyper} in that case would need an additional boundary term to be gauge invariant. To the best of our knowledge the form of such a boundary term has not been found. We will therefore just pick a gauge that reproduces the correct circular Wilson loop expectation value, namely that used in ref.~\cite{Drukker:2005kx}, in which\footnote{Note that the minus sign on the $C_4$ is due to a different choice of orientation with respect to ref.~\cite{Drukker:2005kx}.}
\begin{equation}
\begin{aligned}
\label{eq:C4:hyp_gauge}
C_4 =\;& - \zeta^2 (\zeta^2-1) \sinh^2 u \sin \theta \, du \wedge d\tau \wedge d \theta \wedge d\phi \\
&+ \frac{\zeta \sinh^2 u \sin \theta (\sinh u - \cos \theta \cosh u)}{\cosh u -\cos \theta \sinh u} \, d \zeta \wedge d\tau \wedge d\theta \wedge d\phi \\
&-  \frac{\zeta \sinh u \sin^2 \theta}{\cosh u -\cos \theta \sinh u} \, d \zeta \wedge d\tau \wedge du \wedge d\phi \,.
\end{aligned}
\end{equation}
When $n=1$, the first term in eq.~\eqref{eq:C4:hyp_gauge} vanishes at the horizon $\zeta= \zeta_h =1$. This is no accident: it is necessary to avoid a singularity at the horizon. When $n \ne 1$, this condition is no longer satisfied because $\zeta_h \ne 1$. In appendix~\ref{app_HEE_C4} we find the requisite gauge transformation to make $C_4$ regular at the horizon, with the result 
\begin{equation}
\begin{aligned}
\label{eq:C4:hyp_gauge_corr}
C_4 =\;& - \zeta^2 f_n(\zeta) \sinh^2 u \sin \theta \, du \wedge d\tau \wedge d \theta \wedge d\phi \\
&+ \frac{\zeta \sinh^2 u \sin \theta (\sinh u - \cos \theta \cosh u)}{\cosh u -\cos \theta \sinh u} \, d \zeta \wedge d\tau \wedge d\theta \wedge d\phi \\
&-  \frac{\zeta \sinh u \sin^2 \theta}{\cosh u -\cos \theta \sinh u} \, d \zeta \wedge d\tau \wedge du \wedge d\phi\,.
\end{aligned}
\end{equation}
The only change is in the first line, where $(\zeta^2-1)$ has been replaced by the function $f_n(\zeta)$ defined in eq.~\eqref{eq:hyperb_metric_n}.

The pull-back of $C_4$ in eq.~\eqref{eq:C4:hyp_gauge_corr} to the D3-brane worldvolume is then 
\begin{equation}
\label{eq:C4:hyp_gauge_D3_brane}
P[C_4] =- \left[\zeta^2 f_n(\zeta) (\partial_\zeta u)  -  \zeta \frac{  \sinh u - \cos \theta \cosh u}{\cosh u -\cos \theta \sinh u} \right] \;  \sinh^2 u  \sin \theta \, d\zeta \wedge d\tau \wedge d \theta \wedge d\phi\,.
\end{equation}
In the following we will show that this choice reproduces the correct EE of both the conformal Wilson line and the Coulomb branch D3-branes. If we also split $I_\mathrm{ct} = I_\mathrm{UV}+I_{U(1)}$, with $I_{U(1)}$ defined in eq.~\eqref{eq:u1}, then the complete action of the D3-brane in eq.~\eqref{eq:D3_action_hyper} is
\begin{equation}
\begin{aligned}
\label{eq:S_D3_action_hyp_intermediate}
\left.I_{D3}(n)\right|_{2\pi} =\;&
4\pi  T_{D3} \int  d  \zeta d\tau  \Bigg\{  \zeta^2 \sinh^2 u\sqrt{\zeta ^2  (\partial_\zeta u)^2
	f_n(\zeta )+\frac{\zeta ^2  ( \partial_\tau u)^2}{f_n(\zeta )}-F_{\tau \zeta}^2+ 1} \\
&  -   \kappa F_{\tau \zeta} +\omega_\pm(u)  \left( \zeta^2 f_n(\zeta) (\partial_\zeta u) \sinh^2 u   + \zeta \left( u -\sinh u \cosh u \right)  \right)\Bigg\}  \\
&+ I_{\text{UV}}\,. 
\end{aligned}
\end{equation}
where $\omega_\pm(u)= \pm$ takes into account a possible change in the orientation of the D3-brane due to the explicit parametrization of the embedding $u(\zeta, \tau)$. In the explicit examples below this will be the case only for the soliton solution studied in section~\ref{sec:EE_W_bos}.
To compute the EE using eq.~\eqref{eq:EE_prob_brane_gen}, we need to evaluate the action in eq.~\eqref{eq:S_D3_action_hyp_intermediate} on-shell. We can easily solve $F_{\tau \zeta}$ equation of motion and plug the result back into the action, finding
\begin{equation}
\label{eq:S_D3_action_hyp_final}
\begin{split}
\left.I_{D3}(n)\right|_{2\pi} =\;& 4\pi T_{D3} \int d \zeta d\tau  \Bigg\{ \,  \sqrt{\left(\kappa^2 +   \zeta ^4\sinh ^4 u\right) \left(1+\zeta^2  (\partial_\zeta u )^2 f_n(\zeta )+\frac{\zeta ^2 ( \partial_\tau u)^2}{f_n(\zeta )}\right)}  \\
&  +   \omega_\pm(u)  \left( \zeta^2 f_n(\zeta) (\partial_\zeta u) \sinh^2 u   + \zeta \left( u -\sinh u \cosh u \right)  \right) \Bigg\} \\
&+ I_{\text{UV}}\,.
\end{split}
\end{equation}
The EE is then given by a derivative with respect to $n$ of this expression. In particular, we need to consider the four terms in $S_A^{(1)}$ in eq.~\eqref{eq:EE_prob_brane_gen}. They respectively correspond to:
\begin{description}
	\item[Term 1:]$\partial/\partial n$ of the integrand in eq.~\eqref{eq:S_D3_action_hyp_final}. The integral of this we call $\mathcal{S}$.
		\item[Term 2:]$\partial/\partial n$ of the boundary term $I_{\text{UV}}$ in eq.~\eqref{eq:S_D3_action_hyp_final}. As mentioned below eq.~\eqref{eq:u1}, $I_{\text{UV}}$ is independent of $n$, so this term vanishes.
	\item[Term 3:]$\partial/\partial n$ acting on the lower limit of the $\zeta$-integration, $\zeta_h$. This is non-zero since \(\z_h\) depends on \(n\) through eq.~\eqref{eq:hyperb_metric_n}. We call this term $\mathcal{S}^{(bdy)}_1$.
	\item[Term 4:]the boundary contribution from the equations of motion of $u(\zeta,\tau)$, which we denote by $\mathcal{S}^{(bdy)}_2$.
\end{description}

In summary, the EE has three distinct contributions:
\begin{equation}
\label{eq:ee3split}
S^{(1)}_A = \mathcal{S} + \mathcal{S}^{(bdy)}_1 + \mathcal{S}^{(bdy)}_2\,.
\end{equation}
For $\mathcal{S}$, we take $\partial/\partial n$ of the integrand of eq.~\eqref{eq:S_D3_action_hyp_final}, evaluate the result at $n=1$, and integrate over $\tau$ and $\zeta$, with the result
\begin{equation}
\label{eq:contr_S}
\mathcal{S} = \frac{8\pi }{3}T_{D3}   \int d \zeta d\tau \left[\frac{\left( (\partial_\zeta u)^2-\frac{(\partial_\tau u)^2}{ f^2_1(\zeta )}\right)
	\sqrt{ \kappa^2+\zeta ^4  \sinh ^4 u}}{2 \sqrt{1+\zeta ^2  (\partial_\zeta u)^2
		f_1(\zeta )+   \frac{\zeta ^2 (\partial_\tau u)^2}{f_1(\zeta )} }}  +  \omega_\pm(u)(\partial_\zeta u) \sinh ^2 u\right].
\end{equation}
For the contribution $\mathcal{S}^{(bdy)}_1$ from the limit of integration, we find
\begin{equation}
\label{eq:contr_bdy_S1}
\begin{aligned}
\mathcal{S}^{(bdy)}_1 =\;&   \frac{8\pi^2}{3} T_{D3}  \bigg\{  \sqrt{\left(\kappa^2 +   \zeta ^4\sinh ^4 u_h\right) \left(1+\zeta_h^2  (\partial_\zeta u_h )^2 f_n(\zeta )+\frac{\zeta ^2 ( \partial_\tau u_h)^2}{f_n(\zeta_h )}\right)} \\
& +   \omega_\pm(u_h)\left[  \zeta_h^2 f_n(\zeta_h) (\partial_\zeta u_h) \sinh^2 u_h   + \zeta_h \left( u_h -\sinh u_h \cosh u_h \right)    \right] \bigg\}\,,
\end{aligned}
\end{equation}
with $u_h \equiv u(\zeta_h,\tau)$ and $\partial_\alpha u_h \equiv (\partial_\alpha u) (\zeta_h,\tau)$, where $\alpha=\zeta, \tau$, and $\zeta_h=1$ because $n=1$. For the contribution $\mathcal{S}^{(bdy)}_2$ coming from the equations of motion's boundary term, we find
\begin{equation}
\label{eq_S_bdy}
\begin{aligned}
\mathcal{S}^{(bdy)}_2  = - 4\pi T_{D3}  \int_0^{2\pi} d\tau   \Bigg\{ \,  \sqrt{\frac{\kappa^2 +   \zeta_h ^4\sinh ^4 u_h}{ 1+\zeta_h^2  (\partial_\zeta u_h )^2 f_n(\zeta )+\frac{\zeta_h^2 ( \partial_\tau u_h)^2}{f_n(\zeta_h )}}} \zeta_h^2 (\partial_\zeta u_h ) f_n(\zeta_h)  \\ 
+  \omega_\pm(u_h) \,\zeta_h^2 f_n(\zeta_h) \sinh^2 u_h    \Bigg\} \partial_n u_h \Bigg|_{n =1}&\,.
\end{aligned}
\end{equation}
To evaluate eq.~\eqref{eq_S_bdy}, we need $\partial_n u_h$. In general this can be very complicated to compute because in most cases we do not know the solution to the equations of motion when $n \ne 1$. Luckily, we can show that whenever this term is non-vanishing, it is completely fixed by the on-shell solution at $n=1$. To do so, we start with the following ansatz for the behaviour of the embedding close to the hyperbolic horizon, $\zeta \sim \zeta_h$,
\begin{equation}
\label{eq:anasatz_exp_hor}
u (\zeta,\tau) = u_n^{(0)}(\tau) + u_n^{(1)}(\tau) \sqrt{\zeta-\zeta_h} + u_n^{(2)}(\tau) (\zeta-\zeta_h) +\mathcal{O}\left[(\zeta -\zeta_h)^{3/2}\right].
\end{equation}
The equation of motion at lowest order in a small \((\z-\z_h)\) expansion gives $ \partial_\tau u_n^{(0)}(\tau)  = 0$, which is solved by constant $u_n^{(0)}(\tau) = u_n^{(0)}$. At the next order, the equation of motion gives
\beq
\zeta_ h^2 \,  \partial_\tau^2 u_n^{(1)}(\tau )+\left(1-2 \zeta_h^2\right)^2 u_n^{(1)}(\tau ) =0\,,
\eeq
with general solution
\beq
\label{eq:eq_for_u_1^1}
u_n^{(1)}(\tau) = c_1(n) \sin \left( \frac{\tau}{n} \right) + c_2 (n) \cos \left( \frac{\tau}{n} \right)\,.
\eeq
In principle, we should fix the integration constants $c_1(n)$ and $c_2(n)$ for all $n$. However, we will see that this is not necessary because the final result will depend only on the coefficients evaluated at $n=1$, and not on their derivatives with respect to $n$. In other words, knowing the embedding at $n=1$ is enough. Indeed, plugging our solutions for $u_n^{(0)}(\tau)$ and $u_n^{(1)}(\tau)$ into eq.~\eqref{eq_S_bdy} gives
\begin{equation}
\mathcal{S}^{(bdy)}_2 = -2 \sqrt{2} \pi T_{D3}\int_0^{2\pi} d\tau  \left[ \frac{\frac{1}{\sqrt{8 n^2+1}}+1}{4 n^2}\frac{  \zeta_h^3 \left(2 \zeta_h^2-1\right)  \sqrt{\kappa^2 +\sinh ^4\bigl(u_n^{(0)}\bigr)} \;\bigl(u_n^{(1)}(\tau)\bigr)^2}{\sqrt{\frac{\zeta_h^3 (\partial_\tau u_n^{(1)}(\tau))^2}{2 \zeta_h^2-1}+\left(2 \zeta_h^2-1\right) \zeta_ h
		\bigl(u_n^{(1)}(\tau)\bigr)^2+2}}\right]_{n =1}.
\end{equation}
The limit $n \rightarrow 1$, for which $\zeta_h \rightarrow 1$, is non-singular, and gives
\begin{equation}
\label{eq:S_bdy_2_general}
\mathcal{S}^{(bdy)}_2 = -\frac{2 \sqrt{2} \pi}{3} T_{D3}\int_0^{2\pi} d\tau \,\sqrt{\frac{\kappa^2 + \sinh ^4u_1^{(0)}}{2+ \big(\partial_\tau u_1^{(1)}(\tau)\bigr)^2+ \bigl(u_1^{(1)}(\tau)\bigr)^2} } \, \bigl(u_1^{(1)}(\tau)\bigr)^2.
\end{equation}

Note that $\mathcal{S}^{(bdy)}_2\neq0$ only if $u_1^{(1)}(\tau)\neq0$. In the examples below $u_1^{(1)}(\tau)\neq0$ in all cases except the conformal Wilson line. In other words, $\mathcal{S}^{(bdy)}_2\neq0$ in all of our \textit{non-conformal} examples. This is no surprise. Any configuration that preserves conformal symmetry will map to a $\tau$-independent embedding. The equation of motion eq.~\eqref{eq:eq_for_u_1^1} then forces $u_1^{(1)}(\tau)=0$ and so $\mathcal{S}^{(bdy)}_2=0$. For example, as mentioned in section~\ref{sec:D3_conf_wils}, the Wilson line D3-brane has worldvolume $AdS_2 \times S^2$. For a spherical region centred on the Wilson line, the coordinate transformation in eq.~\eqref{eq:CHM_mapping} produces a D3-brane extended along $\tau$ and wrapping the equator of the hyperbolic plane. In particular, its embedding $u(\zeta,\tau)$ is $\tau$-independent (see eq.~\eqref{eq:D3WLhyp} below). Displacing the sphere and/or introducing a non-conformal brane introduces at least one other scale besides $R$, and hence breaks conformal symmetry. Generically, the coordinate transformation in eq.~\eqref{eq:CHM_mapping} will then produce a $\tau$-dependent embedding, and thus $u_1^{(1)}(\tau)\neq0$ and $\mathcal{S}^{(bdy)}_2\neq0$, as we will see in our non-conformal examples below.

In writing $\mathcal{S}_1^{(bdy)}$ and $\mathcal{S}_2^{(bdy)}$ we implicitly assumed that the D3-brane has precisely one connected component at the horizon $\zeta_h=1$. In Poincar\'e coordinates, this corresponds to the RT surface intersecting the brane only once. This is not always the case. If the D3-brane has multiple disconnected components at the hyperbolic horizon, we must sum over all of them. As we will see, this happens for the spherical soliton in section~\ref{sec:EE_W_bos}. Another possibility is that the D3-brane does not reach the hyperbolic horizon, i.e. the RT surface does not intersect the D3-brane. In that case, $\mathcal{S}_1^{(bdy)}=0$ and $\mathcal{S}_2^{(bdy)}=0$ identically.

In order to compute the $S^{(1)}_A$ we need specific solutions for the embedding $u(\zeta,\tau)$ to plug into eqs.~\eqref{eq:contr_S}, \eqref{eq:contr_bdy_S1} and~\eqref{eq:S_bdy_2_general}. In the following subsections we compute $S^{(1)}_A$ for the four D3-brane embeddings reviewed in section~\ref{sec:D3_brane}.

\subsection{Conformal Wilson line}
\label{sec:EE_Conf_Wilson}

The conformal Wilson line in the $k$-th symmetric representation, reviewed in section~\ref{sec:D3_conf_wils}, is the simplest case we consider. As mentioned in sections~\ref{sec:string} and~\ref{sec:D3_conf_wils}, an exact result for the EE appears in ref.~\cite{Lewkowycz:2013laa}, and a holographic calculation using the Karch-Uhlemann method appears in ref.~\cite{Kumar:2017vjv}. Here, we review the latter computation both for completeness and to highlight the salient steps that will be important for the following cases.

The D3-brane embedding corresponding to the conformal Wilson line was given in eq.~\eqref{eq:Conf_Wils_emb}. In the hyperbolic coordinates of eq.~\eqref{eq:hyperb_metric} it becomes 
\begin{equation}
\label{eq:D3WLhyp}
u(\zeta,\tau) = \sinh^{-1} \left(\kappa/\zeta\right)\,,
\end{equation}
which is manifestly independent of $\tau$,  and $\omega_\pm(u) =+$. As discussed below eq.~\eqref{eq:S_bdy_2_general}, we thus have $\mathcal{S}_2^{(bdy)}=0$. The other two contributions to $S^{(1)}_A$ in eq.~\eqref{eq:ee3split} are
\begin{eqnarray}
\mathcal{S} &= &-\frac{8 \pi }{3}\, T_{D3} \,\int_0^{2\pi} d\tau \int_1^{\infty} d \zeta \, \frac{\kappa^3}{2 \zeta^3 \sqrt{\kappa^2+ \zeta^3}} = \frac{4 \pi^2 }{3} T_{D3} \left(\sinh ^{-1}\kappa- \kappa \sqrt{\kappa^2+1}\right),  \nonumber \\
\mathcal{S}^{(bdy)}_1 &= &\frac{8 \pi^2 }{3} \, T_{D3} \,\sinh^{-1}\kappa \,.
\end{eqnarray}
Denoting the symmetric representation Wilson line EE as $S^{(1)}_A = S_\mathrm{symm}$, we thus have
\begin{equation}
\label{eq:EE_conf_wilson}
S_\mathrm{symm} = \frac{4 \pi^2 }{3} \, T_{D3} \, \left(3\sinh ^{-1}\kappa- \kappa \sqrt{\kappa^2+1}\right) = \frac{2}{3}\, N\, \left(3\sinh ^{-1}\kappa- \kappa \sqrt{\kappa^2+1}\right),
\end{equation} 
where in the second equality we used $T_{D3} = N/(2\pi^2)$. As mentioned above, eq.~\eqref{eq:EE_conf_wilson} agrees with the exact result of ref.~\cite{Lewkowycz:2013laa}. Also note that since this case is conformal, the Wilson line's contribution to the EE is independent of the entangling region's radius $R$.

\subsection{Coulomb branch}
\label{sec:EE_Coul_branch}

The embedding of the Coulomb branch brane was given in eq.~\eqref{eq:Coulb_branch_solution}, and has $\kappa =0$. In the hyperbolic coordinates of eq.~\eqref{eq:hyperb_metric} this becomes
\begin{equation}
\label{eq:coul_branc_hyp_emb}
u(\zeta,\tau)= \cosh^{-1} \left(\frac{R v-\sqrt{\zeta ^2 -1} \cos \tau    }{\zeta }\right)\,.
\end{equation}
As we discussed below eq.~\eqref{eq:S_bdy_2_general}, for this non-conformal D3-brane $u(\zeta,\tau)$ depends on $\tau$.

In this case, all three contributions to $S^{(1)}_A = \mathcal{S} + \mathcal{S}^{(bdy)}_1 + \mathcal{S}^{(bdy)}_2$ in eq.~\eqref{eq:ee3split} are non-zero. For the first contribution, we plug eq.~\eqref{eq:coul_branc_hyp_emb} into eq.~\eqref{eq:contr_S} to find
\begin{equation}
\label{eq:S_coul_branch1}
\begin{split}
\mathcal{S} = \frac{4\pi }{3} T_{D3}  \int d \zeta d\tau \left\{\frac{1}{Rv}\left[ (\partial_\zeta u)^2-\frac{(\partial_\tau u)^2}{ (\zeta^2-1)^2}\right]
\zeta^2\sinh^3 u  +2(\partial_\zeta u) \sinh ^2 u\right\}\,,
\end{split}
\end{equation}
where we used $\omega_\pm(u)=+$. In fact, we can show that this integral reduces to a boundary term as follows. We use
\begin{align}
\partial_\zeta u &=- \frac{1}{\zeta^2 \,\sinh u} \left( Rv + \frac{\cos \tau}{\sqrt{\zeta^2-1}} \right), & \partial_\tau u &= \frac{\sqrt{\zeta^2-1}}{\zeta \sinh u} \sin \tau\,,
\end{align}
to show that the first term of the integrand in eq.~\eqref{eq:S_coul_branch1} can be written as
\begin{equation}
\frac{1}{Rv}\left[ (\partial_\zeta u)^2-\frac{(\partial_\tau u)^2}{ (\zeta^2-1)^2}\right]\zeta^2\sinh^3 u= \varepsilon_{ \alpha \beta} \partial_\alpha J_\beta - (\partial_\zeta u) \sinh^2 u\,, 
\end{equation}
where $\varepsilon_{\alpha \beta}$ is the 2d Levi-Civita symbol with $\varepsilon_{\zeta\tau}=-\varepsilon_{\tau\zeta}=1$, and we defined
\beq
J_\alpha \equiv \frac{\sinh^2 u \, \sin \tau}{Rv \sqrt{\zeta^2 -1}}\, \partial_\alpha u\,.
\eeq
The second term of the integrand in eq.~\eqref{eq:S_coul_branch1} is also a total derivative, namely
\begin{equation}
2(\partial_\zeta u) \sinh ^2 u = - \partial_\zeta (u -\sinh u \cosh u)\,.
\end{equation}
We can thus finally write
\begin{equation}
\label{eq:coul_branch_S_cont}
\mathcal{S} = \frac{4\pi }{3}\, T_{D3}\,  \int  d \zeta d\tau   \left[ \varepsilon_{ \alpha \beta} \partial_\alpha J_\beta  - \frac{1}{2}\partial_\zeta (u -\sinh u \cosh u) \right].
\end{equation}
A boundary only exists if $Rv>1$, i.e. when the RT surface intersects the D3-brane. This means that if $Rv<1$ then $\mathcal{S}=0$.\footnote{This can be easily verified by explicit computation.} For $Rv>0$, we can use Stokes' theorem to evaluate the first term in eq.~\eqref{eq:coul_branch_S_cont} and the divergence theorem for the second term. We thus obtain 
\begin{equation}
\label{eq:S_coul_branch}
\begin{split}
\mathcal{S} & = - \frac{4\pi }{3} \, T_{D3}\, \int_0^{2\pi} d \tau \left[  J_\tau  - \frac{1}{2} (u-\cosh u \sinh u) \right]_{\zeta=1} \\
& =  \frac{4\pi^2 }{3} \, T_{D3}\, \left[ \cosh^{-1} ( Rv  )- \left( Rv + \frac{1}{Rv} \right) \sqrt{(Rv)^2-1}  \right], \qquad Rv>1\,,
\end{split}
\end{equation} 
where we used $u_h\equiv u(1,\tau) = \cosh^{-1}(Rv)$.

The second contribution is straightforward to compute from eq.~\eqref{eq:contr_bdy_S1}, and we find
\begin{equation}
\label{eq:S_bdy_1_coul}
\mathcal{S}^{(bdy)}_1  =  \frac{8\pi^2}{3} T_{D3}  \cosh^{-1} (Rv)\,, \qquad Rv >1\,,
\end{equation}
where we used $f_n(\zeta_h) (\partial_\zeta u_h) \rightarrow 0$ when $\zeta_h \rightarrow 1$. 

The third and final contribution is $\mathcal{S}_{2}^{(bdy)}$ in eq.~\eqref{eq:S_bdy_2_general}. Expanding the embedding $u(\zeta,\tau)$ in eq.~\eqref{eq:coul_branc_hyp_emb} near the horizon as in eq.~\eqref{eq:anasatz_exp_hor}, extracting the coefficients $u_1^{(0)}$ and $u_1^{(1)}(\tau)$, and plugging them into eq.~\eqref{eq:S_bdy_2_general}, we find
\begin{equation}
\label{eq:S_bdy_2_coul}
\mathcal{S}^{(bdy)}_2 =- \frac{4\pi}{3} \, T_{D3}\,\int_0^{2\pi} d\tau \, \frac{\sqrt{(Rv)^2-1}}{Rv}\cos^2 \tau = - \frac{4\pi^2 }{3}\, T_{D3}\, \frac{\sqrt{(Rv)^2-1}}{Rv} \,.
\end{equation}

\begin{figure}
	\begin{center}
		\includegraphics[scale=0.55]{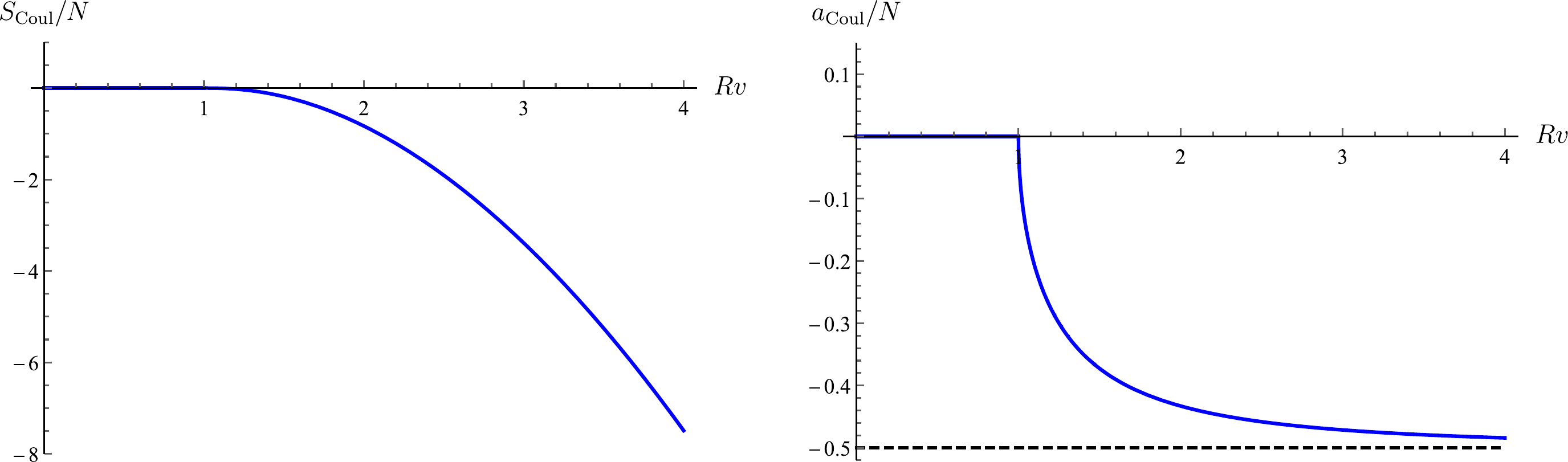}
		\caption{\textbf{Left:} Contribution of the Coulomb branch D3-brane to EE, $S_\mathrm{Coul}$ in eq.~\eqref{eq:S_coul_analytic}, divided by $N$, as a function of $Rv$. \textbf{Right:} The quantity $a_\mathrm{Coul}(R)$ defined in eq.~\eqref{eq:a_coul} as a function of $Rv$. The function is continuous but not analytic at $Rv=1$. }
		\label{fig:EE_coul}
	\end{center}
\end{figure}

Summing all three contributions in eqs.~\eqref{eq:S_coul_branch}, \eqref{eq:S_bdy_1_coul} and \eqref{eq:S_bdy_2_coul}, and denoting the contribution of the Coulomb branch D3-brane to the EE as $S^{(1)}_A = S_\mathrm{Coul}$, we find
\begin{equation}
\label{eq:S_coul_analytic}
S_\mathrm{Coul}=\begin{cases}
0\,, & Rv<1\,, \\
\frac{ 2}{3} \, N \, \left[3 \cosh ^{-1}(Rv)-\left( Rv + \frac{2}{Rv} \right) \sqrt{(Rv)^2-1}\right],  & Rv > 1\,.
\end{cases}
\end{equation}
This is the first original result of this paper. As mentioned in section~\ref{sec:coul_branch}, $S_\mathrm{Coul}$ depends only on the dimensionless combination $Rv$. We plot eq.~\eqref{eq:S_coul_analytic} in figure~\ref{fig:EE_coul} on the left.

In appendix~\ref{app:EE_coul_backReac} we show that eq.~\eqref{eq:S_coul_analytic} agrees with a calculation using the RT method in the fully back-reacted geometry describing the Coulomb branch, as discussed in section~\ref{sec:coul_branch}. This agreement of course is only possible because $\mathcal{S}^{(bdy)}_2$ in eq.~\eqref{eq:S_bdy_2_coul}, i.e. the boundary term coming from the D3-brane's equations of motion, is non-zero.

For a $(3+1)$-dimensional Poincar\'e-invariant QFT describing an RG flow from a UV CFT with central charge $a_\mathrm{UV}$ to an IR CFT with central charge $a_\mathrm{IR}$, the $a$-theorem is the statement that unitarity requires $a_\mathrm{UV} \geq a_\mathrm{IR}$~\cite{Cardy:1988cwa,Komargodski:2011vj}. As discussed in section~\ref{sec:coul_branch}, the Coulomb branch D3-brane describes an RG flow from a UV CFT, namely $\N=4$ SYM with gauge group $SU(N)$, to an IR CFT, namely $\N=4$ SYM with gauge group $SU(N-1)\times U(1)$. In the large-$N$ limit we thus have $a_\mathrm{UV} \approx N^2/4$ and $a_\mathrm{IR} \approx (N-1)^2/4 \approx N^2/4 - N/2$, where we dropped terms of $\mathcal{O}(N^0)$. Hence, the $a$-theorem is obeyed, as expected.

We can show that our $S_\mathrm{Coul}$ in eq.~\eqref{eq:S_coul_analytic} obeys the $a$-theorem, as follows. Eq.~\eqref{eq:S_coul_analytic} vanishes in the UV limit $Rv \ll 1$. In that case the only contribution to the EE comes from the RT calculation in $AdS_5 \times S^5$~\cite{Ryu:2006bv,Ryu:2006ef},
\begin{equation}
\label{eq_EE_N4}
S_{SU(N)} =N^2 \left[ \frac{R^2}{\varepsilon^2} - \log \left( \frac{R}{\varepsilon} \right) + \mathcal{O}(1) \right],
\end{equation}
where $\varepsilon$ is the UV cut-off. Eq.~\eqref{eq_EE_N4} has the form expected for a $(3+1)$-dimensional CFT. The first term on the right-hand side is the area-law term, which is $\varepsilon$-dependent and hence unphysical. The logarithm in the second term also depends on $\varepsilon$, however its coefficient is independent of rescalings of $\varepsilon$ and hence is physical. Indeed, that coefficient is $-4\,a_\mathrm{UV}$.

In the IR limit $Rv \gg 1$ we still have the contribution from the $AdS_5 \times S^5$ background in eq.~\eqref{eq_EE_N4}, but now $S_\mathrm{Coul}$ in eq.~\eqref{eq:S_coul_analytic} also contributes. When $Rv \gg 1$, eq.~\eqref{eq:S_coul_analytic} gives
\begin{equation}
\label{eq:EE_coul_LargeBeha}
S_\mathrm{Coul} \approx 
\frac{2}{3} \,N \,\left[ - (Rv)^2 + 3 \log (Rv) + 3 \log 2 +\frac{3}{2}   \right] + \mathcal{O}\left((Rv)^{-2}\right) \,, \qquad Rv \gg 1\,.
\end{equation}
At leading order, this has the CFT form of eq.~\eqref{eq_EE_N4}, including an area-law term and logarithm, but now with cutoff $1/v$. This is intuitive: in the IR we expect the cutoff to be the inverse W-boson mass, which is indeed $\propto 1/v$. The probe D3-brane's contribution to the logarithm's coefficient is $2N$, hence we reproduce $a_\mathrm{IR} = N^2/4 - N/2$, with the $N^2/4$ from the $AdS_5 \times S^5$ background and the $-N/2$ from the probe D3-brane. Our results thus obey the $a$-theorem exactly as expected.

However, a stronger version of the $a$-theorem exists, namely the entropic $a$-theorem of ref.~\cite{Casini:2017vbe}. For the EE of a spherical region of radius $R$, $S(R)$, we define an effective position-dependent central charge,
\beq
\label{eq:effa}
a(R) = \frac{1}{8} \left[ R^2 \frac{d^2}{d R^2}S(R) - R \frac{d}{d R}S(R)\right],
\eeq
which in the UV limit $R \to 0$ obeys $a(R) \to a_\mathrm{UV}$, and in the IR limit $R \to \infty$ obeys $a(R) \to a_\mathrm{IR}$. The entropic $a$-theorem is then the statement that strong subadditivity of EE and the Markov property of a CFT vacuum require~\cite{Casini:2017vbe},
\begin{equation}
\label{eq:a_entropic_theorem}
a(R) \leq a_\mathrm{UV}\,.
\end{equation}
The effective central charge $a(R)$ in eq.~\eqref{eq:effa} is an entropic $a$-function, defined not only at the fixed points of the RG flow, but for all scales in between. The entropic $a$-theorem is thus a constraint for all $R$, not just the UV and IR. However, when evaluated in the IR limit $R \to \infty$, eq.~\eqref{eq:a_entropic_theorem} reproduces the original $a$-theorem. Notice the entropic $a$-theorem does not require $a(R)$ to be monotonic in $R$, but simply imposes an upper limit.

We can show that our $S_\mathrm{Coul}$ obeys the entropic $a$-theorem as follows. Eq.~\eqref{eq:S_coul_analytic} gives
\beq
\label{eq:a_coul}
a_\mathrm{Coul}(R) = \begin{cases} 0\,, & Rv<1\,, \\ - \dfrac{N}{2} \, \dfrac{\sqrt{(Rv)^2-1}}{Rv}\,, & Rv>1\,,\end{cases}
\eeq
where \(a_\mathrm{Coul}(R) \equiv a(R) - a_\mathrm{UV}\),
and hence for the total EE, $S(R) = S_{SU(N)} + S_\mathrm{Coul}$,
\begin{equation}
\label{eq:entrop_RG_flow_coul}
a(R) = \begin{cases}
\dfrac{N^2}{4}\,, & Rv<1\,,\\[1em]
\dfrac{N^2}{4} -\dfrac{N}{2} \dfrac{\sqrt{(Rv)^2-1}}{Rv}\,,  & Rv > 1\,,
\end{cases}
\end{equation}
which is clearly always $\leq a_\mathrm{UV} = N^2/4$, hence the entropic $a$-theorem is obeyed, as expected.

Figure~\ref{fig:EE_coul} on the left makes clear that our $S_\mathrm{Coul}$ is continuous and decreases monotonically as $Rv$ increases. However, although our entropic $a$-function $a(R)$ in eq.~\eqref{eq:entrop_RG_flow_coul} is also continuous and decreases monotonically as $Rv$ increases, it is not analytic at $Rv=1$, because $S_\mathrm{Coul}$ in eq.~\eqref{eq:S_coul_analytic} is only twice differentiable. The non-analyticity of our $a(R)$ at $Rv=1$ is clear in figure~\ref{fig:EE_coul} on the right, where we plot $a_\mathrm{Coul}(R)/N = a(R)/N - N/4$.

We can also show that $S_\mathrm{Coul}$ obeys the four-dimensional ``area theorem'' of refs.~\cite{Casini:2016udt,Casini:2017vbe}, which in our notation states that $\frac{1}{2R} \p_R S_\mathrm{Coul}(R)$ monotonically decreases as $R$ increases. From eq.~\eqref{eq:S_coul_analytic} we straightforwardly find
\begin{equation}
\label{eq:coulomb_area_theorem}
\frac{1}{2R} \p_R S_\mathrm{Coul}(R) = \begin{cases}
0\,, & R v < 1\,,
\\
- \dfrac{2}{3} N v^2 \left[1 - \dfrac{1}{(R v)^2} \right]^{3/2}, & R v > 1\,,
\end{cases}
\end{equation} 
which indeed decreases monotonically from zero at $R=0$ to \(-\frac{2}{3} N v^2\) at $R \to \infty$, thus satisfying the area theorem. The boundary term $\mathcal{S}_{2}^{(bdy)}$ is crucial for this. If we neglect the contribution of \(\mathcal{S}_2^{(bdy)}\) to the EE, then the right-hand side of eq.~\eqref{eq:coulomb_area_theorem} receives an additive contribution \(\frac{1}{3} N R^{-4}v^{-2} \left[1 - (R v)^{-2}\right]^{-1/2}\) for \(R v > 1\). This diverges to \(+\infty\) as \(Rv \to 1^+\). In that case, as $Rv$ increases through $Rv=1$, $\frac{1}{2R} \p_R S_\mathrm{Coul}(R)$ jumps from zero to $+\infty$. Clearly that would not be a monotonic decrease, and the area theorem would be violated.

\subsection{Screened Wilson line}
\label{sec:EE_Screen_Wilson}

The embedding of the D3-brane describing a screened Wilson line was given in eq.~\eqref{eq:screenedsol}, which in the hyperbolic coordinates of eq.~\eqref{eq:hyperb_metric} is given implicitly by
\begin{equation}
\label{eq:screende_Wilson_embedding}
\zeta \sinh u=\frac{\kappa \left(\sqrt{\zeta ^2-1} \cos \tau +\zeta  \cosh u\right)}{-Rv+\sqrt{\zeta ^2-1} \cos \tau+\zeta  \cosh u}\,.
\end{equation}
As we discussed below eq.~\eqref{eq:S_bdy_2_general}, for this non-conformal D3-brane $u(\zeta,\tau)$ depends on $\tau$.

In this case, all three contributions to $S^{(1)}_A = \mathcal{S} + \mathcal{S}^{(bdy)}_1 + \mathcal{S}^{(bdy)}_2$ in eq.~\eqref{eq:ee3split} are non-zero.   Furthermore, also in this case we have $\omega_\pm(u)=+$. We evaluate $\mathcal{S}$ numerically. Some details of our numerical evaluation appear in appendix~\ref{app_screend_W_line}. The boundary term arising from the derivative of the limits of integration is
\begin{equation}
\mathcal{S}^{(bdy)}_1 = \frac{8 \pi^2}{3} T_{D3}  \left(\frac{Rv \left(\kappa^2+\sinh ^4u_h \right)}{Rv \kappa \cosh u_h+\sinh u_h \left(\kappa-\sinh u_h\right)^2}+ u_h-\cosh u_h \sinh u_h\right),
\end{equation}
where $u_h$ is a function of $\k$ and $Rv$ defined from eq.~\eqref{eq:screende_Wilson_embedding} by setting $\zeta=1$ and solving the resulting equation,
\begin{equation}
\label{eq:screenedWLuh}
\sinh u_h=\frac{\kappa \cosh u_h}{\cosh u_h - Rv}\,.
\end{equation}
Eq.~\eqref{eq:screenedWLuh} has only one real and positive solution, which we will not write explicitly since the expression is rather cumbersome. To compute the boundary term of the equations of motion, $\mathcal{S}^{(bdy)}_2$, we start from $u(\zeta,\tau)$ in eq.~\eqref{eq:screende_Wilson_embedding}, perform the expansion in eq.~\eqref{eq:anasatz_exp_hor} with $n=1$, extract $u_1^{(0)}(\tau)=u_h$ and $u_1^{(1)}(\tau)$ (both of which are non-zero), plug these into eq.~\eqref{eq:S_bdy_2_general} for $\mathcal{S}^{(bdy)}_2$, and perform the integration over $\tau$, obtaining
\begin{equation}
\mathcal{S}^{(bdy)}_2 = -\frac{4 \pi^2}{3}T_{D3}\frac{(Rv)^2 \kappa^2 \sqrt{\frac{\kappa^2+\sinh ^4 u_h}{(Rv)^2 \kappa^2+\left[Rv \kappa \sinh u_h+\cosh u_h (Rv-\cosh u_h)^2\right]^2}}}{ Rv \kappa \sinh	u_h+\cosh u_h (Rv-\cosh u_h)^2 }\,.
\end{equation}

\begin{figure}
	\centering
	\begin{center}
		\includegraphics[scale=0.56]{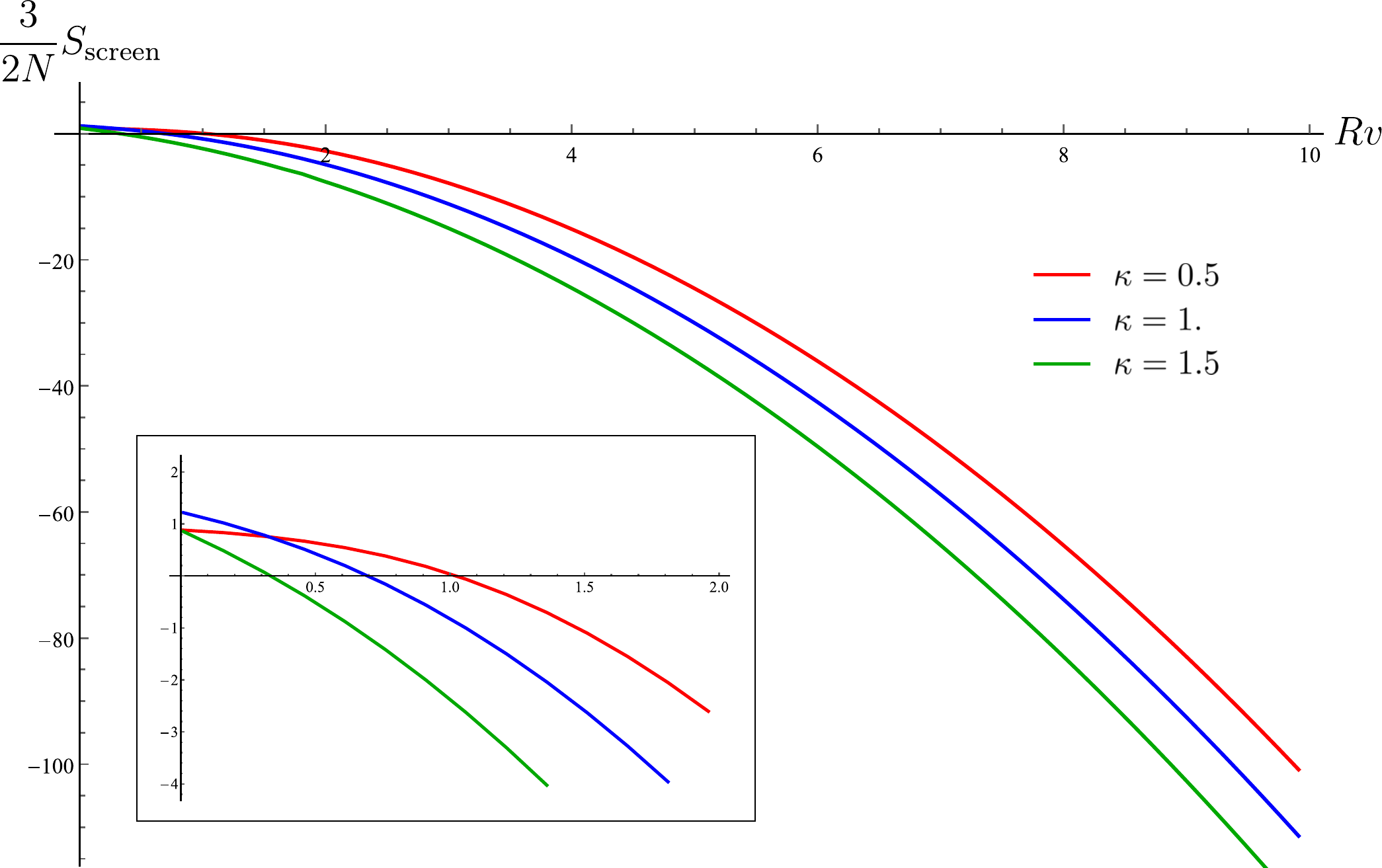}
	\end{center}
	\caption{ Our numerical result for the EE of the screened Wilson line, $S_\mathrm{screen}/(2N/3)$, as a function of $Rv$ for $\kappa=0.5$ (red), $1$ (blue), and $1.5$ (green). The inset shows that $S_\mathrm{screen}/(2N/3)$ approaches a non-zero value in the limit $Rv \to 0$, namely the value for the conformal Wilson line in eq.~\eqref{eq:EE_conf_wilson}.}
	\label{plot:fig_EE_wilson}
\end{figure}

We denote this probe D3-brane contribution to the EE as $S^{(1)}_A = S_\mathrm{screen}$. We show in figure~\ref{plot:fig_EE_wilson} our results for $S_\mathrm{screen}/(2N/3)$ as a function of $Rv$, for several values of $\k$. In all cases this EE decreases monotonically as $Rv$ increases. In contrast, the result for $S_\mathrm{screen}$ in ref.~\cite{Kumar:2017vjv} had a maximum. The discrepancy is due to the boundary term from the equations of motion, $\mathcal{S}^{(bdy)}_2$, which was neglected in ref.~\cite{Kumar:2017vjv}. However, as mentioned in sections~\ref{sec:intro} and~\ref{sec:screend_wils_line}, we know of no physical principle that requires $S_\mathrm{screen}$ to be monotonic in $Rv$.

Since the EE depends on \(R\) and \(v\) only through their product \(Rv\), the \(R = 0\) and \(v \to 0\) limits are equivalent. Hence, at small \(Rv\) we find that \(S_\mathrm{screen}\) approaches the EE of the conformal Wilson line in eq.~\eqref{eq:EE_conf_wilson}, $S_\mathrm{symm}$, which had $v=0$. The inset in figure~\ref{plot:fig_EE_wilson} shows \(S_\mathrm{screen}\) near $Rv \to 0$, which indeed approaches the non-zero value in eq.~\eqref{eq:EE_conf_wilson}.

In the IR limit $Rv \to \infty$, we expect $S_\mathrm{screen}$ to approach the EE of the Coulomb branch in the same limit, $S_\mathrm{Coul}$ in eq.~\eqref{eq:EE_coul_LargeBeha}. As we move towards smaller $Rv$, we expect a different set of finite-$(Rv)$ corrections to those in $S_\mathrm{Coul}$. As mentioned in section~\ref{sec:intro}, we have found a simple and intuitive derivation of these corrections, up to order $1/(Rv)$, as follows.

\begin{figure}
	\begin{center}
		\includegraphics[scale=0.47]{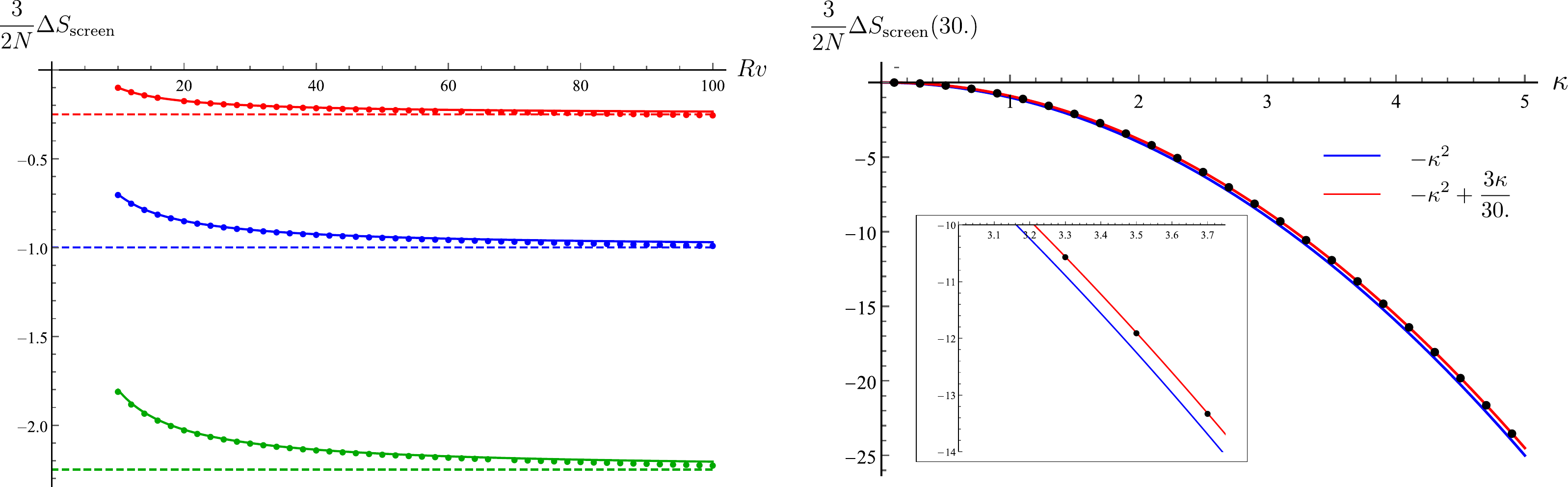}
		\caption{\textbf{Left:} The dots are our numerical results for $\Delta S_\mathrm{screen}$ defined in eq.~\eqref{eq:deltascreen}, divided by $2N/3$, as a function of the region's size $Rv$ for $\kappa=0.5$ (red), $1$ (blue), and $1.5$ (green). The solid lines are the expression in eq.~\eqref{eq:coulreplace} without the $\mathcal{O}\left((Rv)^{-2}\right)$ corrections, and the horizontal dashed lines are $-\k^2$. \textbf{Right:} The dots are our numerical results for $\Delta S_\mathrm{screen}/(2N/3)$ for $Rv = 30$, for various $\k$. The solid blue line is $-\k^2$ while the solid red line is $-\k^2 + 3\k/(Rv)$. The inset is a close-up that clearly shows the latter is a better approximation to our numerical results than the former.}
		\label{plot:EE_wilson_large}
	\end{center}
\end{figure}

As mentioned below eq.~\eqref{eq:EE_coul_LargeBeha}, on the Coulomb branch at large $Rv$ the UV cutoff is $1/v$, which is proportional to the inverse W-boson mass. In the holographic picture of the Coulomb branch in figure~\ref{fig:cartoons}b, the W-boson is a string stretched from the probe D3-brane to the Poincar\'e horizon. In the Maldacena limit the mass of such a string is $v/(2\pi\alpha')$ (in our units where $L\equiv 1$). For the screened Wilson line of figure~\ref{fig:cartoons}c, the mass of such a string clearly increases upon approaching the Wilson line at $\rho=0$. More specifically, such a string's mass is determined by the D3-brane's embedding, $(v+\k/\rho)/(2\pi\alpha')$. In CFT terms, the W-boson acquires a position-dependent mass, hence the UV cutoff becomes position-dependent. We thus introduce the position-dependent effective cutoff,
\beq
\label{eq:veffscreened}
v_\mathrm{eff}(\r) = v+\frac{\k}{\rho}\,.
\eeq
We then find numerically that if we start with the Coulomb branch result $S_\mathrm{Coul}$ at large $Rv$ in eq.~\eqref{eq:EE_coul_LargeBeha}, make the replacement $v \to v_\mathrm{eff}$, and then expand in $Rv \gg 1$, the result agrees precisely with our $S_\mathrm{screen}$ at large $Rv$, up to order $1/(Rv)$. Explicitly, for $Rv \gg1  $,
\begin{eqnarray}
\label{eq:coulreplace}
S_\mathrm{screen}(Rv) & = & S_\mathrm{Coul}(Rv_\mathrm{eff}) + \mathcal{O}\left((Rv_\mathrm{eff})^{-2}\right)   
\nonumber \\ & = & \frac{2}{3} \, N \,\left[ - (R\,v_\mathrm{eff}(R))^2 + 3 \log (R\,v_\mathrm{eff}(R)) + 3 \log 2 +\frac{3}{2}  \right] + \mathcal{O}\left((R\,v_\mathrm{eff}(R))^{-2}\right)   \nonumber\\
& = & S_\mathrm{Coul}(Rv) + \frac{2}{3} \, N\, \left(- 2 \,\k \,Rv - \k^2 + \frac{3\k}{Rv}\right)+ \mathcal{O}\left((Rv)^{-2}\right)\,
\end{eqnarray}
agrees with our numerical results for $S_\mathrm{screen}$ up to order $1/(Rv)$. Indeed, in figure~\ref{plot:EE_wilson_large} on the left we show our numerical results for
\beq
\label{eq:deltascreen}
\Delta S_\mathrm{screen} \equiv S_\mathrm{screen}-S_\mathrm{Coul} + \frac{2}{3} N (2 \k \,Rv)\,,
\eeq
divided by $\frac{2}{3} N$, which clearly approaches $-\k^2$ as $Rv \to \infty$, consistent with eq.~\eqref{eq:coulreplace}. Furthermore, at large but finite $Rv$ we show in figure~\ref{plot:EE_wilson_large} on the right that $\frac{2}{3} N (-\k^2 +  \frac{3\k}{Rv})$ provides an even better approximation to $\Delta S_\mathrm{screen}$ than $\frac{2}{3} N (-\k^2)$ alone.

We have not been able to resolve numerically whether the replacement $v \to v_\mathrm{eff}$ reproduces corrections at $\mathcal{O}\left((Rv)^{-2}\right)$ or higher. However, $S_\mathrm{Coul}$ with $v \to v_\mathrm{eff}(R)$ cannot reproduce $S_\mathrm{screen}$ for all $Rv$. For example, as mentioned above, as $Rv \to 0$ our $S_\mathrm{screen}$ approaches the conformal Wilson line result $S_\mathrm{symm}$ in ~\eqref{eq:EE_conf_wilson}, whereas $S_\mathrm{Coul}$ with $v \to v_\mathrm{eff}(R)$ does not. Nevertheless, the fact that such a simple and intuitive replacement works at large $Rv$ up to order $1/(Rv)$ is surprising. In the next section we will see that a similar replacement works for the spherical soliton EE.

\subsection{Spherical soliton}
\label{sec:EE_W_bos}

In this section, we consider the spherical soliton discussed in section~\ref{sec:W_bos}, whose embedding in the hyperbolic coordinates of eq.~\eqref{eq:hyperb_metric} is given implicitly by
\begin{equation}
\label{eq:embedding_W_bos}
\zeta \sinh u = \frac{\kappa  \left(\sqrt{\zeta ^2-1} \cos \tau +\zeta  \cosh u\right)}{
	Rv -\sqrt{\zeta ^2-1}  \cos \tau -\zeta   \cosh u} \,.
\end{equation}
As we discussed below eq.~\eqref{eq:S_bdy_2_general}, for this non-conformal D3-brane $u(\zeta,\tau)$ depends on $\tau$.

\begin{figure}
	\begin{center}
		\includegraphics[scale=0.6]{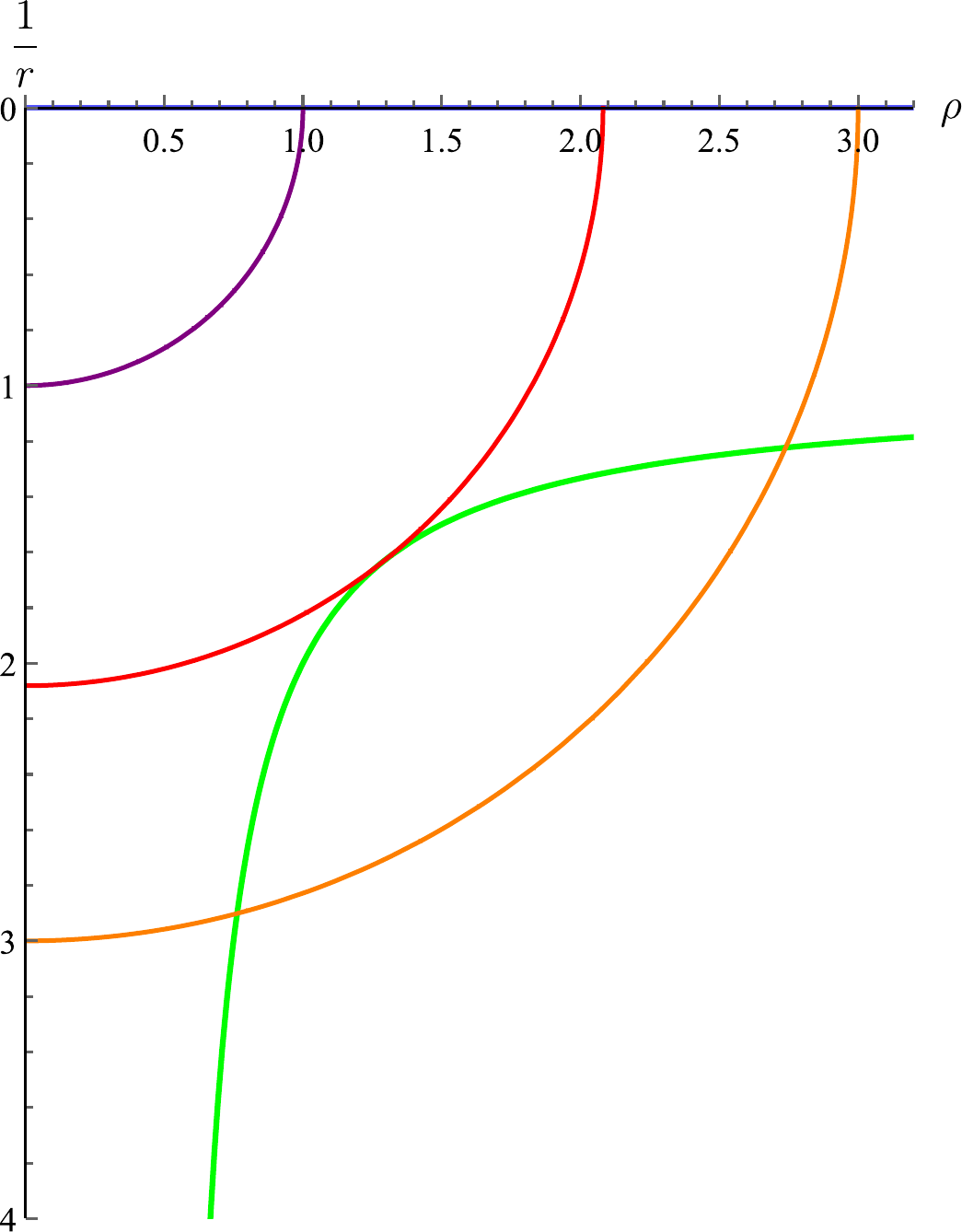}
		\caption{Cross sections of RT surfaces of increasing $Rv$ (purple, red, orange) and of the D3-brane embedding (green) for $\kappa=0.5$. When $Rv<(Rv)_\mathrm{crit}$, with $(Rv)_\mathrm{crit}$ in eq.~\eqref{eq:R_v_lim}, the two do not intersect (purple). When $Rv = (Rv)_\mathrm{crit}$, they are tangent (red). When $Rv > (Rv)_\mathrm{crit}$, they intersect twice (orange). }
		\label{plot:fig_solutions}
	\end{center}
\end{figure}

We discuss several features of eq.~\eqref{eq:embedding_W_bos} in appendix~\ref{app_W_boson}. Here we focus on what happens when the RT surface, $\zeta =1$, intersects the D3-brane. Figure~\ref{plot:fig_solutions} is a cross section of figure~\ref{fig:cartoons}d showing the spherical soliton D3-brane as the green curve. We expect that for sufficiently small $Rv$ the RT surface will not intersect the D3-brane, as depicted by the purple curve in figure~\ref{plot:fig_solutions}. As we increase $Rv$ we expect to find a critical value, $\left(Rv\right)_\mathrm{crit}$, at which the RT surface is tangent to the D3-brane, as depicted by the red curve in figure~\ref{plot:fig_solutions}. For $Rv > \left(Rv\right)_\mathrm{crit}$ we expect the RT surface to intersect the D3-brane at two points, as depicted by the orange curve in figure~\ref{plot:fig_solutions}.

To determine the intersection points, we set $\zeta=1$ in eq.~\eqref{eq:embedding_W_bos}, which gives
\begin{equation}
\label{eq:emb_at_hor}
\sinh u_h = \frac{\kappa \cosh u_h}{Rv -\cosh u_h}\,.
\end{equation}
We indeed find a critical radius,
\begin{equation}
\label{eq:R_v_lim}
(Rv)_\mathrm{crit} =\left(\kappa^{2/3}+1\right)^{3/2},
\end{equation} 
where if $Rv < (Rv)_\mathrm{crit}$, then eq.~\eqref{eq:emb_at_hor} has no real solutions, and so the RT surface does not intersect the D3-brane. If $Rv >(Rv)_\mathrm{crit}$, then the equation has two real solutions, so the RT surface intersects the D3-brane twice, as expected.  We denote these two solutions as $u_1(\kappa,Rv)$ and $u_2(\kappa,Rv)$ such that $0< u_1(\kappa,Rv) < u_2(\kappa,Rv)$. These correspond to different orientations of the D3-brane, namely $\omega_\pm ( u_1) = -1$ and $\omega_\pm( u_2) = +1$. For the Coulomb branch D3-brane, which has $\k=0$, we expect $(Rv)_\mathrm{crit}=1$, which is indeed the case when $\k=0$ in eq.~\eqref{eq:R_v_lim}.

As in the previous two cases, all three contributions to $S^{(1)}_A = \mathcal{S} + \mathcal{S}^{(bdy)}_1 + \mathcal{S}^{(bdy)}_2$ in eq.~\eqref{eq:ee3split} are non-zero. For the first contribution we find
\begin{eqnarray}
\label{eq:integr_EE_W_bos}
\mathcal{S} = \frac{4\pi }{3}  T_{D3} \int d \zeta \, d\tau \,  \Bigg[&\dfrac{1}{ Rv }&\left( (\partial_\zeta u)^2-\frac{(\partial_\tau u)^2}{ f^2_1(\zeta )}\right)
\left|\kappa Rv \cosh u- \sinh
u (\kappa+\zeta  \sinh u)^2\right| \nonumber  \\
&+& 2\,\omega_\pm(u) (\partial_\zeta u) \sinh ^2 u\Bigg].
\end{eqnarray}
The integral in the first line of eq.~\eqref{eq:integr_EE_W_bos} is non-trivial for any value of $Rv$, while the one in the second line, being a boundary term, is non-trivial only for $Rv >(Rv)_\mathrm{crit}$. We evaluate the first line of eq.~\eqref{eq:integr_EE_W_bos} numerically. In appendix~\ref{app_W_boson} we discuss in detail our numerical evaluation for $Rv >(Rv)_\mathrm{crit}$, which is the most subtle case. In the appendix~\ref{app_W_boson}, we also show that the first line of eq.~\eqref{eq:integr_EE_W_bos} diverges in the limit $Rv \rightarrow (Rv)^+_\mathrm{crit}$.\footnote{In the limit from below, namely $Rv \rightarrow (Rv)^-_\mathrm{crit}$, the integral goes to a finite value. } To characterise the divergence, we take $Rv = (Rv)_{\text{crit}}+ \delta$ and expand the first line of eq.~\eqref{eq:integr_EE_W_bos} around $\delta = 0$, obtaining
\beq
\label{eq:S_W_bos_diver}
\mathcal{S}=-\pi\sqrt{\frac{2}{3}} \frac{
	\left(\kappa^{2/3}+1\right)^{5/4} \kappa^{2/3}}{\sqrt{\delta}} + \mathcal{O}(\sqrt{\delta})\,, \qquad Rv \gtrsim (Rv)_\mathrm{crit}\,,
\eeq
which indeed diverges as $1/\sqrt{\delta}$ as $\delta \to 0$. This divergence is cancelled by $\mathcal{S}^{(bdy)}_1$ and $\mathcal{S}^{(bdy)}_2$, as we discuss below. The second line of eq.~\eqref{eq:integr_EE_W_bos} reduces to boundary terms,
\begin{equation}
 2\int d \zeta \, d\tau \, \omega_\pm(u) (\partial_\zeta u) \sinh ^2 u= - 2\pi\sum_{i=1,2} \omega_\pm(u_i) \left( u_i -\sinh u_i \cosh u_i \right).
\end{equation}

For $\mathcal{S}^{bdy}_1$ a straightforward calculation gives
\begin{equation}
\label{eq:I_bdy_1Wboson}
\begin{split}
\mathcal{S}^{(bdy)}_1 =   \frac{8\pi^2}{3} T_{D3} \sum_{i=1,2} &  \Bigg\{ Rv \frac{\left(\kappa^2 +   \sinh ^4 u_i\right) }{ \left|\kappa Rv \cosh u_i - \sinh
	u_i (\kappa+\  \sinh u_i)^2\right|}  \\
&  + \omega_\pm(u_i) \left( u_i -\sinh u_i \cosh u_i \right)    \Bigg\}, \qquad  Rv > (Rv)_{\text{crit}} \,.
\end{split}
\end{equation}
The denominator of the first  term in eq.~\eqref{eq:I_bdy_1Wboson} vanishes at $Rv = (Rv)_{\text{crit}}$, hence $\mathcal{S}^{(bdy)}_1$ diverges there. Again taking $Rv = (Rv)_{\text{crit}}+ \delta$ and expanding in $\delta$, we find
\begin{equation}
\label{eq:I_1_bdy_W_bos_diver}
\mathcal{S}^{(bdy)}_1  = 2 \, \pi\,\sqrt{\frac{2}{3}}  \frac{ \left(\kappa^{2/3}+1\right)^{5/4} \kappa^{2/3}}{\sqrt{\delta }}+\mathcal{O}\left(\sqrt{\delta} \right), \qquad  Rv \gtrsim (Rv)_{\text{crit}}\,.
\end{equation}

For the boundary contribution from the equations of motion, $\mathcal{S}^{(bdy)}_2$, we start from $u(\zeta,\tau)$ in eq.~\eqref{eq:embedding_W_bos}, perform the expansion in eq.~\eqref{eq:anasatz_exp_hor} with $n=1$, extract $u_1^{(0)}(\tau)=u_h$ and $u_1^{(1)}(\tau)$ (both of which are non-zero), plug these into eq.~\eqref{eq:S_bdy_2_general} for $\mathcal{S}^{(bdy)}_2$, and perform the integration over $\tau$, obtaining for $Rv \ge (Rv)_{\text{crit}}$
\begin{equation}
\mathcal{S}^{(bdy)}_2 = -\frac{4 \pi^2}{3}T_{D3} \sum_i \frac{  (Rv)^2 \kappa^2 \sqrt{\frac{\kappa^2+\sinh ^4u_i }{(Rv)^2 \kappa^2+\left[
			\cosh u_i (Rv-\cosh u_i)^2-Rv \kappa \sinh u_i
			\right]^2}}}{\left|\cosh u_i (Rv-\cosh u_i)^2-Rv \kappa \sinh u_i\right|}\,.
\end{equation}
This term also diverges at $Rv = (Rv)_{\text{crit}}$. Once again taking $Rv = (Rv)_{\text{crit}}+ \delta$ and expanding around $\delta = 0$, we find 
\begin{equation}
\label{eq:I_2_bdy_W_bos_diver}
\mathcal{S}^{(bdy)}_2 = -\pi \sqrt{\frac{2}{3}} \frac{
	\left(\kappa^{2/3}+1\right)^{5/4} \kappa^{2/3}}{\sqrt{\delta }} + \mathcal{O}(\sqrt{\delta})\,, \qquad  Rv \gtrsim (Rv)_{\text{crit}}\, .
\end{equation}

Clearly, in $S^{(1)}_A = \mathcal{S} + \mathcal{S}^{(bdy)}_1 + \mathcal{S}^{(bdy)}_2$ the divergences in eqs.~\eqref{eq:S_W_bos_diver},~\eqref{eq:I_bdy_1Wboson}, and~\eqref{eq:I_2_bdy_W_bos_diver} cancel, so that $S^{(1)}_A$ will be finite and continuous at $Rv = (Rv)_{\text{crit}}$. Indeed, denoting the contribution of the spherical soliton D3-brane to the EE as $S^{(1)}_A = S_\mathrm{soliton}$, we find

\begin{figure}[t]
	\begin{center}
		\includegraphics[scale=.7]{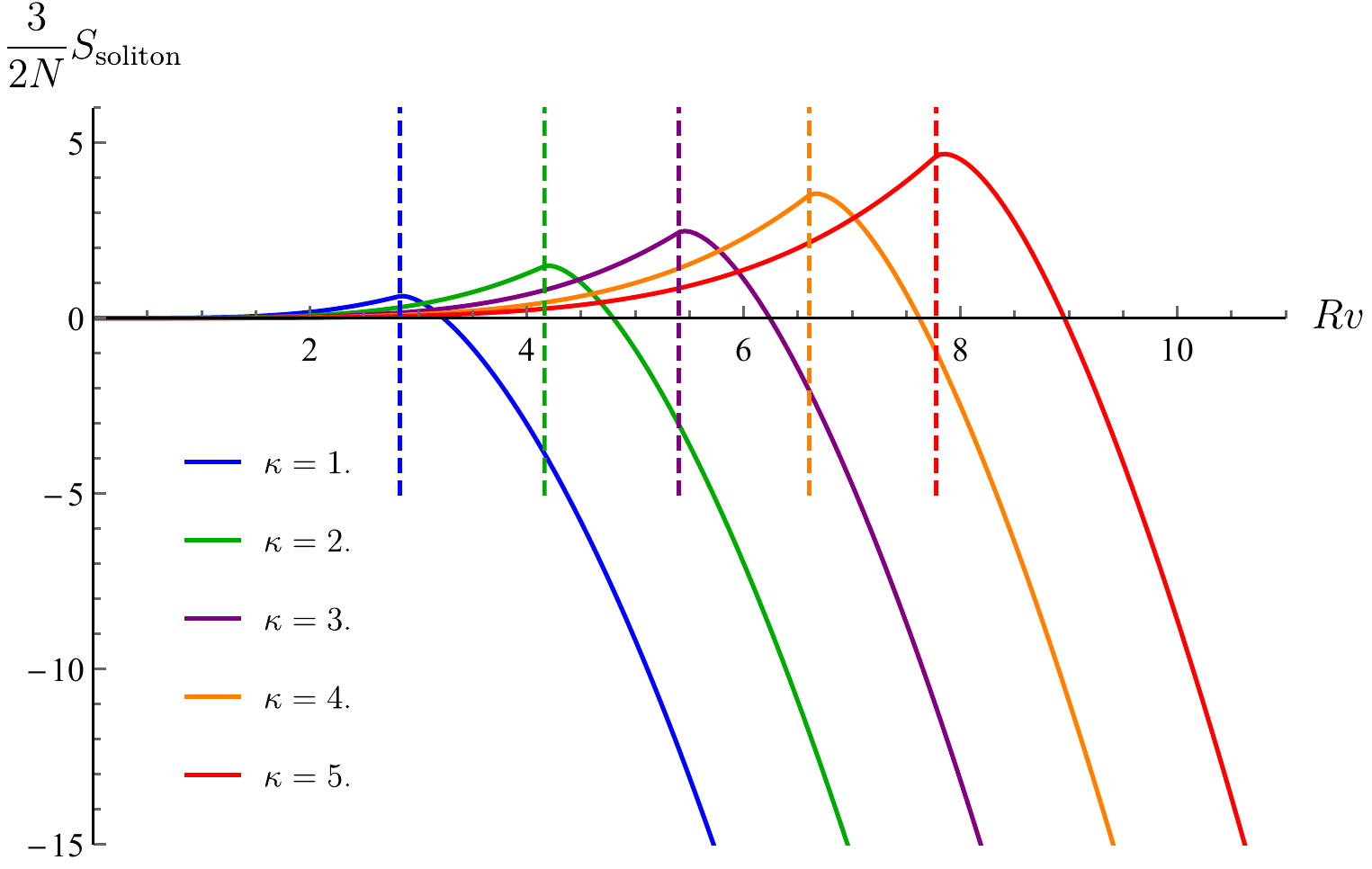}
		\caption{Our numerical results for $S_{\text{soliton}}/(2N /3)$ in eq.~\eqref{eq:eesol} as a function of $Rv$ for (from left to right) $\kappa=1$ (blue), $2$ (green), $3$ (purple), $4$ (orange), $5$ (red). The dashed vertical lines denote the location of the critical radius $Rv = (Rv)_{\text{crit}}$ defined in eq.~\eqref{eq:R_v_lim}. Each curve also has a maximum at $(Rv)_\mathrm{max} \gtrsim (Rv)_{\text{crit}}$.}
		\label{plot:SEE_tot}
	\end{center}
\end{figure}

\begin{align}
\label{eq:eesol}
S_\mathrm{soliton}=\;& \frac{2N}{3 \pi}   \int d\tau d \zeta \,  \Bigg\{ \dfrac{1}{ Rv }\left( (\partial_\zeta u)^2-\frac{(\partial_\tau u)^2}{ f^2_1(\zeta )}\right)
\left|\kappa Rv \cosh u- \sinh
u (\kappa+\zeta  \sinh u)^2\right|  \Bigg\} \nonumber\\
&+  \frac{2 N}{3} \sum_{i=1,2}  \left\{\vphantom{\frac{  R_v^2 \kappa^2 \sqrt{\frac{\kappa^2+\sinh ^4u_i }{R_v^2 \kappa^2+\left(\cosh
				u_i (R_v-\cosh u_i)^2-R_v k \sinh u_i\right)^2}}}{\left|\cosh u_i (R_v-\cosh u_i)^2-R_v \kappa \sinh u_i\right|}} 2Rv \frac{\left(\kappa^2 +   \sinh ^4 u_i\right) }{ \left|\kappa Rv \cosh u_i - \sinh
	u_i (\kappa+\  \sinh u_i)^2\right|}  \right.\nonumber\\
&  + \omega_\pm(u_i) \left( u_i -\sinh u_i \cosh u_i \right) \\
&\left.- \frac{  (Rv)^2 \kappa^2 \sqrt{\frac{\kappa^2+\sinh ^4u_i }{(Rv)^2 \kappa^2+\left[\cosh
			u_i (Rv-\cosh u_i)^2-Rv \kappa \sinh u_i\right]^2}}}{\left|\cosh u_i (Rv-\cosh u_i)^2-Rv \kappa \sinh u_i\right|}   \right\}\Theta(Rv-(Rv)_{\text{crit}}) \nonumber
\,,
\end{align}
where $\Theta$ is the Heaviside step function. Figure~\ref{plot:SEE_tot} shows our numerical result for $S_\mathrm{soliton}/(2N/3)$ as a function of \(R v\) for several $\kappa$ values. It is clearly finite and continuous at $(Rv)_\mathrm{crit}$, which we indicate for each $\kappa$ by a dashed vertical line.

One prominent feature of $S_\mathrm{soliton}$ is very different from the previous cases. While $S_\mathrm{Coul}$ in section~\ref{sec:EE_Coul_branch} and $S_\mathrm{screen}$ in section~\ref{sec:EE_Screen_Wilson} were monotonic in $Rv$, our $S_\mathrm{soliton}$ has a maximum at an $Rv$ value $(Rv)_\mathrm{max}$ slightly larger than $(Rv)_\mathrm{crit}$. In ref.~\cite{Schwarz:2014rxa}, Schwarz interpreted the spherical soliton as an infinitely thin shell with $U(1)$ charge. Although our result for $S_\mathrm{soliton}$ does not show any ``smoking gun'' features characteristic of an infinitely thin interface, our results are consistent with a charge distribution peaked at $(Rv)_\mathrm{max}$: if charges are entangled with each other, then regions with a larger charge density will contribute more to the EE. Furthermore, we will see below that in the limit $\kappa, v \rightarrow +\infty$ our $S_\mathrm{soliton}$ may be interpreted as that of an infinitely thin shell.

\begin{figure}[t]
	\begin{center}
		\includegraphics[scale=.39]{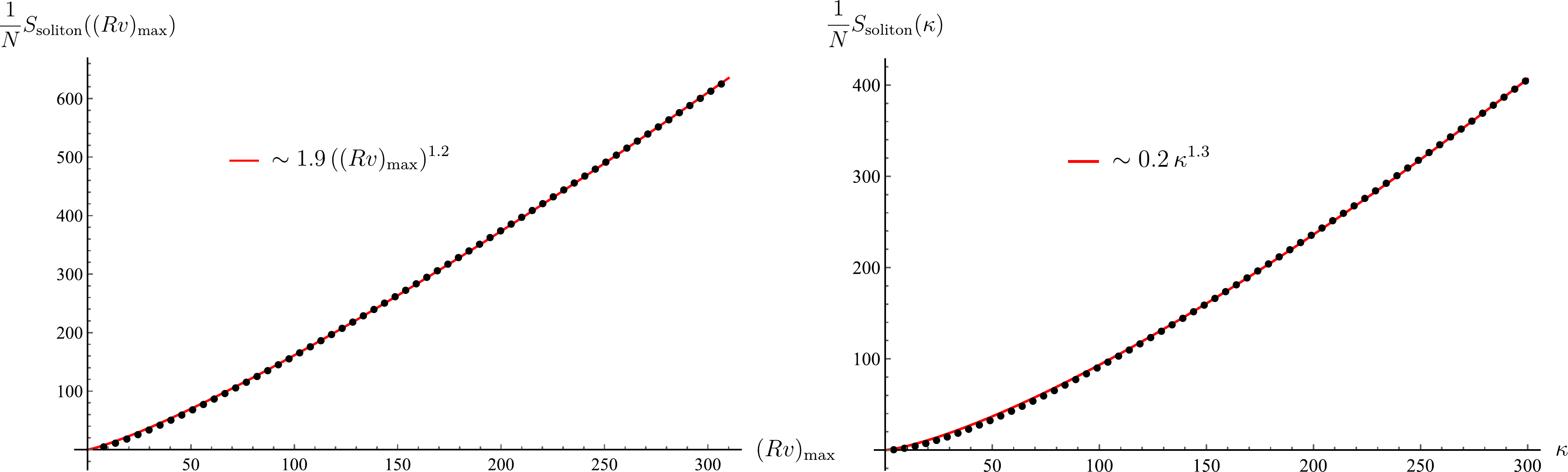}
		\caption{\textbf{Left:} The black dots are our numerical results for $S_{\text{soliton}}/N$ in eq.~\eqref{eq:eesol} at the position of its maximum $(Rv)_\mathrm{max}$ as a function of $(Rv)_\mathrm{max}$. The solid red line is a fit to a function $\propto (Rv)_\mathrm{max}^{1.2}$. \textbf{Right:} The black dots are our numerical results for $S_{\text{soliton}}/N$ at $R=R_0$, with $R_0 = \kappa/v$ the spherical soliton's radius (and we set $v=1$), as a function of $\k$. The solid red line is a fit to a function $\propto \k^{1.3}$.}
		\label{plot:soleemax}
	\end{center}
\end{figure}

As mentioned in section~\ref{sec:intro}, Schwarz in ref.~\cite{Schwarz:2014zsa} asked whether the EE of a sphere coincident with the soliton might scale with surface area at large $\kappa$, similar to a black hole's Bekenstein-Hawking entropy. Recalling from section~\ref{sec:W_bos} that the spherical soliton's radius is $R_0\equiv\kappa/v$ (in our units with $L\equiv1$), Schwarz's question becomes whether at $R=R_0$ and large $\kappa$ we find $S_\mathrm{soliton}\propto \kappa^2$. Our numerical results suggest this is not the case. Figure~\ref{plot:soleemax} on the left shows our numerical results for the value of $S_\mathrm{soliton}/N$ at $(Rv)_\mathrm{max}$ as a function of $(Rv)_\mathrm{max}$. We also show a fit to a function $\propto (Rv)_\mathrm{max}^{1.2}$, which is clearly better than a $(Rv)_\mathrm{max}^2$ fit, suggesting that the scaling is not with surface area. Figure~\ref{plot:soleemax} on the right shows our result for $S_\mathrm{soliton}/N$ at $R_0$ (with $v=1$) as a function of $\kappa$. To answer Schwarz's question: at large $\kappa$ our results fit a function $\propto \kappa^{1.3}$ better than a function $\propto \kappa^2$, suggesting again that the scaling is not with surface area.

\begin{figure}[t]
	\begin{center}
		\includegraphics[scale=.75]{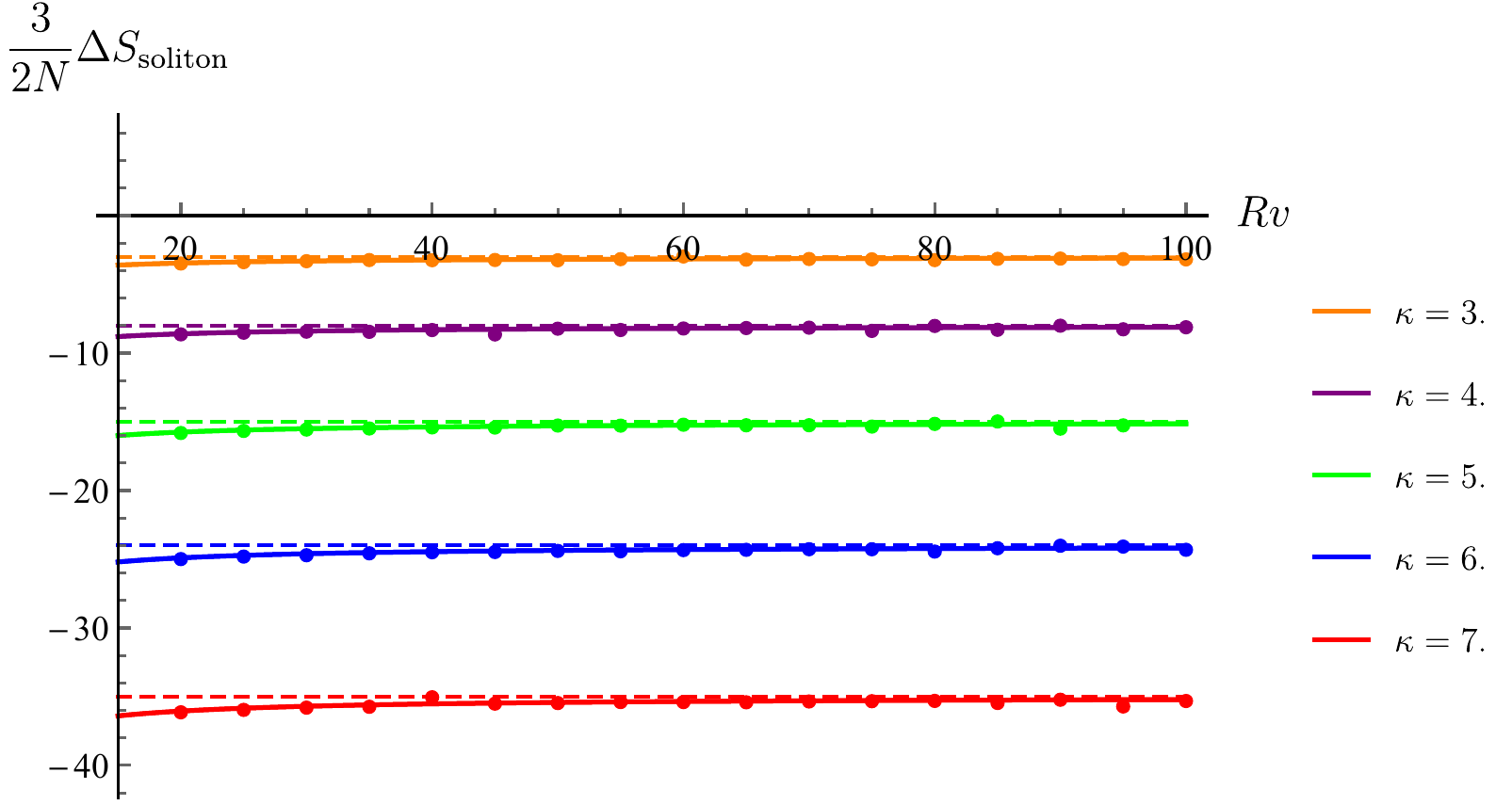}
		\caption{The dots are our numerical results for $\Delta S_\mathrm{soliton}$ in eq.~\eqref{eq:deltasoliton}, divided by $2N/3$, as a function of $Rv$ and for $\kappa=3$ (orange), $4$ (purple), $5$ (green), $6$ (blue), $7$ (red). The solid lines are the prediction from eq.~\eqref{eq:EE_W_boson_linear}. The horizontal dashed lines show the limit \(\lim_{Rv\to \infty}\Delta S_\mathrm{soliton} = - \kappa^2 + 2 \kappa\).}
		\label{plot:large_A_EE_finite}
	\end{center}
\end{figure}

In the IR limit $Rv \to \infty$, we expect $S_\mathrm{soliton}$ to approach the EE of the Coulomb branch in the same limit, $S_\mathrm{Coul}$ in eq.~\eqref{eq:EE_coul_LargeBeha}. As we move towards smaller $Rv$ we expect a different set of finite-$(Rv)$ corrections to those in $S_\mathrm{Coul}$. Numerically we find
\begin{equation}
\label{eq:EE_W_boson_linear}
S_\mathrm{soliton} = S_\mathrm{Coul} +  \frac{2}{3} N \left(2\kappa Rv - \k^2 - \frac{3\k}{Rv}   \right) + \frac{4}{3} N \kappa + \mathcal{O}\left((Rv)^{-2}\right), \qquad Rv \gg 1 \,.
\end{equation}
Indeed, figure~\ref{plot:large_A_EE_finite} shows our numerical results for
\beq
\label{eq:deltasoliton}
\Delta S_\mathrm{soliton} \equiv S_\mathrm{soliton}-S_\mathrm{Coul} - \frac{2}{3} N (2 \k \,Rv)\,,
\eeq
divided by $2N/3$, at large values of \(R v\) and for several values of \(\kappa\). In the figure, the dots show our numerical results, while the curves show $(-\kappa^2 + 2 \kappa-3 \kappa/(R v))$, from the \(\mathcal{O}\left((Rv)^0\right)\) and \(\mathcal{O}\left((Rv)^{-1}\right)\) terms in eq.~\eqref{eq:EE_W_boson_linear}. We find very good agreement between the two, showing that the behaviour of $S_\mathrm{soliton}$ at large $Rv$ is indeed given by eq.~\eqref{eq:EE_W_boson_linear}.

The large-$(Rv)$ behaviour in eq.~\eqref{eq:EE_W_boson_linear} up to order $\mathcal{O}((R v)^{-1})$ consists of two contributions. The first contribution, in round brackets in eq~\eqref{eq:EE_W_boson_linear}, comes from the same ``trick'' we used in eq.~\eqref{eq:coulreplace} to compute the leading $1/(Rv)$ corrections to $S_\mathrm{screen}$. We introduce an effective position-dependent cutoff like that in eq.~\eqref{eq:veffscreened}, but with $\kappa \to -\kappa$,
\beq
\label{eq:veffsoliton}
v_\mathrm{eff}(\r) = v-\frac{\k}{\rho}\,.
\eeq
In the large-$(Rv)$ Coulomb branch result, $S_\mathrm{Coul}$ in eq.~\eqref{eq:EE_coul_LargeBeha}, we then make the replacement $v \to v_\mathrm{eff}$, and expand in $Rv \gg 1$, obtaining the terms in the round brackets in eq.~\eqref{eq:EE_W_boson_linear}. The fact that this ``trick'' works for both $S_\mathrm{screen}$ and $S_\mathrm{soliton}$ is remarkable, and suggests this is neither a trick nor a coincidence, but a genuine physical effect: in these cases, at large $Rv$, the UV cutoff is an inverse W-boson mass that has acquired position dependence given by $v_\mathrm{eff}$ in eq.~\eqref{eq:veffscreened} or~\eqref{eq:veffsoliton}. The second contribution is $\frac{4}{3}N \kappa$. Recalling from below eq.~\eqref{eq:Conf_Wils_emb} that this D3-brane carries $k=\kappa \,4 N/\sqrt{\lambda}$ units of string charge, this term is $\kappa \, 4N/3 = k \sqrt{\lambda}/3$, which is precisely $k \, S_{\square}$ with $S_{\square}=\sqrt{\lambda}/3$ from eq.~\eqref{eq:EE_fund_quark}. In other words, this second contribution is that of a Wilson line in a direct product of $k$ fundamental representations. To summarise, we have shown that the $Rv \gg 1$ result in eq.~\eqref{eq:EE_W_boson_linear} is
\beq
\label{eq:solitonfinal}
S_\mathrm{soliton} (Rv) = S_\mathrm{Coul}(R\,v_\mathrm{eff}) + k \,  S_{\square} + \mathcal{O}\left((Rv)^{-2}\right), \qquad Rv \gg 1\,.
\eeq

We can in fact show that the $k \,  S_{\square}$ contribution comes from the part of the D3-brane that reaches the Poincar\'e horizon. Consider the limits $\kappa \rightarrow \infty$ and $v \rightarrow \infty$ with the soliton's radius $R_0 \equiv \kappa/v $ fixed. In this limit, eq.~\eqref{eq:embedding_W_bos} for the embedding reduces to 
\begin{equation}
\label{eq:emb_infinte_W}
\sinh u = \frac{R_0}{R} \frac{ \sqrt{\zeta ^2-1} \cos \tau +\zeta  \cosh u}{\zeta}\,, 
\end{equation}
which in Poincar\'e coordinates is simply a cylinder $\rho= R_0$. Starting from figure~\ref{fig:cartoons}d, intuitively these limits correspond to sending the Coulomb branch part of the D3-brane to the $AdS_5$ boundary while simultaneously increasing the size of the spike down to the Poincar\'e horizon, until all that remains in the limit is a cylinder of radius $R_0$ extending from the $AdS_5$ boundary to the Poincar\'e horizon. This cylinder carries $k$ units of string charge, hence we interpret this solution as a uniform cylindrical distribution of $k$ strings that have expanded into a D3-brane via the Myers effect~\cite{Emparan:1997rt,Myers:1999ps}.

This limit drastically alters the EE. In particular, we will show that  the EE in this limit diverges at $R_0$, and when $R \to \infty$ reproduces the $k \,  S_{\square}$ term in eq.~\eqref{eq:solitonfinal} (and none of the other terms). Such behaviour suggests this solution is dual to an infinitely thin spherical shell of charge at $R_0$ that at $R \to \infty$ produces the EE of a Wilson line in the direct product of $k$ fundamental representations. A similar divergent behaviour in the EE can also be found in boundary conformal field theories when the entangling region approaches the boundary~\cite{Seminara:2018pmr,Bastianello:2019yyc,Bastianello:2020hqs}. 

\begin{figure}[t]
	\begin{center}
		\includegraphics[scale=.75]{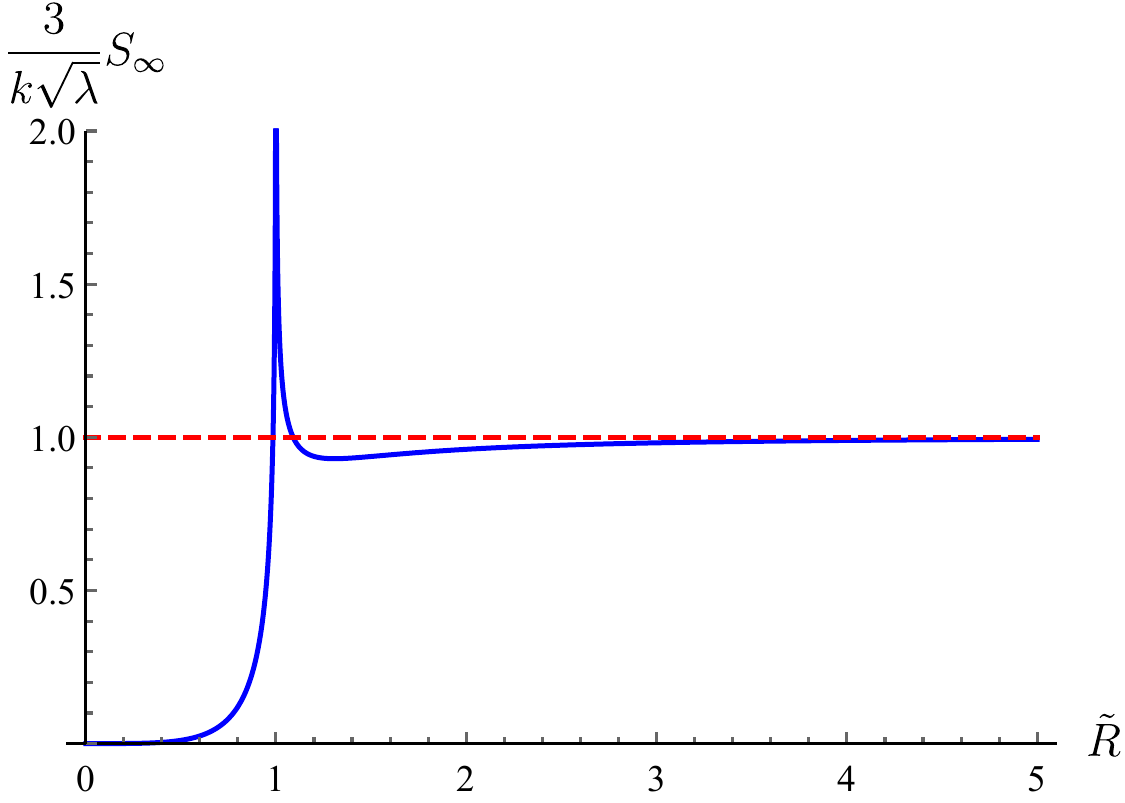}
		\caption{Our result for $S_\infty/(k \sqrt{\lambda}/3)$ from eq.~\eqref{eq:soliton_ee_large_mass} as a function of $\tilde R \equiv R/R_0$, with $R_0 \equiv \kappa/v$ fixed as $\kappa \to \infty$ and $v \to \infty$. Clearly $S_\infty$ diverges at $\tilde R=1$, and approaches $k\sqrt{\lambda}/3$ as $\tilde R \to \infty$, as indicated by the horizontal dashed line.}
		\label{plot:SEE_infinte_mass}
	\end{center}
\end{figure}

To demonstrate these features, we return to eq.~\eqref{eq:S_D3_action_hyp_final} for the D3-brane action in the hyperbolic black hole background with arbitrary $n$, plug in the embedding given by eq.~\eqref{eq:emb_infinte_W}, take large $\k$, and retain only the leading contribution, which is linear in $\kappa$. The result is
\begin{equation}
\label{eq:largechargeaction}
\begin{split}
\left. I_{D3}(n)\right|_{2\pi} =  4\pi \, T_{D3} \, \kappa \int d\tau \int d \zeta  \,  \sqrt{ 1+\zeta^2  (\partial_\zeta u )^2 f_n(\zeta )+\frac{\zeta ^2 ( \partial_\tau u)^2}{f_n(\zeta )}} \,.
\end{split}
\end{equation}
Crucially, the term involving $C_4$ is subleading when $\kappa \to \infty$, so we can safely ignore it. Taking $\partial/\partial n$ and $n=1$ then gives the expected form, $S_A^{(1)} = \mathcal{S} + \mathcal{S}_1^{(bdy)}+\mathcal{S}_2^{(bdy)}$, with
\begin{eqnarray}
\mathcal{S} & = & \frac{4\pi }{3} T_{D3} \, \kappa   \,\int d\tau \int  d \zeta \frac{\left( (\partial_\zeta u)^2-\frac{(\partial_\tau u)^2}{ f^2_1(\zeta )}\right)
}{ \sqrt{\left(1+\zeta ^2  (\partial_\zeta u)^2
		f_1(\zeta )+\frac{\zeta ^2 (\partial_\tau u)^2}{f_1(\zeta )}\right) }} \nonumber \\
& = & \frac{2\pi^2}{3}\, T_{D3}\, \kappa \left[\frac{\left(\tilde R^2-3\right) \log \left(\frac{| \tilde R -1| }{\tilde R+1}\right)-6\tilde R}{ \tilde R} + \frac{2 \left(4 \tilde R^2-5\right)}{\tilde R \sqrt{\tilde R^2-1}} \Theta
\left(\tilde R-1\right) \right],
\end{eqnarray}
where we introduced $\tilde R \equiv R/R_0$. We also find
\begin{equation}
\begin{split}
\mathcal{S}_1^{(bdy)} &=  \frac{8\pi^2}{3} T_{D3} \kappa   \frac{\tilde R}{ \left|\tilde R \cosh u_h- \sinh
	u_h \right|} = \frac{8\pi^2}{3} T_{D3} \kappa   \frac{\tilde R}{ \left|\tilde R^2-1\right|\sinh u_h} \\
& =  \frac{8\pi^2}{3} T_{D3} \kappa   \frac{\tilde R}{ \sqrt{\tilde R^2-1}}\,, \qquad \tilde R > 1\,,
\end{split}
\end{equation}
\begin{equation}
\mathcal{S}_2^{(bdy)} = -\frac{4 \pi  }{3}T_{D3} \int_0^{2\pi} d\tau \,\frac{\cos ^2\tau}{ \tilde R \sqrt{\tilde R^2-1}}\Theta
\left(\tilde R-1\right) = -\frac{4 \pi^2  }{3} \,  T_{D3}\frac{1}{\tilde R \sqrt{\tilde R^2-1}}\Theta
\left(\tilde R-1\right).
\end{equation}
Denoting the contribution of this D3-brane to the EE as $S_A^{(1)} = S_\infty$, we thus find
\begin{equation}
\label{eq:soliton_ee_large_mass}
S_\infty= \frac{1}{3}\,N\k\, \left\{\frac{\left(\tilde R^2-3\right) \log \left(\frac{| \tilde R -1| }{\tilde R+1}\right)-6\tilde R}{ \tilde R} + \frac{12 \sqrt{\tilde R^2-1}}{\tilde R} \Theta
\left(\tilde R-1\right) \right\},
\end{equation}
where the prefactor is $N\k/3 = k \sqrt{\lambda}/3$. Figure~\ref{plot:SEE_infinte_mass} shows $S_\infty/(k \sqrt{\lambda}/3)$ as a function of $\tilde R$. Clearly $S_\infty$ diverges at $\tilde R = 1$, and $S_\infty \to k \sqrt{\lambda}/3$ at large $\tilde R$, as advertised.

In the next section we compute the VEV of the Lagrangian for the screened Wilson line and spherical soliton. For the latter we find large-$R$ behaviour that can also be interpreted as the contribution of a Wilson line in the direct product of $k$ fundamental representations. Furthermore, for the spherical soliton in the limits $\kappa \rightarrow \infty$ and $v \rightarrow \infty$ with $R_0$ fixed we find behaviour similar to that of $S_\infty$, namely a divergence at $R_0$ and the large-$(Rv)$ limit of a Wilson line in a direct product of $k$ fundamental representations.

\section{Lagrangian and Stress-Energy Tensor}
\label{sec:lagran_dens}

In a gauge theory the one-point functions of single-trace gauge-invariant operators provide a natural way to characterise the spatial profiles of objects like a Wilson line, screened or not, or a spherical soliton. In this section we will consider two such operators. The first is an exactly marginal scalar operator, namely the $\N=4$ SYM Lagrangian density, which we denote $\mathcal{O}_{F^2} \equiv \frac{1}{N} \text{Tr} F_{mn}F^{mn} + \dots$, where the $\ldots$ represents supersymmetric completion. This operator is holographically dual to the dilaton. The second is the stress-energy tensor, $T_{mn}$, which in a CFT can acquire a non-zero vacuum expectation value in the presence of a conformal defect of codimension two or higher. This operator is holographically dual to the metric. In each case we will consider a single probe D3-brane's effect on the dilaton or metric, so in the dual CFT, where the breaking $SU(N) \to SU(N-1)\times U(1)$ and large $N$, we will compute the order $N$ contribution to $\langle\mathcal{O}_{F^2}\rangle$ or $\langle T_{mn} \rangle$, respectively.

Evaluating $\left\langle\mathcal{O}_{F^2}\right\rangle$ is straightforward for all cases discussed in this paper. However, for $\langle T_{mn} \rangle$ a subtlety arises: in general we need to take into account the D3-brane's back-reaction on $C_4$. As a simplification, we will only consider $\langle T_{mn}\rangle$ for the spherical soliton in the limit $\kappa \to \infty$ and $v \to \infty$ with $R_0=\kappa/v$ fixed, where the contribution of $C_4$ is negligible, as mentioned below eq.~\eqref{eq:largechargeaction}.

\subsection{Expectation value of $\mathcal{O}_{F^2}$}
\label{sec:Fvev}
We follow refs.~\cite{Danielsson:1998wt, Callan:1999ki, Fiol:2012sg} and compute $\langle \mathcal{O}_{F^2}\rangle$ holographically by computing the linearised perturbation of the dilaton field generated by the D3-brane, and then reading $\langle \mathcal{O}_{F^2}\rangle$ from its near-boundary behavior, per the standard AdS/CFT dictionary.

Let us first recall existing results for $\langle \mathcal{O}_{F^2}\rangle$ for the fundamental-representation Wilson line, the symmetric-representation Wilson line, and the screened Wilson line. For the fundamental-representation Wilson line, the result of refs.~\cite{Danielsson:1998wt, Callan:1999ki} is
\begin{equation}
\label{eq:O_F_fund_rep}
\langle\mathcal{O}_{F^2} \rangle_{\square} = \frac{\sqrt{\lambda}}{16 \pi^2} \frac{1}{\rho^4}\,.
\end{equation}
For conformal Wilson lines, the dependence on $\rho$ is fixed simply by dimensional analysis to be $1/\rho^4$, as in eq.~\eqref{eq:O_F_fund_rep}. The nontrivial information is therefore the dimensionless coefficient, $\frac{\sqrt{\lambda}}{16 \pi^2} $. For the conformal Wilson line in a symmetric representation the result of ref.~\cite{Fiol:2012sg} is
\begin{equation}
\label{eq:O_F_Wil_sim_k_rep}
\langle\mathcal{O}_{F^2} \rangle_\mathrm{symm} = \frac{N }{4 \pi^2} \frac{\kappa \sqrt{1+\kappa^2}}{\rho^4} =\frac{k \sqrt{\lambda}}{16 \pi^2} \sqrt{1+\frac{k^2 \lambda}{16 N^2}} \frac{1}{\rho^4}\,,
\end{equation}
which in the limit $k \ll N$ reduces to $k \langle\mathcal{O}_{F^2} \rangle_{\square}$. For the screened Wilson line, ref.~\cite{Kumar:2017vjv} was able to reduce the result to the integral, with $z=1/r$,
\begin{align}
\label{eq:O_F_Wil_screened_rep}
\langle\mathcal{O}_{F^2} \rangle_{\text{screen}} =\frac{3 N}{16 \pi^2} \frac{1}{\rho^4} \int_0^{\frac{1}{\rho v}}& dz \, \left( 1- \rho v z \right) z \\
&\!  \times \left\{ \left[  z^2  + \left(1 - \frac{\kappa z}{1- \rho v z}\right)^2\right]^{-\frac{5}{2}} - \left[ z^2 + \left(1 + \frac{\kappa z}{1 - \rho v  z}\right)^2\right]^{-\frac{5}{2}}  \right\} . \nonumber
\nonumber
\end{align}
At small and large $\rho v$ we can perform this integral, with the results
\beq
\langle\mathcal{O}_{F^2} \rangle_{\text{screen}} \approx \begin{cases} \dfrac{N }{4 \pi^2} \dfrac{\kappa \sqrt{1+\kappa^2}}{\rho^4}\,, & \quad \rho v \ll 1\,, \\[1em] \dfrac{ N  }{4 \pi^2}  \dfrac{\kappa^2}{\rho^4}\,, & \quad \rho v \gg 1\,.
\end{cases}
\label{eq:O_Fsq_screened}
\eeq
As expected, in both the UV and IR limits, $\rho v \ll 1$ and $\rho v \gg 1$, respectively, we find $1/\rho^4$, as required by scale invariance. For the dimensionless coefficient, the UV limit reproduces the conformal Wilson line result in eq.~\eqref{eq:O_F_Wil_sim_k_rep}, while the IR limit produces a factor $\propto \k^2$.

For the screened Wilson line, ref.~\cite{Evans:2019pcs} provided evidence for the screening in the form of a quasi-normal mode spectrum, a feature characteristic of any screened impurity. Eq.~\eqref{eq:O_Fsq_screened} provides more detailed information, as follows. Consider $(3+1)$-dimensional Maxwell theory, which is a CFT, and in which a point electric charge $Q$ produces an electric field $\propto Q/\rho^2$, and hence $F^2 \propto Q^2/\rho^4$. Eq.~\eqref{eq:O_Fsq_screened} has the same form, including in particular the IR limit $\rho v \gg 1$, where $\kappa$ plays a role analogous to $Q$. Clearly, in the $U(1)$ sector the Wilson line is not screened in the IR, but rather survives and appears as a point-like electric charge with strength $\kappa$.\footnote{Recall that our IR includes $\N=4$ SYM with gauge group $U(1)$, so eq.~\eqref{eq:O_Fsq_screened} includes contributions from both a $U(1)$ Maxwell field and its scalar superpartners.} We thus learn that what appears in the UV as a Wilson line of $SU(N)$ becomes in the IR, where $SU(N) \to SU(N-1) \times U(1)$, a point charge of the $U(1)$ sector, and is completely screened in the $SU(N-1)$ sector.

\begin{figure}[t]
	\begin{center}
		\includegraphics[scale=.6]{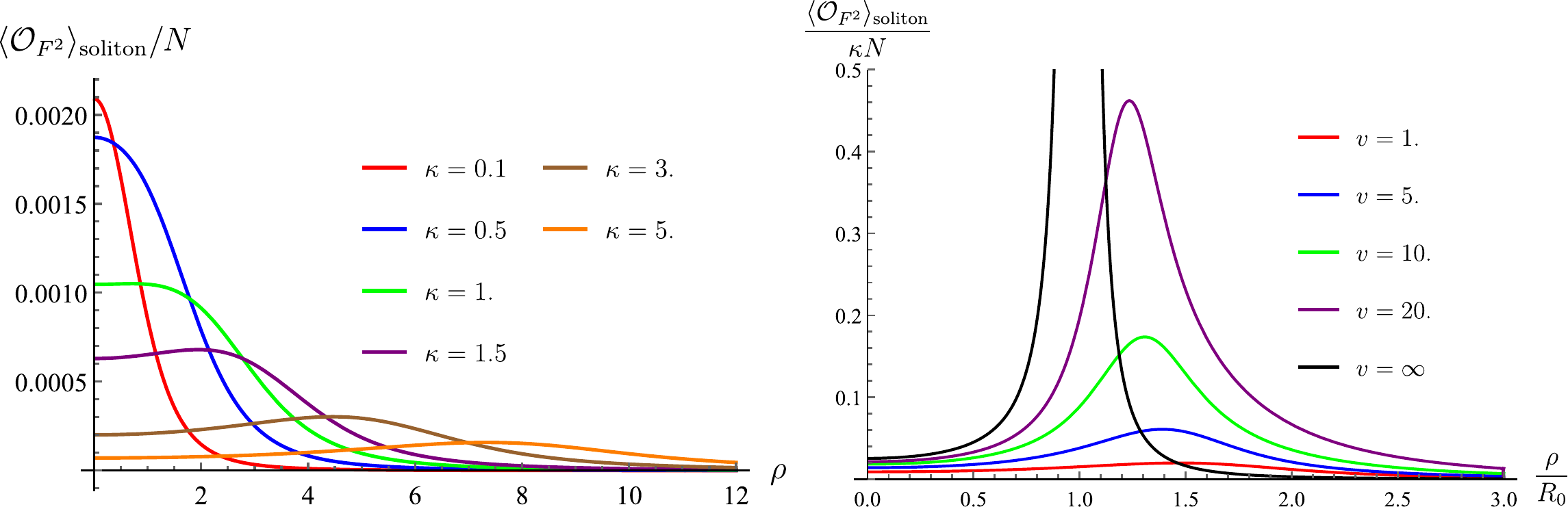}
		\caption{\textbf{Left:} Our numerical results for $\langle\mathcal{O}_{F^2}\rangle_\mathrm{soliton}/N$ from eq.~\eqref{eq:screenedof} as a function of $\rho/R_0$ with $R_0=\k/v$ for $v=1$ and $\k=0.1$ (red), $0.5$ (blue), $1$ (green), $1.5$ (purple), $3$ (brown), and $5$ (orange). \textbf{Right:} Our numerical results for $\langle\mathcal{O}_{F^2}\rangle_\mathrm{soliton}/(N\k)$ from eq.~\eqref{eq:screenedof} as a function of $\rho/R_0$ with $\k=v$ and $v=1$ (red), $5$ (blue), $10$ (green), and $20$ (brown). The black curve corresponds to the analytical solution in the $v \rightarrow \infty$ limit given by eq.~\eqref{eq:OF_v_inf}.}
		\label{plot:OF}
	\end{center}
\end{figure}

Let us now consider the Coulomb branch spherical soliton. This case has not been considered in the literature, so here our results are novel. The embedding for the D3-brane dual to the spherical soliton in eq.~\eqref{eq:solitonsol} is of the same form as the embedding for the D3-brane dual to the screened Wilson line in eq.~\eqref{eq:screenedsol}, but with $\kappa \to -\kappa$. As a result, we can obtain an integral for $\langle \mathcal{O}_{F^2} \rangle_\mathrm{soliton}$ simply by sending $\k \to -\k$ in eq.~\eqref{eq:O_F_Wil_screened_rep} and taking the region of integration to be the complement,
\begin{align}
\label{eq:screenedof}
\langle\mathcal{O}_{F^2}\rangle_\mathrm{soliton}
= \frac{3 N}{16 \pi^2} \frac{1}{\rho^4} \int_{\frac{1}{ \rho v}}^\infty & d z \, \left( 1-\rho v z\right)z   \\
& \! \times\left\{ \left[ z^2 + \left( 1 + \frac{\kappa z}{ \rho v z-1}\right)^2 \right]^{-\frac{5}{2}}- \left[z^2 + \left( 1 - \frac{\kappa z}{ \rho v  z -1}\right)^2 \right]^{-\frac{5}{2}} \right\}. \nonumber
\end{align}
In general we must evaluate this integral numerically. Figure~\ref{plot:OF} on the left shows our numerical results for $\langle\mathcal{O}_{F^2}\rangle_\mathrm{soliton}/N$ as a function of $\rho/R_0$, for various values of $\kappa$ and $v=1$, and on the right shows our numerical results for $\langle\mathcal{O}_{F^2}\rangle_\mathrm{soliton}/(N\k)$ as a function of $\rho/R_0$, with $\k=v$, for various values of $v$. We can also obtain analytical results for $\langle\mathcal{O}_{F^2}\rangle_\mathrm{soliton}$ in various limits, as follows.

In the large-$(\rho v)$ limit, following arguments similar to those in ref.~\cite{Kumar:2017vjv}, we find
\begin{equation}
\label{eq:OF_large_rho}
\langle\mathcal{O}_{F^2}\rangle_{\text{soliton}} =  \frac{ N }{4 \pi^2} \frac{\kappa^2 + \kappa}{\rho^4}\,, \qquad \rho v \to \infty\,.
\end{equation}
Similarly to $S_\mathrm{soliton}$ in the large-$R v$ limit of eq.~\eqref{eq:EE_W_boson_linear}, $\langle\mathcal{O}_{F^2}\rangle_{\text{soliton}}$ includes contributions of order $\k^2$ and $\k$. The order $\k^2$ contribution to $\langle\mathcal{O}_{F^2}\rangle_{\text{soliton}}$ is identical to that of $\langle\mathcal{O}_{F^2}\rangle_{\text{screen}}$ at large $\rho v$ in eq.~\eqref{eq:O_Fsq_screened}. The order $\k$ contribution is precisely that of a Wilson line in a direct product of $k$ fundamental representations~\cite{Kumar:2017vjv}. We thus interpret the order $\k^2$ term as a contribution from the $U(1)$ sector, where the spherical soliton looks like a point charge, and the order $\k$ term as a contribution from the $SU(N-1)$ sector, where the spherical soliton looks like a Wilson line in the direct product of $k$ fundamental representations.

When we approach the origin inside the spherical soliton, $\rho v \to 0$, we find that $\langle{\cal O}_{F^2}\rangle_\mathrm{soliton}$ approaches a constant $\propto v^4$, unlike the screened Wilson line's $1/\rho^4$ behaviour in eq.~\eqref{eq:O_Fsq_screened}. The power of $v$ is fixed  by dimensional analysis, but comes with a coefficient that is a nontrivial function of $\kappa$,
\begin{equation}
\langle {\cal O}_{F^2}\rangle_\mathrm{ soliton}=
\frac{N}{64\pi^2}\frac{v^4 \kappa}{(1+\kappa^2)^7} \left[p_{10}(\kappa)-\frac{p_8(\kappa)}{\sqrt{1+\kappa^2}}\left(\sinh^{-1}\kappa +\sinh^{-1}\tfrac{1}{\kappa}\right)\right], \qquad \rho v \to 0\,,
\end{equation}
where $p_{8}(\k)$ and $p_{10}(\k)$ are degree eight and ten polynomials, respectively,
\begin{align}
p_8(\kappa) &= 105 \kappa^2 (40 \kappa^6- 204 \kappa^4+ 165 \kappa^2-20)\,,
\nonumber
\\
p_{10}(\kappa) &=  \kappa^{10}- 120 \kappa^9+ 1072 \kappa^8+ 
8790 \kappa^7 - 15624 \kappa^6- 25179 \kappa^5+ 
23380 \kappa^4\\\nonumber&\phantom{=} + 10572 \kappa^3- 4955 \kappa^2- 384 \kappa +30\,.
\end{align}
In the small and large $\kappa$ limits, we find,
\beq
\label{eq:solitonofsmallrhosmallkappa}
\langle\mathcal{O}_{F^2} \rangle_{\rm soliton}= \begin{cases} \dfrac{15N}{32\pi^2} v^4 \kappa\,, & \quad \kappa \to 0\,, \\[1em]  \dfrac{N}{4\pi^2} \dfrac{v^4} {\kappa^3}\,,& \quad\kappa \to \infty\,,
\end{cases} \qquad \qquad \rho v \to 0\,.
\eeq
As $\kappa \to 0$ the spherical soliton's charge and size vanish, and correspondingly $\langle\mathcal{O}_{F^2} \rangle_{\rm soliton}$ at $\rho v =0$ in eq.~\eqref{eq:solitonofsmallrhosmallkappa} vanishes as well. As $\kappa \to \infty$, the spherical soliton's radius $R_0  = \k/v \to \infty$ as we expect to recover the results of the $SU(N)$ $\N=4$ SYM conformal vacuum, where $\langle\mathcal{O}_{F^2} \rangle=0$. Indeed, in that limit $\langle\mathcal{O}_{F^2} \rangle_{\rm soliton}$ in eq.~\eqref{eq:solitonofsmallrhosmallkappa} vanishes.

As discussed in section~\ref{sec:EE_W_bos} for the EE, our results for $\langle\mathcal{O}_{F^2} \rangle_{\rm soliton}$ in general do not exhibit ``smoking gun'' features characteristic of an infinitely thin interface. However, figure~\ref{plot:OF} clearly shows that in general $\langle\mathcal{O}_{F^2} \rangle_{\rm soliton}$ has a maximum, consistent with a charge distribution peaked near $R_0$. In particular, figure~\ref{plot:OF} on the left shows that if $\k$ is small then $\langle{\cal O}_{F^2}\rangle_\mathrm{soliton}$ is peaked at $\rho=0$, and as $\k$ grows the peak moves to $\rho \neq 0$. Clearly, some critical value $\k_\mathrm{crit}$ exists where the peak first leaves $\rho=0$. We can determine $\k_\mathrm{crit}$ by computing $\partial^2/\partial\rho^2$ of $\langle{\cal O}_{F^2}\rangle_\mathrm{soliton}$ at $\rho=0$ and setting it to zero. This results in a transcendental equation for $\kappa$ with numerical solution $\k_\mathrm{crit} \approx 0.901$. A natural interpretation is that when $\k<\k_\mathrm{crit}$, the spherical soliton is a lump at $\rho=0$, and as we increase $\k$ through $\k_\mathrm{crit}$ the lump turns into an interface or bubble at $\rho \neq 0$. A natural question is whether any other observables exhibit similar behaviour.

As discussed at the end of section~\ref{sec:EE_W_bos}, when $\kappa \rightarrow \infty$ and $v \rightarrow \infty$ with $\kappa/v \equiv R_0$ fixed, the D3-brane becomes a cylinder of radius $R_0$ extending from the $AdS_5$ boundary to the Poincar\'e horizon. We saw that $S_\mathrm{soliton}$ then exhibited a divergence at $R = R_0$ and at $R \to \infty$ approached the result for a Wilson line in the direct product of $k$ fundamental representations. The same occurs in $\langle\mathcal{O}_{F^2} \rangle_{\rm soliton}$: in the same limits, we find
\begin{equation}
\label{eq:OF_v_inf}
\begin{split}
\langle\mathcal{O}_{F^2}\rangle_{\text{soliton}} & =   \kappa \, \frac{3 N }{16 \pi^2}  \int_{0}^\infty d z \, \frac{z^2}{\rho R_0 } \left\{ \left[ z^2 + \left( \rho - R_0 \right)^2\right]^{-\frac{5}{2}} - \left[ z^2 + \left( \rho + R_0\right)^2 \right]^{-\frac{5}{2}}\right\} \\
& = \kappa \, \frac{ N }{4 \pi^2}  \frac{1}{(\rho^2-R_0^2)^2} = k \, \frac{ \sqrt{\lambda} }{16 \pi^2}  \frac{1}{(\rho^2-R_0^2)^2} \,,
\end{split}
\end{equation}
which we show in figure~\ref{plot:OF} on the right as a black line. Clearly $\langle\mathcal{O}_{F^2}\rangle_{\text{soliton}}$ in eq.~\eqref{eq:OF_v_inf} diverges at $\rho=R_0$, and as $\rho \to \infty$ approaches $k$ times $\langle\mathcal{O}_{F^2}\rangle_{\square}$ in eq.~\eqref{eq:O_F_fund_rep}, as expected.

\subsection{Stress-energy tensor}

In this section we consider the one-point function of the stress-tensor, $\langle T_{mn}\rangle$, for the spherical soliton. To compute it we will need to find the linearised back-reaction of the D3-brane on the $AdS_5$ metric. This can be done via the method of ref.~\cite{DHoker:1999bve}. One obtains the linearised correction to the metric by integrating the graviton propagator against the probe D3-brane's stress-energy tensor, and then extracts $\langle T_{mn}\rangle$ from the metric's near-boundary behaviour, per the usual AdS/CFT dictionary~\cite{deHaro:2000vlm,Bianchi:2001kw}.\footnote{Note that we compute the D3-brane's linearised back-reaction on the metric only in a near-boundary limit, to extract $\langle T_{mn}\rangle$. Computing the D3-brane's contribution to EE using the RT formula requires computing the linearised back-reaction on the metric everywhere in the bulk, not just near the boundary.}

A subtlety is that the D3-brane also sources \(C_4\). In general, we would need solve for the linearised back-reaction of the D3-brane on both the metric and \(C_4\), since they satisfy coupled equations of motion (see e.g. ref.~\cite{Chang:2013mca}). We will avoid this issue by considering only the limits introduced at the end of section~\ref{sec:EE_W_bos}, $\kappa \to \infty$ and $v \rightarrow \infty$ with the spherical soliton's radius $R_0 \equiv \kappa/v$ fixed. In these limits the D3-brane becomes a cylinder of radius $R_0$ extending from the $AdS_5$ boundary to the Poincar\'e horizon, and only sources components of \(C_4\) orthogonal to the non-zero components in the background solution in eq.~\eqref{eq:C4_gauge_Poinc}. The linearised equations of motion for the metric and \(C_4\) then decouple, making it sufficient to consider the D3-brane's back-reaction on the metric only. Similarly to $S_\mathrm{soliton}$ and $\langle\mathcal{O}_{F^2}\rangle_{\text{soliton}}$, in these limits we expect $\langle T_{mn}\rangle$ to diverge at $R_0$, and as $\rho \to \infty$ to look like the result for a Wilson line in a direct product of $k$ fundamental representations, as computed in refs.~\cite{Friess:2006fk,Lewkowycz:2013laa}. We will see that is indeed the case.

Details of our computation appear in appendix~\ref{app:stress_tensor}. Here we merely present the result,
\begin{align}
\label{eq:stress_tensor_W}
\begin{split}
\left< T_{t_E t_E} \right> &= \k \,\frac{N}{2 \pi^2}  \frac{2}{3}\frac{1}{ \left(R_0^2-\rho ^2\right)^2}\,, \\
\left<T_{\rho \rho} \right> &=  \k \,\frac{N}{2 \pi^2}   \left(\frac{\log \frac{\left(R_0 + \rho\right)^2}{\left(R_0-\rho\right)^2}}{12
	\rho ^3 R_0}-\frac{1}{3 \rho ^2 \left(R_0^2-\rho ^2\right)}\right),  \\
\left<T_{\theta \theta} \right> &=  \k \,\frac{N}{2 \pi^2} \left(\frac{ \left(R_0^2-3 \rho ^2\right)}{6 \left(R_0^2-\rho ^2\right)^2}-\frac{\log \frac{\left(R_0 + \rho\right)^2}{\left(R_0-\rho\right)^2}}{24 \rho  R_0}\right), \\
\left<T_{\phi \phi}\right>  &=  \left<T_{\theta \theta}\right> \sin^2 \theta\,,
\end{split}
\end{align}
with all other components vanishing. This one-point function is finite as $\rho \rightarrow 0$,
\begin{equation}
\begin{aligned}
\left< T_{t_E t_E}\right> &=  \kappa   T_{D3}\left(\frac{2}{3 R_0^4} + \frac{4\rho ^2}{3 R_0^6}\right) + \mathcal{O}(\rho^4)\,, &  \left<T_{\rho \rho}\right>& = -\kappa   T_{D3}\left(\frac{4 \rho ^2}{15 R_0^6}+\frac{2}{9 R_0^4}\right) + \mathcal{O}(\rho^4) \,, \\
\left<T_{\theta \theta}\right> &= -\kappa     T_{D3}\frac{2  \rho ^2 }{9 R_0^4} + \mathcal{O}(\rho^4)\,, & \left<T_{\phi \phi}\right> &= \left<T_{\theta \theta}\right>\sin^2 \theta \,.
\end{aligned}
\end{equation}
However, at $\rho = R_0$ it diverges, as expected,
\begin{equation}
\begin{aligned}
\left< T_{t_E t_E} \right> &=   \frac{\kappa   T_{D3}}{6 R_0^2 (\rho -R_0)^2} + \mathcal{O}\left( \frac{1}{\rho-R_0} \right)\,, & \left< T_{\rho \rho} \right> &= \frac{\kappa   T_{D3}}{6 R_0^3 (\rho -R_0)} + \mathcal{O}\left[\log \left( \frac{1}{\rho - R_0} \right) \right]\,,\\
\left< T_{\theta \theta} \right> &= -  \frac{\kappa   T_{D3}}{12 (\rho -R_0)^2} +  \mathcal{O}\left( \frac{1}{\rho-R_0} \right)\,,&\left< T_{\phi \phi}\right>  &= \left< T_{\theta \theta} \right> \sin^2 \theta \, . 
\end{aligned}
\end{equation}

In the large distance limit $\rho \rightarrow \infty$, we expect to recover the result for $\langle T_{mn}\rangle$ in the presence of a one-dimensional conformal defect, which takes the form~\cite{Kapustin:2005py} 
\begin{align}
\label{eq:conformal_defect_stress_tensor}
\left< T_{t_E t_E} \right> &= \frac{h}{\rho^4}\,, & \left< T_{i j} \right> &= - h \frac{\delta_{ij}-2 n_i n_j}{\rho^4} \,, & \left< T_{t j} \right> = 0\,,
\end{align}
where $h$ is a constant and $n_i = x_i / \rho$ with $x_i$, $i=1,2,3$, the spatial Cartesian coordinates on the boundary. Indeed, from eq.~\eqref{eq:stress_tensor_W} we find the leading order behaviour at large \(\rho\)
\begin{align}
\left< T_{t_E t_E} \right> &= \frac{2}{3}   \frac{\kappa   T_{D3}   }{  \rho ^4} \,,& \left< T_{\rho \rho} \right> &=  \frac{2}{3}   \frac{\kappa   T_{D3}   }{  \rho ^4}    \,, & \left< T_{\theta \theta} \right> &=  -  \frac{2}{3}   \frac{\kappa   T_{D3}   }{  \rho ^2}\,, &  \left< T_{\phi \phi} \right> &= \left<  T_{\theta \theta}  \right>\sin^2 \theta\,, 
\end{align}
which agrees with eq.~\eqref{eq:conformal_defect_stress_tensor} after changing from spherical to Cartesian coordinates, with \(h=2\kappa T_\mathrm{D3}/3\). Using \(\k = k \sqrt{\lambda}/4N\) and \(T_\mathrm{D3} = N/2\pi^2\), this becomes
\begin{equation}
k \frac{\sqrt{\lambda}}{12 \pi^2} = k\,  h_\square\,,
\end{equation}
where \(h_\square =\sqrt{\lambda}/12\pi^2\) is the value of \(h\) for a fundamental representation Wilson line at large \(N\) and strong coupling~\cite{Friess:2006fk,Lewkowycz:2013laa}.

We have thus shown that in the limits $\kappa \to \infty$ and $v \rightarrow \infty$ with $R_0 \equiv \kappa/v$ fixed, all three of $S_\mathrm{soliton}$, $\langle\mathcal{O}_{F^2}\rangle_{\text{soliton}}$, and $\langle T_{mn}\rangle$ diverge at $R_0$ and at $\rho \to \infty$ approach the result for a Wilson line in the direct product of $k$ fundamental representations. Our interpretation is that in these limits the spherical soliton becomes an infinitely thin shell of $U(1)$ charge at $R_0$ that at large distances looks like a point charge in the $U(1)$ sector and Wilson line in the direct product of $k$ fundamental representations in the $SU(N-1)$ sector.

\section{Summary and Outlook}
\label{sec:concl}

We used holography to compute the EE of a spherical region of radius $R$ in large-$N$, strongly coupled $\N=4$ SYM theory on the Coulomb branch where an adjoint scalar VEV $\propto v$ breaks the gauge group from $SU(N)$ to $SU(N-1) \times U(1)$. We also considered cases with a screened symmetric-representation Wilson line or a spherical soliton separating $SU(N)$ inside from $SU(N-1) \times U(1)$ outside. Each case is supersymmetric, BPS, and non-conformal. The objects are described holographically by probe D3-branes in $AdS_5 \times S^5$. We used the Karch-Uhlemann method~\cite{Karch:2014ufa}, i.e. Lewkowycz and Maldacena's generalised gravitational entropy~\cite{Lewkowycz:2013nqa} applied to probe branes, to compute the probe's contribution to the EE directly from the probe D3-brane action, without ever computing the D3-brane's back-reaction on the bulk fields.

For the Coulomb branch vacuum, we found the closed-form result for the EE, $S_{\rm Coul}$ in eq.~\eqref{eq:S_coul_analytic}. In this case, a fully back-reacted solution is available, so we could compare to a calculation using the RT minimal-area prescription, in a probe limit. We found perfect agreement. This agreement relied crucially on a certain boundary term in the probe D3-brane's action, our $\mathcal{S}^{(bdy)}_2$ in eq.~\eqref{eq:S_bdy_2_coul}, that was neglected by Karch and Uhlemann~\cite{Karch:2014ufa} and in previous calculations using their method~\cite{Vaganov:2015vpq,Kumar:2017vjv,Rodgers:2018mvq}. 

Our result for $S_{\rm Coul}$ vanishes when $Rv<1$ and is a monotonically decreasing function of $Rv$ for $Rv > 1$. Our $S_{\rm Coul}$ is consistent with the $a$-theorem~\cite{Cardy:1988cwa,Komargodski:2011vj}, its more stringent entropic cousin~\cite{Casini:2017vbe}, and the entropic area theorem~\cite{Casini:2016udt,Casini:2017vbe}. Our $S_{\rm Coul}$ also has a square root non-analyticity at $Rv = 1$, being continuous and monotonic in $Rv$ but having a discontinuous third derivative with respect to $Rv$ at $Rv=1$.

Such behaviour is strongly reminiscent of the potential experienced by a probe eigenvalue at a large-$N$ matrix model saddle point, where the eigenvalues are distributed along the branch cut of the matrix model's resolvent function.  On this branch cut, the force on a probe eigenvalue is vanishing (or the potential is flat). Outside the branch cut the force is non-zero and dictated by the non-analytic resolvent function. A natural question for future research is whether this similarity is more than coincidence and could be made precise, for example by generalising the calculation to include multiple branes, or groups of branes, including back-reaction where necessary. More directly, in the field theory on hyperbolic space we could attempt to compute an effective action for Coulomb branch VEVs using perturbative methods and/or supersymmetric localisation.

Similar to our result for $S_{\rm Coul}$, our numerical result for the screened Wilson line EE, $S_{\rm screen}$, was a monotonically decreasing function of $Rv$. In contrast, a previous attempt to calculate the same EE in ref.~\cite{Kumar:2017vjv} claimed to find a maximum. Crucially, ref.~\cite{Kumar:2017vjv} overlooked $\mathcal{S}^{(bdy)}_2$, the effect being the apparent, but false, maximum. This once again highlights the importance of $\mathcal{S}^{(bdy)}_2$. In this case, no known physical principle, such as a monotonicity theorem like the $c$- or $a$-theorems, requires the EE to be monotonic in $Rv$. Monotonicity theorems have been proven for RG flows on certain defects or impurities in CFTs~\cite{Affleck:1991tk,Friedan:2003yc,Jensen:2015swa,Casini:2016fgb,Casini:2018nym}. However, explicit examples of RG flows on line defects in $(3+1)$-dimensional CFTs show that the spherical EE does not have to be monotonic in $R$~\cite{Kobayashi:2018lil}. Moreover, in our case a \textit{bulk} RG flow, from $SU(N)$ to $SU(N-1) \times U(1)$, actually triggers the RG flow on our Wilson line. Our result raises the question of whether a monotonicity theorem for EE could be proven in such cases. At the very least, our results do not rule out the possibility of such a monotonicity theorem.

For the screened Wilson line we also computed the probe's contribution to the one-point function of the field theory's Lagrangian density, which we denoted $\langle\mathcal{O}_{F^2}\rangle_\mathrm{screen}$. At large distances the result had the same form as a point charge in Maxwell theory. This revealed more detailed information about the screening, namely that what appears in the UV as a Wilson line of $SU(N)$ appears in the IR as a point charge in the $U(1)$ sector, while being completely absent, i.e. screened, in the $SU(N-1)$ sector.

Our numerical result for the EE of the spherical soliton, $S_{\rm soliton}$ in eq.~\eqref{eq:eesol}, has a maximum just outside the soliton's radius, $R_0 = \kappa/v$, where $\kappa$ determines the soliton's $U(1)$ charge. Here again $\mathcal{S}^{(bdy)}_2\neq 0$ is crucial: without it, the EE appears to diverge as $R \to R_0^+$. Schwarz in refs.~\cite{Schwarz:2014rxa,Schwarz:2014zsa} proposed that the spherical soliton is an infinitesimally thin shell, i.e. a phase bubble or domain wall. In general we did not find clear evidence for an infinitesimally thin shell. At least, our results were consistent with a shell of finite thickness around $R_0$. In fact, we computed $\langle\mathcal{O}_{F^2}\rangle_\mathrm{soliton}$ in this case, and found a critical value $\k_\mathrm{crit}\approx 0.901$ such that when $\k< \k_\mathrm{crit}$ the spherical soliton appears to be a lump localised at the origin, rather than a shell, bubble, or domain wall. A natural question is whether VEVs of other single-trace operators display similar behaviour. However, in the limits $v \to \infty$ and $\k \to \infty$ with $R_0$ fixed, we found that all three of $S_{\rm soliton}$, $\langle\mathcal{O}_{F^2}\rangle_\mathrm{soliton}$, and the stress-energy tensor diverge at $R_0$, consistent with an infinitesimally thin shell. This is nicely consistent with the form of the D3-brane solution in these limits. In particular, the D3-brane becomes a cylindrical shell with radius $R_0$ and $k$ units of string charge, stretching from the $AdS_5$ boundary to the Poincar\'e horizon.

Schwarz in refs.~\cite{Schwarz:2014rxa,Schwarz:2014zsa} also asked whether $S_{\rm soliton}$ at $R_0$ scales for large $\k$ with the surface area, i.e. like $R_0^2$, similar to a black hole's Bekenstein-Hawking entropy. Our numerical results suggest this is not the case: we found that $S_{\rm soliton}$ at $R_0$ scales for large $\k$ as $R_0^{1.3}$. Nevertheless, the spherical soliton has many features similar to an asymptotically flat extremal black hole, such as a mass and radius proportional to its charge and a spectrum of quasi-normal modes~\cite{Kumar:2020hif}. A natural question is thus to what extent this soliton, and other similar QFT objects~\cite{Popescu:2001rf,Forgacs:2003yh,Bolognesi:2005rk,Bolognesi:2010xt,Manton:2011vm}, can model aspects of black hole physics, and in particular whether any could indeed have EE that scales with their surface area.

A tantalizing fact is that in the limits $v \to \infty$ and $\k \to \infty$ with $R_0$ fixed, the probe D3-brane becomes a cylindrical shell with radius $R_0$ and $k$ units of string charge, stretching from the $AdS_5$ boundary to the Poincar\'e horizon. The D3-brane then resembles a supertube~\cite{Mateos:2001qs}, a cylindrical D2-brane solution with non-zero entropy whose microstates have been counted explicitly~\cite{Palmer:2004gu,Bak:2004kz}, and are simply BPS zero-mode deformations of the supertube's shape. A natural question is whether our D3-brane cylinder has similar microstates, and if so, do they contribute to $S_{\rm soliton}$, or any other entropy?

Our result for $S_{\rm Coul}$ at large $R$ has the form of a $(3+1)$-dimensional CFT with cutoff $1/v$. This makes sense, since in the IR the cutoff should be the inverse W-boson mass, which is indeed $\propto 1/v$. Remarkably, however, we find that the large-$R$ asymptotics of $S_{\rm screen}$ and $S_{\rm soliton}$ can be determined from $S_{\rm Coul}$ as follows. As mentioned in section~\ref{sec:intro}, the shape of the D3-branes in figures~\ref{fig:cartoons}c and d suggest that the W-boson mass, and hence the cutoff, become position-dependent. We thus introduce an effective value of $v$,
\begin{eqnarray}
v_{\rm eff}^\pm(R)\,=\,v\pm\frac{\kappa}{R}\,,
\end{eqnarray}
with the plus sign for the screened Wilson line and the minus sign for the spherical soliton. We found that, for $Rv \gg 1$,
\begin{equation}
\begin{split}
\label{eq:largeR}
&S_{\rm screen}(Rv) = S_{\rm Coul}( Rv_{\rm eff}^+)+\mathcal{O}\left((Rv)^{-2}\right)\,,\\
&S_{\rm soliton}(Rv) = S_{\rm Coul}( Rv_{\rm eff}^-)+k \,S_{\Box}+\mathcal{O}\left((Rv)^{-2}\right),
\end{split}
\end{equation}
with $S_{\Box}$ the EE of a conformal Wilson line in the fundamental representation, eq.~\eqref{eq:EE_fund_quark}. We thus find that at large $Rv$ both $S_{\rm screen}$ and $S_{\rm soliton}$ approach $S_{\rm Coul}$. Indeed, in figure~\ref{fig:cartoons} all three D3-branes look the same at large $Rv$. However, the behaviour in eq.~\eqref{eq:largeR} captures some of the $1/R$ corrections, some of which are difficult to explain otherwise. For example, although $S_{\rm Coul}$ at large $R$ has the form of a $(3+1)$-dimensional CFT, including an area law term $\propto (Rv)^2$ and a $\log (Rv)$ term, the corrections include a term $\propto Rv$ that does not look like a CFT or an impurity entropy, which would be independent of $R$. Eq.~\eqref{eq:largeR} captures the term $\propto (Rv)$, and others, in a simple and intuitive way.

For $S_{\rm soliton}$, we showed that the term $k \,S_{\Box}$ in eq.~\eqref{eq:largeR} comes from the part of the D3-brane that reaches the Poincar\'e horizon. Indeed, when $v \to \infty$ and $\kappa \to \infty$ with $R_0$ fixed, that part of the D3-brane becomes the cylindrical shell with charge $k$ and radius $R_0$. We found also that in these limits $\langle\mathcal{O}_{F^2}\rangle_\mathrm{soliton}$ and the stress tensor's one-point function at large distances take the form of $k$ charges of the $U(1)$ plus a Wilson line in a direct product of $k$ fundamental representations of $SU(N-1)$, consistent with eq.~\eqref{eq:largeR} in these limits.

Although we focused on probe D3-branes in $AdS_5 \times S^5$, the Karch-Uhlemann method, upgraded to include $\mathcal{S}^{(bdy)}_2$, is of course applicable to practically any probe object in any holographic space-time. We thus have an enormous number of possibilities for future research. An obvious starting point is to re-visit previous calculations using the Karch-Uhlemann method where $\mathcal{S}^{(bdy)}_2$ was neglected~\cite{Vaganov:2015vpq,Kumar:2017vjv,Rodgers:2018mvq}. More generally, RG flows and other forms of conformal symmetry breaking are typically easier to study using probe branes, whose equations of motion are usually easier to solve than Einstein's equation. As a result, the Karch-Uhlemann method, with $\mathcal{S}^{(bdy)}_2$ included, could potentially be used to address questions about EE, such as its behaviour with RG flows, defects, non-perturbative objects like baryons and solitons, and so on, and more broadly questions about its relation to monotonicity theorems, the emergence of probe actions from probe sectors of QFT, and much more. We intend to pursue these and many related questions in the future, using this paper as a foundation.

\section*{Acknowledgements}

We would like to thank Zohar Komargodski, Christoph Uhlemann and Konstantin Zarembo for useful discussions. A.~C. is supported by the Royal Society award RGF/EA/180098. S.~P.~K. acknowledges support from STFC grant ST/P00055X/1. A. O’B. is a Royal Society University Research Fellow. A.~P. is supported by SFI and the Royal Society award RGF/EA/180167. The work of R.~R. was supported by the D-ITP consortium, a program of the Netherlands Organisation for Scientific Research (NWO) that is funded by the Dutch Ministry of Education, Culture and Science (OCW). J.~S. is supported by the Royal Society award RGF/EA/181020.

\appendix

\section{HEE of the Coulomb Branch from Back-Reaction}
\label{app:EE_coul_backReac}

In this appendix we consider the Coulomb branch D3-brane introduced in section~\ref{sec:coul_branch}, and compute its contribution to the EE of a spherical region of radius \(R\), using the RT formula, eq.~\eqref{eq:Ryu_Tak}. More precisely, we consider the solution of type IIB supergravity describing the back-reaction of the D3-brane and calculate the EE using the RT formula in a probe limit. We then compare to our result computed directly in the probe limit eq.~\eqref{eq:S_coul_analytic} (rather than including back-reaction and then taking the probe limit). We find perfect agreement.

If we begin with $N$ coincident D3-branes, with low-energy worldvolume theory $SU(N)$ $\N=4$ SYM, then a generic point on the Coulomb branch is described by multiple separate stacks of D3-branes, say with $N_I$ D3-branes in each stack, breaking $SU(N) \to S\left[\prod_{I} U (N_I) \right]$ with $\sum_I N_I =N$. In this case, if we arrange the six adjoint scalars as a vector $\vec{\Psi}$, then
\begin{equation}
\langle\vec \Psi \rangle \propto \left( \begin{matrix}
\vec d_1 \, \mathbb{I}_{N_1} && 0 && \dots && 0 \\
0 && \vec d_2 \, \mathbb{I}_{N_2} && 0 && \dots \\
0 && 0 &&\ddots && 0 
\end{matrix} \right),
\end{equation}
with constants $\vec{d}_1$, $\vec{d}_2$, etc. The fully back-reacted solution of type IIB supergravity describing such generic points on the Coulomb branch is known~\cite{Klebanov:1999tb}. Its metric is
\begin{subequations}
	\begin{equation}
	d s^2 = H(\vec{y})^{-1/2} \h_{mn} d x^m d x^n + H(\vec{y})^{1/2} \delta_{ij} d y^i d y^j\,,
	\end{equation}
	\begin{equation}
	H(\vec{y}) \equiv 1 + 4 \pi g_s \a'^2 \sum_I \frac{N_I}{|\vec{y} - \vec{d}_I|^4}\,,
	\end{equation}
\end{subequations}
where $m,n$ label the four field theory directions, $i,j=1,\ldots,6$ and $g_s$ is the string coupling. We place \(N-M\) D3-branes at \(\vec{y}=0\) and \(M\) at \(\vec{y} = \vec{y}_0\). In the decoupling limit \(\a' \to 0\) with \(\vec{y}/\a'\) fixed, the metric becomes
\begin{align}
d s^2 &= L^{-2} \left(
\frac{N-M}{N|\vec{y}|^4} + \frac{M}{N|\vec{y} - \vec{y}_0|^4}
\right)^{-1/2} \h_{mn} d x^m d x^n
\nonumber \\ &\phantom{=}
+ L^2 \left( \frac{N - M}{N|\vec{y}|^4} + \frac{M}{N|\vec{y} - \vec{y}_0|^4} \right)^{1/2} \d_{ij} d y^i d y^j\,,
\label{eq:2stackmetric}
\end{align}
where we also used \(L^2 = \a' \sqrt{4 \pi g_s N}\). Let us switch to polar coordinates in the \(\vec{y}\) directions, with radius \(|\vec{y}| = r\) and polar angle \(\chi\), orienting the axes such that \(\vec{y}_0\) points in the \(\chi=0\) direction. We also define \(z = L^2/r\) and \(|\vec{y}_0| = Lv\). The metric in eq.~\eqref{eq:2stackmetric} then becomes
\begin{subequations}
	\begin{equation}
	\label{eq:back_reac_metric}
	d s^2 = \frac{L^2}{z^2} f(z,\chi)^{-\frac{1}{2}} \left(-d t^2 + d \r^2 + \r^2 d \Omega_2^2 \right) + L^2 f(z,\chi)^{\frac{1}{2}} \left[
	\frac{ d z^2}{z^2} +  d \chi^2 +  \sin^2 \chi \, d  \Omega_4^2
	\right],
	\end{equation}
	\begin{equation}
	f(z,\chi) \equiv 1 - \frac{M}{N} + \frac{M}{N} \left(1 + \frac{v^2 z^2}{L^2} - \frac{2 vz}{L} \cos\chi \right)^{-2}\,.
	\end{equation}
\end{subequations}

The RT surface is extended along \(z\), \(\chi\), and the $S^2$ and $S^4$, and thus is given by \(\r = \r(z,\chi)\). After integration over the $S^2$ and $S^4$, the area of this surface is
\begin{equation}
\label{eq:area_fun_back}
\mathcal{A} = \frac{32 L^8  \pi^3}{3}   \int_0^\pi d\chi  \int_\epsilon^{z_*} d z \frac{\r^2 \sin^4 \chi}{z^4} \sqrt{f(z,\chi)}
\sqrt{
	(\p_\chi \r)^2 +  z^2 \left[f(z,\chi) + (\p_z \r)^2 \right]
}\,,
\end{equation}
where $\epsilon$ is a small-$z$ cut-off, and $z_*$ denotes the maximal extension of the surface into the bulk, which in general will depend on $\chi$. 

We will work in the probe limit $\eta \equiv M/N \ll 1$. For \(\h = 0\), the metric~\eqref{eq:back_reac_metric} reduces to that of \(AdS_5 \times S^5\). The RT surface is given by \(\r =\sqrt{ R^2 - z^2}\)~\cite{Ryu:2006bv,Ryu:2006ef}, whose maximal extension into the bulk is $\left.z_*\right|_{\eta=0}=R$. The \(\mathcal{O}(\eta)\) term in a small \(\eta\) expansion of the area functional eq.~\eqref{eq:area_fun_back} is the sum of three contributions, \( \mathcal{A}=\mathcal{A}_1+ \mathcal{A}_2 + \mathcal{A}_3\). Their origins are, respectively:
\begin{enumerate}
	\item $\mathcal{A}_1$ comes from the linearised change in the embedding \(\r(z,\chi)\) due to non-zero \(\eta\). The equation of motion for \(\r(z,\chi)\) reduces this contribution to a boundary term.
	\item $\mathcal{A}_2$ comes from the change in \(z_*\) due to non-zero \(\eta\).
	\item $\mathcal{A}_3$ comes from the explicit dependence of \(f(z,\chi)\) on $\eta$.
\end{enumerate}
We will consider each contribution in turn. 

\textbf{Contribution 1. }If we write \(\r(z,\chi) = \sqrt{R^2 - z^2} + \h \, \d\r(z,\chi) + \mathcal{O}(\h^2)\), then we find the change in the area from \(\d \r(z,\chi)\) to be
\begin{equation}
\begin{aligned}
\mathcal{A}_1 &= - \frac{32 \pi^3 L^8}{3 R} \int_0^\pi d \chi \int_\e^{z_*} dz \, \sin^4 \chi \, \p_z \left[ \left(\frac{R^2}{z^2} - 1 \right) \d\r(z,\chi) \right]\\
&= - \frac{32 \pi^3 L^8}{3 R} \int_0^\pi d \chi \sin^4 \chi \,  \left[ \left(\frac{R^2}{z^2} - 1 \right) \d\r(z,\chi) \right]_{z=\e}^{z=R}\,.
\label{eq:coulomb_branch_area_1}
\end{aligned}
\end{equation}
The prefactor of \(\d \r(z,\chi)\) vanishes at \(z=R\). If we require \(\d \r(z,\chi)\) to be regular, then the contribution to eq.~\eqref{eq:coulomb_branch_area_1} from \(z=R\) vanishes. On the other hand, the prefactor diverges as \(\e^{-2}\) near \(z=0\), and so we could get a contribution from small \(z\), depending on the near-boundary behaviour of \(\d\r(z,\chi)\). However, we find that the leading-order term in a near-boundary expansion of \(\d\r(z,\chi)\) is \(\mathcal{O}(z^3)\). Concretely, \(  \d \r(z,\chi) = - z^3 \frac{v}{L R} \cos \chi + \mathcal{O}(z^4) \), so the small-\(z\) contribution to~\eqref{eq:coulomb_branch_area_1} vanishes as well. Thus, \(\mathcal{A}_1 = 0\).

\textbf{Contribution 2.} The fundamental theorem of calculus implies that the change in eq.~\eqref{eq:area_fun_back} due to a small change in \(z_*\) is given by $\eta \frac{d z_*}{d \eta}$ times the \(\h=0\) integrand, evaluated at \(z = z_*|_{\eta=0} = R\). The \(\eta=0\) integrand is proportional to $\sqrt{R^2 - z^2}$, so \(\mathcal{A}_2 = 0\). It is perhaps unsurprising that this boundary term vanishes, since it arises purely from our choice of coordinates on the RT surface. In particular, since the area functional is diffeomorphism invariant we could instead choose coordinates in which the only boundary is the physical one at $z=0$, so we expect that all boundary contributions at $z_*$ must vanish.

\textbf{Contribution 3.} Since the first two contributions vanish, the \(\mathcal{O}(\eta)\) contribution to the entanglement entropy comes entirely from \(\mathcal{A}_3\). By expanding the integrand in eq.~\eqref{eq:area_fun_back} to linear order in $\eta$ and evaluating it on the unperturbed solution  $\rho(z) = \sqrt{R^2 - z^2}$, we find
\begin{align}
\mathcal{A}_3 = \frac{16 \pi^3 L^8 M v}{3 N R} \int_0^\pi d \chi \int_0^{R} dz &\frac{(2 R^2 - z^2)\sqrt{R^2 - z^2}}{z^2} \frac{\sin^4 \chi}{
	\left( L^2 + v^2 z^2 - 2 L v z \cos \chi \right)^2
}
\\
&\times
\Bigl\{
4 L (L^2 + v^2 z^2) \cos \chi - v z \left[4 L^2 + v^2 z^2 + 2 L^2 \cos(2\chi) \right]
\Bigr\}\,.\nonumber
\end{align}
The EE is then obtained by performing the integrals over $z$ and $\chi$. The result is
\begin{equation}
\label{eq:HEE_coul_back_reac}
S_{\mathrm{Coul}}= \frac{\mathcal{A}_3}{4 G_N} =\begin{cases}
0\,, \quad &R v < 1\,,
\\
\dfrac{2 }{3 } \, N M  \, \left[
3 \cosh^{-1} \left( R v \right) - \left(R v + \dfrac{2 }{R v} \right)\sqrt{ \left(R v \right)^2 -1 } 
\right], \quad & R v > 1\,,
\end{cases}
\end{equation}
where we have chosen units in which \(L\equiv1\). In these units, \(1/G_N = 2N^2/\pi^4\). When $M=1$, the result in eq.~\eqref{eq:HEE_coul_back_reac} agrees perfectly with the one obtained by employing the Karch-Uhlemann method eq.~\eqref{eq:EE_prob_brane_gen}, namely eq.~\eqref{eq:S_coul_analytic}.

\section{Modification of $C_4$ for $n \ne 1$}
\label{app_HEE_C4}

In this appendix we discuss the gauge choice that we require for the Ramond-Ramond 4-form $C_4$ to be non-singular at the hyperbolic horizon.

Consider $C_4$ as given by eq.~\eqref{eq:C4:hyp_gauge}. When $n=1$, the hyperbolic horizon is located at $\zeta=\zeta_h=1$, and the first term in $C_4$ vanishes at the horizon. This behavior is a requirement for regularity at the horizon. If we change $n$, then the location of the horizon changes to $\zeta_h \ne 1$. This modification introduces a singularity in $C_4$ that must be cancelled.

To make the discussion above more precise, it is convenient to make the following coordinate transformation from $(\zeta,\tau)$ to $(x,y)$,
\beq
\label{eq:coordinates_hor}
x  = \sqrt{\zeta^2 - \zeta_h^2}\cos \tau\,, \qquad  y = \sqrt{\zeta^2 - \zeta_h^2}\sin \tau\, .
\eeq
Here we take $\zeta_h$ to be arbitrary, i.e. not necessarily equal to one. The hyperbolic horizon $\zeta=\zeta_h$ now corresponds to $x=0$ and $y=0$. Using
\beq
d\zeta = \frac{x \, dx}{\sqrt{x^2 + y^2 + \zeta^2_h}} +  \frac{y \, dy}{\sqrt{x^2 + y^2 + \zeta^2_h}}\,, \qquad d\tau = \frac{x dy}{x^2 + y^2} - \frac{y dx}{x^2 + y^2} \,,
\eeq
eq.~\eqref{eq:C4:hyp_gauge} becomes
\begin{equation}
\begin{split}
C_4  =\;& - x \frac{(x^2 + y^2 + \zeta^2_h)(x^2 + y^2 + \zeta^2_h-1)}{x^2 + y^2}\sinh^2 u \sin\theta \, du \wedge dy \wedge d\theta \wedge d\phi  \\
&+ y \frac{(x^2 + y^2 + \zeta^2_h)(x^2 + y^2 + \zeta^2_h-1)}{x^2 + y^2}\sinh^2 u \sin\theta \, du \wedge dx \wedge d\theta \wedge d\phi  \\
&+  \frac{\sinh^2 u \sin \theta (\sinh u - \cos \theta \cosh u)}{\cosh u -\cos \theta \sinh u} \, d x \wedge d y \wedge d\theta \wedge d\phi \\
&- \frac{\sinh u \sin^2 \theta}{\cosh u -\cos \theta \sinh u} \, dx \wedge dy \wedge du \wedge d\phi\,.
\end{split}
\end{equation}
If $\zeta_h\neq1$, then the first two terms are singular at the horizon $x=0$ and $y=0$. Specifically, the singular terms are
\begin{equation}
\begin{split}
\left. C_4 \right|_\mathrm{sing}   =\;&  - x \frac{\zeta^2_h(\zeta^2_h-1)}{x^2 + y^2}\sinh^2 u \sin\theta \, du \wedge dy \wedge d\theta \wedge d\phi 
\\
&+   y \frac{ \zeta^2_h( \zeta^2_h-1)}{x^2 + y^2}\sinh^2 u \sin\theta \, du \wedge dx \wedge d\theta \wedge d\phi \,,
\end{split}
\end{equation}
which in terms of $(\zeta,\tau)$ become
\begin{equation}
\label{eq:C_4_sing}
\begin{split}
\left. C_4 \right|_\mathrm{sing}   =   -\zeta_h^2(\zeta^2_h-1)\sinh^2 u \sin\theta \, du \wedge d\tau \wedge d\theta \wedge d\phi  \,.
\end{split}
\end{equation}
Crucially, $\left. C_4 \right|_\mathrm{sing}$ is exact everywhere but at the horizon, so we can perform a (singular) gauge transformation to remove the singularity. Starting from the singular $C_4$ in eq.~\eqref{eq:C4:hyp_gauge}, we thus perform a gauge transformation that simply subtracts the $\left. C_4 \right|_\mathrm{sing}$ in eq.~\eqref{eq:C_4_sing}. The result is the non-singular $C_4$ in eq. \eqref{eq:C4:hyp_gauge_corr}, which we use throughout our calculations.

\section{Details of the HEE Computation}
\label{app_HEE}

In this section we discuss the integral $\mathcal{S}$ in eq.~\eqref{eq:contr_S} in more detail.  It consists of the sum of two pieces. The term with $\text{sign}(\partial_\zeta u)$ is a total derivative and thus straightforward to compute. This appendix is devoted to the other piece, 
\begin{equation}
\label{eq:int_S_1}
\mathcal{S}_1 \equiv  \frac{4\pi }{3}T_{D3}   \int d \zeta d\tau \, \frac{\left( (\partial_\zeta u)^2-\frac{(\partial_\tau u)^2}{ f^2_1(\zeta )}\right)
	\sqrt{ \kappa^2+\zeta ^4  \sinh ^4 u}}{ \sqrt{1+\zeta ^2  (\partial_\zeta u)^2
		f_1(\zeta ) + \frac{\zeta ^2 (\partial_\tau u)^2}{f_1(\zeta )} }} \,,
\end{equation}
which in most of the cases needs to be evaluated numerically.
The general strategy is to make a coordinate transformation that maps the embedding from hyperbolic coordinates back to Poincar\'e coordinates. The embeddings are much simpler in the Poincar\'e patch, and the problem is more tractable. However, a subtlety arises as we will see shortly. For later convenience, we note the inverse transformation of eqs.~\eqref{eq:CHM_mapping},
\begin{subequations}
	\begin{align}
	\label{eq:inverse_CHM}
	\zeta &= \frac{\sqrt{R^4+2  R^2 \left(-\rho ^2+t^2+z^2\right)+\left(\rho ^2+t^2+z^2\right)^2}}{2 R z}\,, \\[1em]
	\sinh u &= \frac{2 \rho  R}{\sqrt{R^4+2 R^2 \left(-\rho ^2+t^2+z^2\right)+\left(\rho ^2+t^2+z^2\right)^2}}\,,\\[1em]
	\cos \tau &= \frac{R^2 -\rho ^2-t^2-z^2}{\sqrt{R^4+2 R^2 \left(-\rho ^2+t^2-z^2\right)+\left(\rho ^2+t^2+z^2\right)^2}}\,,
	\end{align}
\end{subequations}
where we have set $L\equiv1$, and for notational simplicity we drop the subscript \(E\) on the Euclidean time in this appendix.

Let us begin by considering the integral $\mathcal{S}_1$ in eq.~\eqref{eq:int_S_1}. For all the non-conformal embeddings that we study, its integrand diverges at the horizon $\zeta=1$ like $1/(\zeta^2-1)$. The divergence in the $\zeta$-integration can be easily regularised, e.g. by introducing a cut-off. Subsequently performing the integral over the Euclidean time $\tau$, one finds that the leading-order piece in $\zeta^2-1$ vanishes, and we can safely take the cutoff to zero. Nonetheless, this singular behaviour makes the numerics unstable in the hyperbolic coordinates. For this reason, our strategy will be to subtract and add the singular term. The integrand minus the singular piece will be computed numerically in Poincar\'e coordinates, while the additional singular term will be treated in hyperbolic coordinates as outlined above.

Employing the ansatz eq.~\eqref{eq:anasatz_exp_hor}, one can straightforwardly find the singular part of the integral $\mathcal{S}_1$ in eq.\eqref{eq:int_S_1} for a generic non-conformal embedding. The result is
\begin{equation}
\label{eq:sing_piece_gener}
\mathcal{S}_1^{(\mathrm{sing})}  = \frac{4 \pi }{3} T_{\text{D3}}\int d\tau d\zeta \, \frac{\sqrt{\kappa^2+\sinh ^4 u^{(0)}_1} \left(\left(u^{(1)}_1\right)^2-\left(\partial_\tau u^{(1)}_1\right)^2 \right)}{ \sqrt{2}  \sqrt{2+\left(\partial_\tau u_1^{(1)}\right)^2+\left(u^{(1)}_1\right)^2}}\frac{1}{\zeta^2-1}+\mathcal{O}\left(\frac{1}{\sqrt{\zeta^2 -1}}\right)\,.
\end{equation}
We now illustrate our method of subtraction of this singular part for the Coulomb branch D3-brane. The resulting numerical evaluation of the EE agrees perfectly with the analytical evaluation found in section~\ref{sec:EE_Coul_branch}. We will then give some details of the subtraction and numerical evaluation for the screened Wilson line and the spherical soliton.

\subsection{Coulomb branch}
\label{app_int_coul_branch}

Recall that the Coulomb branch D3-brane reaches the hyperbolic horizon if $R v > 1$. In this case, expanding the solution eq.~\eqref{eq:coul_branc_hyp_emb} near the horizon we find \(u_1^{(0)} = \cosh^{-1}(Rv)\) and \(u_1^{(1)}(\t) = - \sqrt{\frac{2}{(Rv)^2-1}} \cos \t\). The integral in eq.~\eqref{eq:sing_piece_gener} becomes
\begin{equation}
\label{eq:integrand_sing_coul_branch}
\mathcal{S}_1^{(\mathrm{sing})}  = \frac{4 \pi }{3} T_\text{D3}\frac{\sqrt{(R v)^2-1}}{R v}  \int_0^{2\pi} d\tau \int_{1}^{\tilde \zeta(\tau)} d \zeta\,  \frac{    \cos 2 \tau}{  \zeta^2 -1}\,,
\end{equation}
where $\tilde \zeta (\tau) = \frac{R v - \cos \tau \sqrt{(R v)^2 -\sin^2 \tau}}{\sin^2 \tau} $ is the upper limit on the integration over \(\z\), arising from the condition that \(\cosh u(\z,\t)\) in eq.~\eqref{eq:coul_branc_hyp_emb} is $\geq 1$. In order to deal with the singularity at $\zeta=1$, we introduce a small cut-off such that the lower limit of integration is $\zeta = 1+ \epsilon$. Evaluating eq.~\eqref{eq:integrand_sing_coul_branch} for non-zero \(\e\), and then sending \(\e \to 0\), we find $\mathcal{S}_1^{(\mathrm{sing})}  = 0$.

\begin{figure}
	\begin{center}
		\includegraphics[scale=0.4]{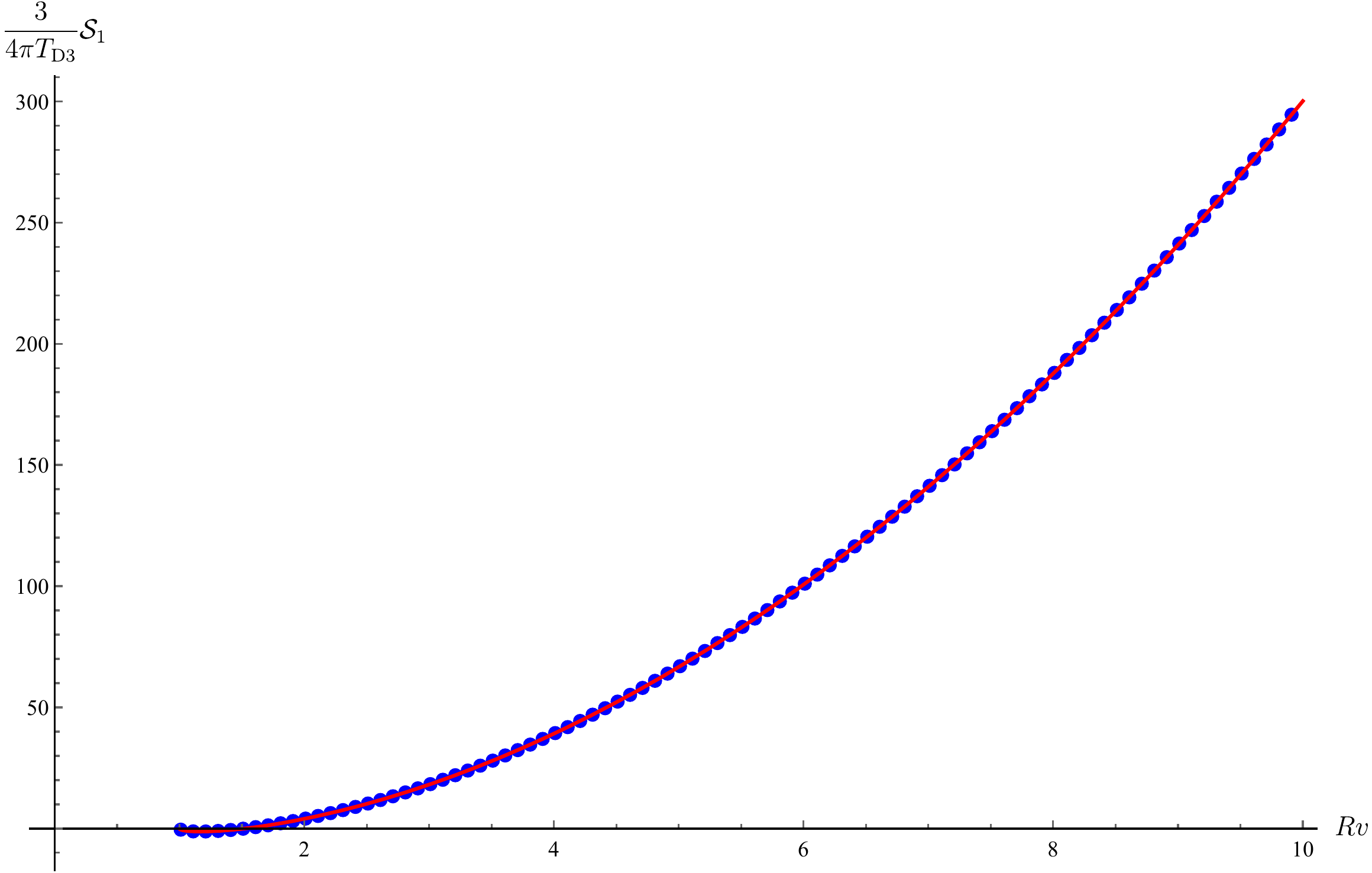}
		\caption{Comparison between the numerical integration of eq.~\eqref{eq:int_poinc_coul} (blue dots) and the analytical expression eq.~\eqref{eq:integr_an_coul} (red curve). The numerical values agree perfectly with the analytical result.  }
		\label{plot:coul_branch_int_check}
	\end{center}
\end{figure}

Now consider changing coordinates in \(\mathcal{S}_1\) from hyperbolic to Poincar\'e coordinates. The integral picks up the following Jacobian
\begin{equation}
\left| \frac{\partial \tau}{\partial \rho}  \frac{\partial \zeta}{\partial t} -  \frac{\partial \tau}{\partial t} \frac{\partial \zeta}{\partial \rho} \right|=\frac{\sinh u}{R \, z}\,.
\end{equation}
We transform the divergent piece eq.~\eqref{eq:integrand_sing_coul_branch} to Poincar\'e coordinates and subtract it from the first integrand in eq.~\eqref{eq:S_coul_branch}. The result is
\begin{eqnarray}
\label{eq:int_poinc_coul}
\mathcal{S}_1 = && \dfrac{4 \pi }{3} T_\text{D3}\int_0^{+\infty}  dt \, d\rho \,\, 16 R v \rho  \left\{ -\frac{\sqrt{\frac{(R v)^2-1}{{\cal F}}} \left({\cal F}-4 R^2 v^2 \left(2 t^2+1\right)\right)}{\left({\cal F}-4 R^2 v^2\right)^2} \right. \\ && \left. + \frac{2 R^3 v^3 \rho  \left(\frac{{\cal F} \left(-2 {\cal F}+f_-^2+f_+^2\right)}{4 R^2 v^2}+4 R^2 v^2 {\cal F}+{\cal F}^2-4 {\cal F}
	f_-+2 (3 {\cal F}-2 {\cal F} f_+)+f_-^2+f_+^2\right)}{{\cal F}^2 \left({\cal F}-4
	(R v)^2\right)^2}\right\}, \nonumber
\end{eqnarray}
where we have defined 
\begin{align}
f_+ &\equiv (R v+\rho )^2+t^2+1\,, & f_- &\equiv (R v-\rho )^2+t^2+1\,,     & {\cal F} &\equiv f_+ f_- \,.
\end{align}
Evaluating the \(t\) and \(\r\) integrals numerically, we obtain perfect agreement with the analytical result
\begin{equation}
\label{eq:integr_an_coul}
\mathcal{S}_1 = \frac{4 \pi^2 T_{\text{D3}}}{3}\frac{ \left(R^2 v^2-1\right)^{3/2}-R v \cosh ^{-1}(R v)}{R v}
\end{equation}
found in section~\ref{sec:EE_Coul_branch}. Concretely, in figure~\ref{plot:coul_branch_int_check} we show the numerical result (blue dots) and the analytical answer (red curve). Evidently they agree very well.

\subsection{Screened Wilson line}
\label{app_screend_W_line}

For the screened Wilson line we need to compute the following integral
\begin{equation}
\label{eq:I_1_Wils}
\begin{split}
\mathcal{S}_1 &= \frac{4\pi }{3} T_{\text{D3}}  \int d\tau \int d \zeta \left[\frac{\left( (\partial_\zeta u)^2-\frac{(\partial_\tau u)^2}{ (\zeta^2-1)^2}\right)
	\left[\kappa R v \cosh u + \sinh
	u (\kappa - \zeta  \sinh u)^2\right]}{ R v } \right],
\end{split}
\end{equation}
which is obtained by substituting the solution in eq.~\eqref{eq:screende_Wilson_embedding} into eq.~\eqref{eq:int_S_1}. In this case the integral in eq.~\eqref{eq:sing_piece_gener} becomes
\begin{equation}
\label{eq:integrand_sing_scren_W}
\mathcal{S}_1^{(\mathrm{sing})}  = \frac{4 \pi }{3}T_\text{D3}\mathbf{C}_{\mathrm{screen}}(\kappa,R v)  \int_0^{2\pi} d\tau \int_{1}^{\infty} d \zeta  \,\frac{    \cos 2 \tau}{  \zeta^2 -1}\,,
\end{equation}
where
\begin{align}
\mathbf{C}_{\mathrm{screen}}(\kappa,R v) &\equiv \frac{c_0^2 \sqrt{\kappa^2+\sinh ^4 u_0}}{ \sqrt{1+ c_0^2}}\,, &  c_0 &\equiv \frac{ R v \, \kappa}{R v \kappa \sinh u_0+\cosh u_0 (R v-\cosh u_0)^2}\,.
\end{align}
It is straightforward to show that again $\mathcal{S}_1^{(\mathrm{sing})}=0$.

The Jacobian for the transformation to Poincar\'e coordinates takes the form
\begin{equation}
\label{eq:jacob_rho}
\left| \frac{\partial \tau}{\partial \rho}  \frac{\partial \zeta}{\partial t} -  \frac{\partial \tau}{\partial t} \frac{\partial \zeta}{\partial \rho} \right|= \frac{(v z(\rho)-1)^2 \left| R v \, \kappa \cosh u +\sinh u (\kappa-\zeta  \sinh u)^2\right|}{R v \, \kappa^2  \, z(\rho)^2}\,,
\end{equation}
where $z=z(\rho)$ is the embedding as given by eq.~\eqref{eq:screenedsol} (recall that $r=1/z$ for $L\equiv 1$). By transforming the integral in eq.~\eqref{eq:I_1_Wils} to Poincar\'e coordinates and subtracting the singular part eq.~\eqref{eq:integrand_sing_scren_W}, we obtain
\begin{equation}
\label{eq:I_1_Wils_Poinc}
\mathcal{S}_1 = \frac{8 \pi }{3} T_{\text{D3}}\int_0^{+\infty} dt\, d\rho \, \bigg[ F_{\kappa,R v}(\rho,t) - F^{\mathrm{sing}}_{\kappa,R v}(\rho,t)  \bigg]\,.
\end{equation}  
The functions $F_{\kappa,R v}$ and $F^{\mathrm{sing}}_{\kappa,R v}$ are 
\begin{equation}
\begin{split}
F_{\kappa,R v} (\rho,t) \equiv \;& 16 R^3 v^3 \kappa^3 \rho ^2 \left[\frac{\rho ^4 \left(\sqrt{4 R^2v^2 \kappa^2 \rho ^2+f_{\mathrm{screen}}^2}-2 R^2 v^2\right)^2}{f_{\mathrm{screen}}^2 \left(4 R^2 v^2 \rho ^2-f_{\mathrm{screen}}^2 (\rho +1)^2\right)^2}\right. \\
&+\frac{\rho ^4}{f_{\mathrm{screen}}^2 (\rho +1)^2 \left(4 R^2 v^2 \rho ^2-f_{\mathrm{screen}}^2 (\rho +1)^2\right)} \\ & \left.+\frac{R^2 v^2 \left(f_{\mathrm{screen}}^2 (\rho +1)^3-2 \rho ^2 \left(\rho  \sqrt{4 R^2 v^2 \kappa^2 \rho ^2+f_{\mathrm{screen}}^2}+2
	R^2 v^2\right)\right)^2}{f_{\mathrm{screen}}^4 (\rho +1)^2 \left(4 R^2 v^2 \rho ^2-f_{\mathrm{screen}}^2 (\rho +1)^2\right)^2}\right],
\end{split}
\end{equation}
\begin{equation}
\begin{split}
F^{\mathrm{sing}}_{\kappa,R v} (\rho,t) \equiv \;\mathbf{C}_{\mathrm{screen}}(\kappa,R v) \frac{4 R^3 v^3  \left[R^2 v^2 \left(t^2+1\right)+\rho ^2 \left(\kappa^2 (2 \rho
	+1)+\frac{1}{(\rho +1)^2}\right)\right]}{ (\rho +1)^2 \left(f_{\mathrm{screen}}^2-\frac{4 R^2 v^2 \rho ^2}{(\rho +1)^2}\right)^2 f_{\mathrm{screen}}} &  \\
\times\left(f_{\mathrm{screen}}^2-\frac{4 R^2 v^2 \left(2 R^2 v^2 (\rho +1)^2 t^2+\rho ^2\right)}{(\rho +1)^2}\right)& \,,
\end{split}
\end{equation}
where we defined
\begin{equation}
f_{\mathrm{screen}} \equiv \sqrt{\frac{2 R^2 v^2 \rho ^2 \left(\kappa^2 (\rho +1)^2 \left(t^2-1\right)+t^2+1\right)}{(\rho +1)^2}+R^4 v^4 \left(t^2+1\right)^2+\frac{\rho ^4 \left(\kappa^2 (\rho +1)^2+1\right)^2}{(\rho +1)^4}} \,.
\end{equation}
The results and plots in section~\ref{sec:EE_Screen_Wilson} were obtained by numerically evaluating the integral in eq.~\eqref{eq:I_1_Wils_Poinc}.

\subsection{Spherical soliton}
\label{app_W_boson}

\begin{figure}[t]
	\begin{center}
		\includegraphics[scale=.47]{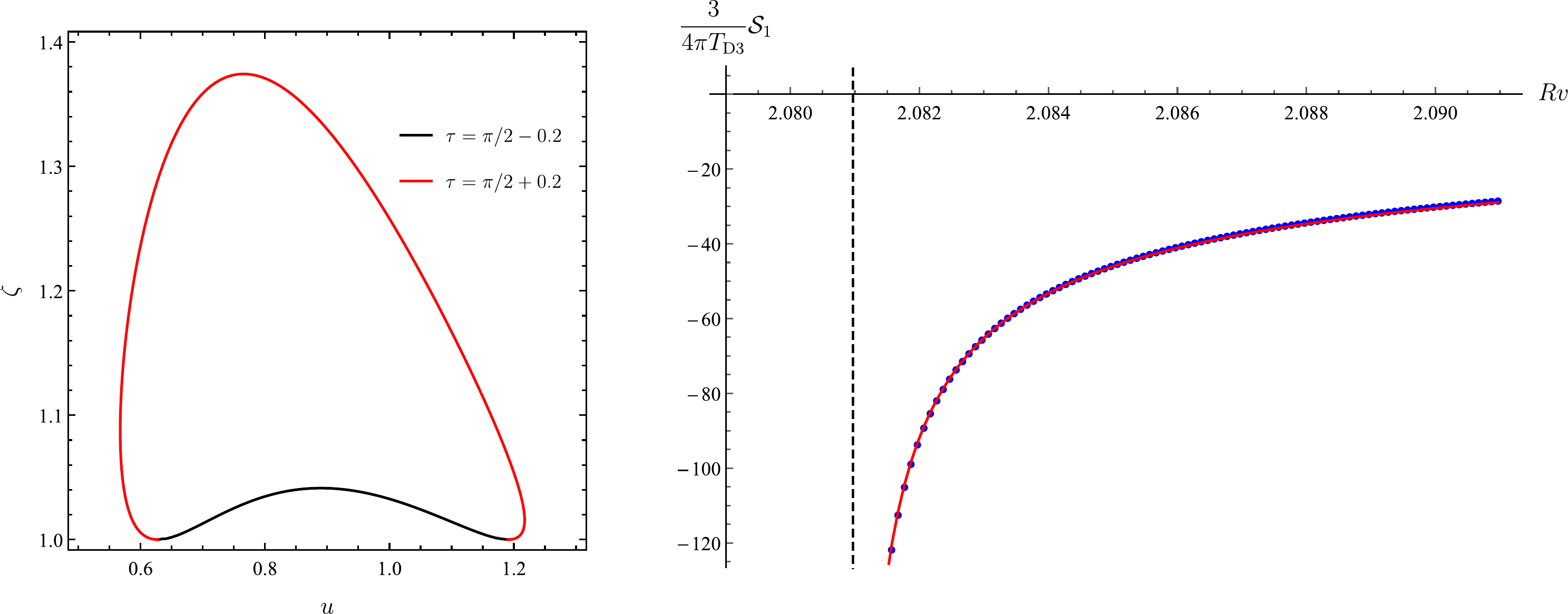}
		\caption{\textbf{Left}: Plot of the embedding eq.~\eqref{eq:embedding_W_bos} at $R v=3$ and $\kappa=1$. The two curves (red and black) correspond to two different values of $\tau$. The embedding reaches the horizon $\zeta=1$ at two different values, $u_1$ and $u_2$. \textbf{Right}: Behaviour of the integral $\mathcal{S}_1$ of eq.~\eqref{eq:int_W_bos_poinc} near $(Rv)_\mathrm{crit}$ at $\kappa=0.5$. The blue dots correspond to the numerical integration while the red curve represents the analytical function eq.~\eqref{eq:W_div_an}.  }
		\label{plot_W_bos_emb}
	\end{center}
\end{figure}

Finally, let us discuss the integral in eq.~\eqref{eq:integr_EE_W_bos} for the spherical soliton case. When $R v < (Rv)_\mathrm{crit}$, where $(Rv)_\mathrm{crit}$ is defined in eq.~\eqref{eq:R_v_lim}, the D3-brane does not reach the horizon $\zeta=1$. In this case, performing the integral numerically is straightforward. 

Let us therefore focus on the more subtle case $R v > (Rv)_\mathrm{crit}$. As anticipated in section~\ref{sec:EE_W_bos}, the D3-brane embedding has two disconnected boundaries located at the hyperbolic horizon $\zeta=1$, corresponding to the two loci where the D3-brane intersects the RT surface. In figure~\ref{plot_W_bos_emb}, we plot the embedding for $\kappa=1$ and $R v=3$ for two different values of $\tau = \pi/2 \pm 0.2$ (in red and black, respectively). The embedding reaches the horizon $\zeta=1$ at two distinct points, $u_1$ and $u_2$, which are independent of $\tau$. Their values are given by the solution of eq.~\eqref{eq:emb_at_hor}. Note also the qualitatively different behaviour of the two curves in the $(\zeta, u)$ plane. Finally, we observe that $u_1$ and $u_2$ approach the same point in the limit $R v \rightarrow (Rv)_\mathrm{crit}$. The reason is very intuitive: at the critical value the RT surface is exactly tangent to the D3-brane.

We now discuss the evaluation of the first integral in eq.~\eqref{eq:integr_EE_W_bos}, which reads
\begin{equation}
\label{eq:I_1_W_boson}
\begin{split}
\mathcal{S}_1 &= \frac{4\pi }{3} T_{\text{D3}}  \int d\tau \int d \zeta \left[\frac{\left( (\partial_\zeta u)^2-\frac{(\partial_\tau u)^2}{ (\zeta^2-1)^2}\right)
	\left|R v \, \kappa \cosh u - \sinh
	u (\kappa + \zeta  \sinh u)^2\right|}{ R v } \right].
\end{split}
\end{equation}
Applying the same procedure as above, we first find the singular part of the integrand. Since there are two distinct points at the horizon, there are two singular parts, namely
\begin{align}
\label{eq:integrand_sing_W_bos}
\mathcal{S}_i^{(\mathrm{sing})}  &= \frac{4 \pi }{3}T_\text{D3}\mathbf{C}^{(i)}_{\mathrm{soliton}}(\kappa,R v)  \int_0^{2\pi} d\tau \int_{1}^{\zeta_{\text{max}}(\tau)} d \zeta \, \frac{    \cos 2 \tau}{  \zeta^2 -1}\,, & i&=1,2\,.
\end{align}
Here $\zeta_{\text{max}}(\tau)$ is the largest value of $\zeta$ that satisfies the embedding equation, as illustrated on the left side of figure~\ref{plot_W_bos_emb}. It can be found numerically by solving $1/\partial_\zeta u =0$. The coefficients $\mathbf{C}^{(i)}_{\mathrm{soliton}}$ are given by
\begin{align}
\mathbf{C}^{(i)}_{\mathrm{soliton}}(\kappa,R v)&\equiv \frac{c_i^2 \sqrt{\kappa^2+\sinh ^4 u_i}}{ \sqrt{1+ c_i^2}} \,, & c_i &\equiv \frac{ R v \, \kappa}{R v \, \kappa \sinh u_i-\cosh u_i (R v-\cosh u_i)^2}\,.
\end{align}
Unlike the previous cases, the integrals $\mathcal{S}_i^{(\mathrm{sing})}$ are now non-zero. We straightforwardly find
\begin{equation}
\label{eq:integrand_sing_W_bos_eval}
\mathcal{S}_{\mathrm{tot}}^{\mathrm{sing}}\equiv \sum_{i=1,2}\mathcal{S}_i^{(\mathrm{sing})} =-\frac{4\pi }{3}T_{\text{D3}} \sum_{i=1,2}\mathbf{C}^{(i)}_{\mathrm{soliton}}(\kappa,R v) \int_ {0}^{2\pi} d\tau \,\tanh^{-1}\left[\zeta_ {\text {max}}(\tau)\right] \cos 2\tau\,. 
\end{equation}
Finally, we apply our usual trick: we subtract the singular piece in the form of eq.~\eqref{eq:integrand_sing_W_bos} and add it again in the form of eq.~\eqref{eq:integrand_sing_W_bos_eval}. Transforming the full integral and the subtracted piece to Poincar\'e coordinates, we find
\begin{equation}
\label{eq:int_W_bos_poinc}
\begin{split}
\mathcal{S}_1 = \frac{8\pi  L^4}{3}T_{D3} & \Bigg\{ \int_{0}^{+\infty} d t  \int_{0}^{\rho_*(t)}  d \rho \,  \bigg[ F_{\kappa,R v}(\rho,t) - F^{\mathrm{sing},(1)}_{\kappa,R v}(\rho,t)  \bigg] + \\
& + \int_{0}^{+\infty} d t  \int_{\rho_*(t)}^{+\infty}  d \rho \,  \bigg[ F_{\kappa,R v}(\rho,t) - F^{\mathrm{sing},(2)}_{\kappa,R v}(\rho,t)  \bigg] \Bigg\}+ \mathcal{S}_{\mathrm{tot}}^{\mathrm{sing}}\,.
\end{split}
\end{equation}
In the above, we defined
\begin{equation}
\begin{split}
F_{\kappa,R v}(\rho,t)=\;  &\frac{4 R v \kappa^3 (\rho +1)^2 \left(-\sqrt{4 R^2 v^2 \kappa^2 (\rho +1)^2+f_{\mathrm{soliton}}^2}+R^2 v^2+\kappa^2 (\rho +1)^2+\frac{(\rho +1)^2}{\rho ^2}\right)}{f_{\mathrm{soliton}}^2
	\left(\frac{f_{\mathrm{soliton}}^2 \rho ^2}{4 R^2 v^2 (\rho +1)^2}-1\right)^2}   \\
& + \frac{R v \kappa^3 \left(-2 (\rho +1)^2 \sqrt{4 R^2 v^2 \kappa^2 (\rho +1)^2+f_{\mathrm{soliton}}^2}+4 R^2 v^2 (\rho +1)+\frac{f_{\mathrm{soliton}}^2 \rho ^3}{\rho +1}\right)^2}{f_{\mathrm{soliton}}^4 \rho ^2
	\left(\frac{f_{\mathrm{soliton}}^2 \rho ^2}{4 R^2 v^2 (\rho +1)^2}-1\right)^2}\,,
\end{split}
\end{equation}
\begin{equation}
\begin{split}
F^{\mathrm{sing},(i)}_{\kappa,R v}(\rho,t) \equiv \frac{ 2 R^2 v^2 \rho ^2 \left(2 \kappa (\rho +1) \sqrt{\frac{f_{\mathrm{soliton}}^2}{4 R^2 v^2 \kappa^2 (\rho +1)^2}+1}-2 R v\right)^2+4 R^2 v^2 (\rho +1)^2-f_{\mathrm{soliton}}^2 \rho ^2}{4 R v f_{\mathrm{soliton}} (\rho +1)^3 \left(\frac{f_{\mathrm{soliton}}^2 \rho ^2}{4 R^2 v^2 (\rho +1)^2}-1\right)^2}&  \\
\times \left|2 \kappa^2 (\rho +1)^2-2 R v \kappa \sqrt{\frac{f_{\mathrm{soliton}}^2}{4 R^2 v^2 \kappa^2 (\rho +1)^2}+1}\right| \mathbf{C}^{(i)}_{\mathrm{soliton}}(\kappa,R v) &  \,,
\end{split}
\end{equation}
\begin{equation}
f_{\mathrm{soliton}} \equiv \sqrt{\frac{2 R^2 v^2 (\rho +1)^2 \left(\kappa^2  \rho ^2 \left( t^2-1\right)+ t^2+1\right)}{\rho ^2}+R^4 v^4 \left( t^2+1\right)^2+\frac{(\rho +1)^4 \left(\kappa^2  \rho ^2+1\right)^2}{ \rho ^4}}\,. 
\end{equation}
Note that the integral over $\rho$ in eq.~\eqref{eq:int_W_bos_poinc} splits into two parts, $(0,\rho_*(t))$ and $(\rho_*(t),+\infty)$. This is because we have two boundaries at the horizon with different $\mathbf{C}^{(i)}_{\mathrm{soliton}}(\kappa,R v)$. The value $\rho_*(t)$ corresponds to $\zeta_{\text{max}}$ in Poincar\'e coordinates, and it is given by the unique positive root of the equation
\begin{equation}
\label{eq_rho_star_eq}
R v \sqrt{\frac{f_{\mathrm{soliton}}^2}{4 R^2 v^2 \kappa^2 (\rho +1)^2}+1}- \kappa (\rho +1)^2=0 \,.
\end{equation}

Before concluding this appendix, let us discuss the behaviour of the integral eq.~\eqref{eq:int_W_bos_poinc} near the critical point $(Rv)_\mathrm{crit} $. The integral must be evaluated numerically, and we find that it diverges at $(Rv)_\mathrm{crit}$. Numerically, we see that the divergent piece reads 
\begin{equation}
\label{eq:W_div_an}
\mathcal{S}_1^{(div)}= -\frac{\sqrt{\frac{2}{3}} \pi  \left(\kappa^{2/3}+1\right)^{5/4} \kappa^{2/3}}{\sqrt{R-(Rv)_\mathrm{crit}}} + \dots \, .
\end{equation}
On the right-hand side of figure~\ref{plot_W_bos_emb} we plot the numerical result of the integral eq.~\eqref{eq:int_W_bos_poinc} (blue dots) for $\kappa=0.5$ close to $(Rv)_\mathrm{crit}$. The red curve is given by eq.~\eqref{eq:W_div_an}. We see that the numerics agrees very well with the expression for the divergent behaviour. Eq.~\eqref{eq:W_div_an} is the divergence we write in eq.~\eqref{eq:S_W_bos_diver}.

Notice that the divergence in eq.~\eqref{eq:W_div_an} matches exactly the one coming from the boundary term $\mathcal{S}^{(bdy)}_2$ in eq.~\eqref{eq:I_2_bdy_W_bos_diver}. This demonstrates the cancellation of the divergences at $(Rv)_\mathrm{crit}$ once all the contributions, i.e. $\mathcal{S}_1^{(bdy)}$, $\mathcal{S}_2^{(bdy)}$ and $\mathcal{S}$, are taken into account, as mentioned below eq.~\eqref{eq:I_2_bdy_W_bos_diver}.

\section{Stress-Energy Tensor One-Point Function}
\label{app:stress_tensor}

In this appendix we present details of the derivation of the one-point function of the stress tensor for the spherical soliton solution in the infinite $\kappa$ and $v$ limit whose embedding is eq.~\eqref{eq:emb_infinte_W}. Our strategy will be to find the leading-order back-reaction of the probe D3-brane on the Euclidean $AdS_5$ metric
\begin{equation}
\label{eq:metric_adsz}
ds^2 = \frac{1}{z^2} \bigg(d t_E^2 + d\rho^2  + \rho^2 d\theta^2+ \rho^2 \sin^2 \theta \, d\phi^2 + dz^2 \bigg)\,.
\end{equation}
From the linearised back-reacted metric $g_{\mu\nu}=g_{\mu\nu}^{(0)}+h_{\mu\nu}$ we may then extract the one-point function of the stress tensor by employing the standard AdS/CFT dictionary as outlined in ref.~\cite{deHaro:2000vlm}. To find the first-order correction we will use the following result of ref.~\cite{DHoker:1999bve}
\begin{equation}
\label{eq:h_linear}
h_{\mu\nu}(z) = \int d^{d+1}x' \, \sqrt{g} \, G_{\mu\nu ; \mu' \nu'}(x, x') T^{\mu' \nu'}(x')\, .
\end{equation}
Here, $G_{\mu\nu ; \mu' \nu'}$ is the graviton propagator. It takes the form
\begin{equation}
\label{eq:grav_prop}
G_{\mu\nu ; \mu' \nu'} = \left( \partial_\mu \partial_{\mu'} u \, \partial_\nu \partial_{\nu'}u + \partial_\mu \partial_{\nu'} u  \,\partial_\nu \partial_{\mu'}u\right) G(u) + g_{\mu\nu}g_{\mu'\nu'} H(u)\,,
\end{equation}
where the variable $u$ (not to be confused with the hyperbolic coordinate) is defined as 
\begin{eqnarray}
\label{eq:u_spherical_coord}
u &\equiv &\frac{(x-x')^2}{2 z z'} \\
& =&\frac{(t_E-t_E')^2 + (z-z')^2+\rho^2+\rho'^2-2\rho \rho' \left[ \cos(\theta-\theta')+\sin\theta \sin\theta' \left( \cos(\phi-\phi')-1\right) \right]}{2 z z'}\,. \nonumber
\end{eqnarray}
The partial derivatives $\partial_{\mu}$ and $\partial_{\mu'}$ are taken with respect to $x$ and $x'$, respectively. For $d=4$, i.e. the present case of interest, the functions $G(u)$ and $H(u)$ in eq.~\eqref{eq:grav_prop} are 
\begin{equation}
\begin{aligned}
G(u) &=  -\frac{1}{8 \pi^2} \left\{\frac{(1+u)\left(2(1+u)^2-3\right)}{\left(u(2 +u)\right)^\frac{3}{2}} -2\right\},\\
H(u) &= \frac{1}{12 \pi^2} \left\{\frac{(1+u)\left(6(1+u)^4-9 (1+u)^2+2\right)}{\left(u(2 +u)\right)^\frac{3}{2}} -6(1+u)^2\right\}.
\end{aligned}
\end{equation}

From the DBI part of the action eq.~\eqref{eq:D3brane_general}, we find the on-shell D3-brane stress tensor in the Euclidean space-time. In the $\kappa \rightarrow + \infty$ limit it reads
\begin{equation}
\sqrt{g}\, T^{\mu \nu} = T_{D3}
\left(\begin{array}{ccccc}
\kappa \sin \theta & 0 & 0 & 0 & 0 \\
0 & 0 & 0 & 0 & 0\\
0 & 0 & 0 & 0 & 0 \\
0 & 0 & 0 & 0 & 0 \\
0 & 0 & 0 & 0 & \kappa \sin \theta
\end{array} \right) \delta(\rho- R_0)\,,
\end{equation}
where $R_0 \equiv \kappa/v$. Since we are interested in the expansion for small $z$, it is convenient to re-express the integral in eq.~\eqref{eq:h_linear} in terms of the variable $v \equiv 1/u$ (not to be confused with the adjoint scalar VEV). Moreover, we change integration variables from $t_E'$ to $v$, which produces the following Jacobian
\begin{align}
\left|\frac{\partial t_E'}{ \partial v}\right|&=\frac{\sqrt{z z'}}{\sqrt{2}v^{3/2}}\frac{1}{\sqrt{1-v/v_m}}\,,  & v_m &\equiv \frac{2 z z'}{(\boldsymbol{x -x'})^2 + (z-z')^2}\,,
\end{align}
where $\boldsymbol{x}$ collectively denotes the three spatial field theory directions.

Let us illustrate the computation by considering the simplest piece, which is the one that contains $H(u)$. Explicitly,
\begin{equation}
I^H_{\mu\nu} = 2 g_{\mu\nu}\int d^4 x' \int_0^{v_m} dv \, \frac{\sqrt{z z'}}{\sqrt{2}v^{3/2}}\frac{ H(1/v) T}{\sqrt{1-v/v_m}}\,, \qquad T \equiv \sqrt{g} T^{\mu\nu}g_{\mu\nu} = \frac{2 T_{D3}\kappa}{z'^2} \delta(\rho'-R_0)\,.
\end{equation}
We now change variables once more to $\varpi = v/v_m $, and expand the integrand to 5th order in $v_m$. This corresponds to an expansion in small $z$. Performing the $\varpi$-integral gives
\begin{equation} \label{eq:IH_intermediate}
I^H_{\mu\nu} = \kappa T_{D3}\sqrt{z} \int_0^\pi d\theta'\sin \theta' \int_0^{2\pi} d\phi'\int_0^\infty dz' \, \frac{\tilde v_m^{3/2}}{z'^{3/2}} \frac{ (  35 \tilde v^3_m-15 \tilde v_m^2+6 \tilde v_m-4)}{96 \sqrt{2} \pi  }\,,
\end{equation}
where now 
\begin{equation}
\begin{split}
\tilde v_m & = \frac{2 z z'}{\rho^2+ R_0^2 -2 \rho R_0 [\cos(\theta-\theta')+ \sin \theta \sin \theta' (\cos(\phi-\phi')-1)] + (z-z')^2} \\
& =\frac{2 z z'}{\rho^2+ R_0^2 -2 \rho R_0 \cos\theta' + (z-z')^2}\,.
\end{split}
\end{equation}
We used spherical symmetry to set $\theta = 0$ in the second step. Evaluating the integrals in eq.~\eqref{eq:IH_intermediate}, we find
\begin{equation}
I^H_{\mu\nu} = -\kappa T_{D3} g_{\mu\nu}\left[\frac{ z^2 \log \left(\frac{(\rho +R_0)^2}{(R_0-\rho )^2}\right)}{12 \rho  R_0}+\mathcal{O}\left(z^6\right)\right].
\end{equation}
The contributions from the piece containing $G(u)$ in eq.~\eqref{eq:grav_prop} can be found in a similar way. The derivation is straightforward but tedious.\footnote{We found the following identity useful for evaluating those terms
	\begin{equation}
	\partial_\mu \partial_{\nu'} u = - \frac{1}{z z'} \left[ \delta_{\mu \nu'}+ \frac{1}{z'} (x-x')_\mu \delta_{\nu' z }+ \frac{1}{z}(x' - x)_{\nu'} \delta_{\mu z}- u \delta_{\mu z}\delta_{\nu' z} \right] .
	\end{equation}
}
We will skip it and only present the final result for the correction $h_{\mu\nu}$:
\begin{equation}
\begin{aligned}
\label{eq:metric_w_inf_firstform}
h_{t_E t_E} &= -\frac{\kappa   T_{D3} \log \left[\frac{(\rho
		+R_0)^2}{(R_0-\rho )^2}\right]}{12 \rho  R_0}+ \frac{\kappa   T_{D3} z^2}{\left(R_0^2-\rho ^2\right)^2} + \mathcal{O}(z^4)\,,\\
h_{\theta \theta} & =   - \kappa   T_{D3} \frac{ \rho \log \left[\frac{(\rho +R_0)^2}{(R_0-\rho )^2}\right]}{12 R_0} + \kappa   T_{D3} \frac{ \frac{4 \rho  R_0 \left(\rho ^2+R_0^2\right)}{\left(R_0^2-\rho
		^2\right)^2}+\log \left[\frac{(R_0-\rho )^2}{(\rho +R_0)^2}\right]}{16 \rho  R_0}z^2+ \mathcal{O}(z^4) \,,\\
h_{\phi\phi} & = h_{\theta \theta} \sin^2 \theta \,,\\
h_{\rho \rho} & =  -\kappa   T_{D3} \frac{\log \left[\frac{(\rho +R_0)^2}{(R_0-\rho
		)^2}\right]}{12 \rho  R_0} +\kappa   T_{D3} \frac{  \frac{\log \left[\frac{(\rho +R_0)^2}{(R_0-\rho
			)^2}\right]}{R_0}-\frac{4 \rho  \left(R_0^2-3 \rho ^2\right)}{\left(\text{R0}^2-\rho
		^2\right)^2}}{8 \rho ^3}z^2 + \mathcal{O}(z^4) \,,\\
h_{zz} & = \kappa   T_{D3}  \frac{ \log \left[\frac{(\rho +R_0)^2}{(R_0-\rho )^2}\right]}{6 \rho  R_0}-\kappa   T_{D3} \frac{8  }{3 \left(R_0^2-\rho ^2\right)^2}z^2 + \mathcal{O}(z^4) \,,\\
h_{\rho z} & = \kappa   T_{D3} \left(\frac{2}{\rho  \left(R_0^2-\rho ^2\right)}-\frac{\log \left[\frac{(\rho
		+R_0)^2}{(R_0-\rho )^2}\right]}{8 \rho ^2 R_0}\right)z + \mathcal{O}(z^3)\,.
\end{aligned}
\end{equation}
In order to straightforwardly obtain the stress tensor from results presented in ref.~\cite{deHaro:2000vlm}, we would like to switch to Fefferman-Graham gauge
\begin{equation}
ds^2 = \frac{1}{z^2} \Bigg[ dz^2 + \bigg(\delta_{mn} + h_{mn} (z ,\boldsymbol{x})\bigg) dx^m dx^n \Bigg]\,,
\end{equation}
where we assumed that the metric is time-independent. To this end we perform an infinitesimal diffeomorphism
\begin{equation}
h_{\mu\nu} \rightarrow h_{\mu\nu} + \nabla_\mu \xi_\nu + \nabla_\nu \xi_\mu
\end{equation}
to remove the off-diagonal terms $h_{\rho z}=h_{z\rho}$ and $h_{zz}$. Any $\xi_\mu$ with components
\begin{equation}
\begin{aligned}
\xi_\rho   &= a_\rho^{(2)}(\rho) \frac{z^2}{\rho^3} + \mathcal{O}(z^3)\,, \\
\xi_z  &= a_z^{(1)}(\rho) \frac{z}{\rho^2} + a_z^{(3)}(\rho) \frac{z^3}{\rho^4} + \mathcal{O}(z^4)\,,
\end{aligned}
\end{equation}
and 
\begin{equation}
\begin{aligned}
a_z^{(1)}(\rho) &=- \kappa   T_{D3} \frac{\rho   \log \left[\frac{(\rho +R_0)^2}{(R_0-\rho )^2}\right]}{24 R_0}\,,\\
a_\rho^{(2)}(\rho)  &= \frac{1}{48} \kappa   T_{D3} \rho   \left(\frac{4 \rho }{\rho ^2-R_0^2}+\frac{\log \left[\frac{(\rho
		+R_0)^2}{(R_0-\rho )^2}\right]}{R_0}\right)\,, \\
a_z^{(3)}(\rho)  &=\frac{1}{3} \kappa   T_{D3}  \frac{\rho ^4}{\left(R_0^2-\rho ^2\right)^2}\,,
\end{aligned}
\end{equation}
will do the job. This transformation puts the metric in eq.~\eqref{eq:metric_w_inf_firstform} in the form
\begin{equation}
\begin{aligned}
\label{eq:metric_w_inf_FGform}
h_{t_E t_E} &= \frac{\kappa   T_{D3} }{3 \left(R_0^2-\rho ^2\right)^2} z^2 + \mathcal{O}(z^4) \,, \\
h_{\theta \theta} &=   \kappa   T_{D3}  \left(\frac{ \left(R_0^2-3 \rho ^2\right)}{12 \left(R_0^2-\rho ^2\right)^2}-\frac{\log \left[\frac{(\rho +R_0)^2}{(R_0-\rho )^2}\right]}{48 \rho  R_0}\right) z^2 + \mathcal{O}(z^4)\,, \\
h_{\phi \phi} &= h_{\theta \theta}\sin^2 \theta\,, \\
h_{\rho \rho} &= \kappa   T_{D3}   \left(\frac{\log \left[\frac{(\rho +R_0)^2}{(R_0-\rho )^2}\right]}{24
	\rho ^3 R_0}-\frac{1}{6 \rho ^2 \left(R_0^2-\rho ^2\right)}\right) z^2 + \mathcal{O}(z^4)\,.
\end{aligned}
\end{equation}
The field theory stress-energy tensor is then given by~\cite{deHaro:2000vlm}
\begin{equation}
\langle T_{mn}\rangle = 2 \, h_{mn}^{(4)}\,,
\end{equation}
where we have written the metric as
\begin{equation}
h_{mn} (z, \boldsymbol{x})= h^{(0)}_{mn}(\boldsymbol{x})+ h^{(2)}_{mn}(\boldsymbol{x})z^2 + h^{(4)}_{mn}(\boldsymbol{x}) z^4 + \mathcal{O}(z^6)\,.
\end{equation}
Substituting eq.~\eqref{eq:metric_w_inf_FGform}, we thus find the result quoted in the main text in eq.~\eqref{eq:stress_tensor_W}.

\bibliographystyle{jhep}
\bibliography{biblio}

\end{document}